\documentclass[fleqn,usenatbib]{mnras}

\usepackage{newtxtext,newtxmath}

\usepackage[T1]{fontenc}

\DeclareRobustCommand{\VAN}[3]{#2}
\let\VANthebibliography\thebibliography
\def\thebibliography{\DeclareRobustCommand{\VAN}[3]{##3}\VANthebibliography}

\usepackage{graphicx}	
\usepackage{amsmath}	
\usepackage{xcolor}
\usepackage{caption}
\usepackage{xspace}

\usepackage{enumitem}
\setlist[itemize]{leftmargin=*, labelsep=0.5em}

\newcommand{\noopsort}[1]{}

\usepackage{placeins}

\newcommand{\Msun}{\ensuremath{\mathrm{M}_{\sun}}\xspace}
\newcommand{\Teff}{\ensuremath{T_{\mathrm{eff}}}\xspace}
\newcommand{\logh}{\ensuremath{\log{(\text{H}/\text{He})}}\xspace}
\newcommand{\cms}{cm\,s\textsuperscript{$-2$}}
\newcommand{\kms}{km\,s\textsuperscript{$-1$}}
\newcommand{\SN}{\ensuremath{\text{S}/\text{N}}\xspace}
\newcommand{\pwd}{\ensuremath{p_\mathrm{WD}}\xspace}
\newcommand{\pwdDESI}{\ensuremath{p_\mathrm{WD}^\mathrm{DESI}}\xspace}
\newcommand{\pwdGaia}{\ensuremath{p_\mathrm{WD}^\mathrm{Gaia}}\xspace}
\newcommand{\pbarwdGaia}{\ensuremath{\bar{p}_\mathrm{WD}^\mathrm{Gaia}}\xspace}
\newcommand{\bprp}{\ensuremath{G_{\mathrm{BP}}-G_{\mathrm{RP}}}\xspace}
\newcommand{\edit}[1]{\textcolor{black}{#1}}

\title[DESI DR1 white dwarf catalogue]{The DESI Data Release 1 white dwarf catalogue}

\author[A. Swan et al.]{Andrew Swan,$^{1}$\thanks{E-mail: andrew.swan.astro@gmail.com}
Boris~T.~G\"ansicke,$^{1}$
Paula~Izquierdo,$^{1}$
Detlev Koester,$^{2}$
Christopher~J.~Manser,$^{1,3}$
\newauthor
Laura K. Rogers,$^{4}$
Siyi Xu,$^{4}$
J.~Aguilar,$^{5}$
S.~Ahlen,$^{6}$
C.~Allende~Prieto,$^{7,8}$
L.~{Beraldo e Silva},$^{9}$
\newauthor
D.~Bianchi,$^{10,11}$
D.~Brooks,$^{12}$
F.~J.~Castander,$^{13,14}$
T.~Claybaugh,$^{5}$
A.~de la Macorra,$^{15}$
A.~Dey,$^{4}$
\newauthor
A.~Font-Ribera,$^{16,17}$
J.~E.~Forero-Romero,$^{18,19}$
E.~Gaztañaga,$^{13,14,20}$
N.~Gentile Fusillo,$^{21}$
\newauthor
S.~{Gontcho A Gontcho},$^{22}$
G.~Gutierrez,$^{23}$
J.~Guy,$^{5}$
K.~Honscheid,$^{24,25,26}$
K.~Inight,$^{1}$
M.~Ishak,$^{27}$
R.~Joyce,$^{4}$
\newauthor
S.~E.~Koposov,$^{28,29}$
M.~Landriau,$^{5}$
L.~Le~Guillou,$^{30}$
T.~S.~Li,$^{31}$
M.~Manera,$^{17,32}$
A.~Meisner,$^{4}$
\newauthor
R.~Miquel,$^{16,17}$
J.~Moustakas,$^{33}$
S.~Nadathur,$^{20}$
J.~Najita,$^{4}$
I.~P\'erez-R\`afols,$^{34}$
W.~J.~Percival,$^{35,36,37}$
\newauthor
F.~Prada,$^{38}$
G.~Rossi,$^{39}$
E.~Sanchez,$^{40}$
D.~Schlegel,$^{5}$
M.~Schubnell,$^{41,42}$
H.~Seo,$^{43}$
J.~Silber,$^{5}$
\newauthor
D.~Sprayberry,$^{4}$
G.~Tarl\'{e},$^{44}$
B.~A.~Weaver,$^{4}$
R.~Zhou,$^{5}$
H.~Zou,$^{45}$
\\
}

\date{Accepted 2026 September 2. Received 2026 September 1; in original form 2026 June 12}

\pubyear{\the\year{}}

\begin{document}
\label{firstpage}
\pagerange{\pageref{firstpage}--\pageref{lastpage}}
\maketitle

\begin{abstract}
The Dark Energy Spectroscopic Instrument is conducting a redshift survey, mapping the universe in three dimensions to measure the history of cosmic expansion. As well as extragalactic objects, it is targeting millions of Milky Way stars, including white dwarfs. Using Data Release~1 we assemble the largest catalogue of spectroscopically-confirmed white dwarfs to date, whose straightforward selection function enables statistically-robust population studies. We visually inspect and fit models to spectra of 63\,968 objects, finding 44\,409 white dwarfs. We present their spectral classifications, atmospheric parameters and radial velocities. We assess survey completeness and uniformity, identify potential spectral contamination caused by flux from nearby sources entering the fibre, and assign confidence scores to our classifications to facilitate selection of statistical and observational samples. We present spectra representing most white dwarf classes, common and exotic. We conclude with recommendations and warnings regarding the use of the catalogue.
\end{abstract}








\section{Introduction}
\label{sectionIntro}

White dwarfs are dense stellar remnants, the evolutionary endpoints of the vast majority of stars. They are central to many branches of astronomy, being excellent calibration sources \citep{Bohlin2014,GentileFusillo2019infrared,Elms2024}, reliable chronometers \citep{Winget1987,Fontaine2001}, frequent participants in binary evolution \citep{Paczynski1976,Belloni2023}, accretors \citep{Osaki1974, Norton2008}, progenitors of supernovae \citep{Shen2018,Munday2025SNprogenitor} with important cosmological applications \citep{Riess1998,Perlmutter1999}, probes of fundamental physics \citep{Bainbridge2017,Burdge2019sevenMinutes,Hollands2023magnetic,Alberino2026}, and hosts to planetary systems \citep{Gansicke2012,Jura2014review,Veras2016,Veras2021,Williams2024,Vanderburg2020planet}.

As Earth-sized stars that are continually cooling, white dwarfs are faint and historically have been challenging to observe. Substantial samples were identified from proper-motion catalogues \citep{Giclas1980} and objective-prism spectroscopy of faint blue objects \citep{Reimers1990}. The proliferation of fragmented samples inevitably spurred the creation of databases. The earliest contained just 166 objects \citep{Eggen1965}, but the tally exceeded 2000 when the Villanova catalogue was published \citep{McCook1999}. The Montreal White Dwarf Database \citep[MWDD;][]{Dufour2017MWDD} is two orders of magnitude larger, mostly owing to discoveries from the Sloan Digital Sky Survey (SDSS) and candidates identified using the \textit{Gaia} spacecraft.

SDSS revolutionised the field by delivering tens of thousands of spectroscopically confirmed white dwarfs, with the total exceeding 36\,000 by the 16th data release \citep{Kleinman2004, Eisenstein2006, Kleinman2013, Kepler2015, Kepler2016, Kepler2019, Kepler2021}. These facilitated statistical analyses of white dwarf properties, primarily masses \citep[e.g.][]{Tremblay2011}, but also magnetic field distributions \citep{Vanlandingham2005}. SDSS also led to many discoveries of rare classes of white dwarfs, including: those hosting circumstellar discs originating from rocky planetesimals or evaporating planets \citep{Gansicke2006,Gansicke2019}; those with atmospheres dominated by carbon or oxygen \citep{Dufour2007carbon, Gansicke2010, Kepler2016oxygen} that trace the extremes of white dwarf formation and evolution; the magnetic DAHe stars whose spectra show Zeeman-split Balmer lines in emission \citep{Gansicke2020DAHe}; ancient, ultra-cool stars with unusual colours caused by molecular interactions in their dense atmospheres \citep{Gates2004}; and cool DZ stars whose spectra reveal the detailed bulk compositions of accreted asteroids \citep{Koester2011}.

The \textit{Gaia} mission has further transformed the field, as its astrometric and photometric precision makes calculating absolute magnitudes straightforward, enabling identification of white dwarfs directly from the Hertzsprung--Russell diagram \citep[HR; ][]{Babusiaux2018}. White dwarf \textit{candidates} have been catalogued for \textit{Gaia}~DR2 \citep[][hereafter GF19]{GentileFusillo2019catalogue} and DR3 \citep[][hereafter GF21]{GentileFusillo2021}, and their tally now exceeds 359\,000. However, they require spectroscopic confirmation, as other objects (e.g.~quasars, subdwarfs and main-sequence stars) can be found in the same region of the \textit{Gaia} HR~diagram. Low-resolution \textit{Gaia} XP spectra are available for around 100\,000 of those objects and allow rough estimates of their properties \citep{Vincent2024}. However, detailed insight into the population of \textit{Gaia} white dwarfs has had to await the higher-quality spectroscopy being obtained by the current generation of large multi-object spectroscopic surveys: SDSS-V \citep{Kollmeier2025}, WEAVE \citep{Dalton2012,Jin2024}, 4MOST \citep{deJong2019} and DESI \citep{DESI2022, DESI2016science}.

The Dark Energy Spectroscopic Instrument (DESI) has conducted the largest white dwarf spectroscopic survey to date. Further, it provides homogeneous spectra with a reproducible selection function, due to its magnitude-limited nature and uniform instrumentation and data processing. The Early Data Release (EDR) white dwarf catalogue \citep{Manser2024DESIEDR} contains 2706 stars, of which about 60~per cent were newly confirmed. That dataset demonstrated the scientific potential of the survey, where its minimal bias enabled a measurement of the fraction of cool He-dominated white dwarfs that host planetary systems \citep{Manser2024HePlanets}.

The first major data release \citep[DR1;][]{DESI2025DR1} provides spectra for around 18\,700\,000 objects, among which we have identified over 44\,000 white dwarfs. In this paper, we present their spectral types, stellar parameters and radial velocities in the largest catalogue of spectroscopically-confirmed white dwarfs yet produced. We also showcase some applications of DESI data in studies of planetary material accreted by some of the most metal-enriched stars catalogued here \citep{Izquierdo2026DxZ}, the frequency of accretion of planetary material in wide binaries (Xu et al. submitted), the contribution of planetary hydrogen to white dwarf spectral evolution (Izquierdo et al. submitted) and cataclysmic variables discovered by machine learning techniques \citep{Inight2026}.

We also note recent work on subsets of the DR1 white dwarf population by independent teams unaffiliated with the DESI collaboration. One study uses supervised machine classification to assign spectral types to 21\,344 objects, and to identify binaries among them \citep{Munday2026}. Another two conduct a detailed model atmosphere analysis of 19\,321 white dwarfs at the younger and hotter end of the cooling track \citep{Kilic2026hot}\edit{, and 25\,642 stars at the older and cooler end \citep{Kilic2026cool}}. A fourth analyses 137 magnetic white dwarfs with Zeeman-split Balmer lines \citep{Amorim2026magnetic}\edit{, identified during classification of the whole population \citep{Amorim2026catalogue}}.

\section{Data}
\label{sectionData}

\subsection{The DESI survey}
\label{subsectionDESIsurvey}

The main science goal of DESI is to determine the nature of dark energy. To do so, it is mapping the three-dimensional structure of the universe, measuring baryon acoustic oscillations and redshift space distortions across multiple cosmological epochs via a spectroscopic survey of about 40~million galaxies and quasars \citep{DESI2016science}. Results so far include improved constraints on several cosmological quantities \citep{DESI2025cosmologicalConstraints}, and mounting evidence that dark energy is time-dependent, challenging the standard $\Lambda$CDM cosmological paradigm \citep{DESI2025DR2results1,DESI2025DR2results2}.

DESI is installed at the 4-m Mayall telescope on Kitt Peak, Arizona, where it can take over 100\,000 spectra per night under optimal conditions. The focal plane has a diameter of 3.2\,deg and hosts 5000 fibres, each controlled by a robotic positioner \citep{DESI2016instrument,Miller2024,Poppett2024}. These feed a bank of 10 identical spectrographs, covering 3600--{9824\,\AA} with resolving power $R\approx2300$--5500 \citep{DESI2022}. Exposures are taken simultaneously across all three spectrograph arms: B (blue), R (red) and Z (near-infrared). Each exposure delivers up to 15\,000 spectra across the 30 CCDs, and is reduced using a dedicated software pipeline \citep{Guy2023}, discussed further in Section~\ref{subsectionDESIspectra}. However, except where obvious from context, in this paper we use the term \textit{exposure} to refer to a per-target data product, i.e. the set of three reduced spectra from the B, R and Z~arms of the spectrograph fed by a single fibre.

In DESI terminology, there are several \textit{programs} that take spectra for various \textit{surveys}. Observing conditions are monitored in real time and determine which program will take the next exposure: in good conditions the \texttt{dark} or \texttt{bright} programs are selected, depending on lunar phase and position \citep{DESI2016science, Schlafly2023}, while the \texttt{backup} program takes over in poor weather or twilight \citep{Dey2025}. Following commissioning (\texttt{cmx}), three validation surveys (\texttt{SV1}, \texttt{SV2}, \texttt{SV3}) were undertaken \citep{DESI2024validation}. The main survey covers five target classes: the Milky Way Survey \citep[\texttt{MWS};][]{AllendePrieto2020, Cooper2023}, the Bright Galaxy Survey \citep[\texttt{BGS};][]{RuizMacias2020,Hahn2023}, luminous red galaxies \citep[\texttt{LRG};][]{Zhou2020,Zhou2022}, emission line galaxies \citep[\texttt{ELG};][]{Raichoor2020,Raichoor2023}, and quasars \citep[\texttt{QSO};][]{Yeche2020,Chaussidon2023}. A small amount of time is allocated to separate projects, e.g. the halo of M31 \citep{Dey2023M31}, all classified as \texttt{special} surveys under the \texttt{other} program. Cosmological targets are observed under the \texttt{dark} program, the \texttt{bright} program covers \texttt{BGS} and \texttt{MWS} targets, and the \texttt{backup} program focuses on \texttt{MWS} stars \citep{Myers2023, Dey2025}. The same object may be targeted by multiple surveys or programs. The basic observational unit of a program is a \textit{tile}, a defined assignment of fibres to targets. Tile centres are pre-defined sky locations, but fibre assignment is performed on the fly.

Each program has a different signal-to-noise ratio ($\SN$) goal, which is defined in terms of \textit{effective exposure time} under nominal conditions \citep{Guy2023}. The \texttt{backup}, \texttt{bright} and \texttt{dark} programs aim for 60\,s, 180\,s and 1000\,s, respectively \citep{Schlafly2023, Dey2025}. In sub-optimal conditions, a tile may require longer or multiple exposures to reach the effective exposure time target.

Commissioning and survey validation operations started on 2020 December 14 and ended on 2021 June 10. The DESI EDR comprises all data from the \texttt{cmx}, \texttt{SV1}, \texttt{SV2}, \texttt{SV3} and \texttt{special} programs during that period \citep{DESI2024EDR}. The main survey started on 2021 May 14 (thus overlapping slightly with survey validation) and was completed on 2026 April 15. DR1 includes all data up to 2022 June 13, when observations were halted by the Contreras fire \citep{DESI2025DR1}.

\subsubsection{White dwarf candidates observed by DESI}
\label{subsectionDESIwhiteDwarfs}

Our DR1 white dwarf catalogue contains a total of 63\,968 sources with 126\,610 exposures, which we identify using the WD\,J$xxxxxx.xx{\pm}xxxxxx.xx$ naming scheme introduced by GF19. Tables~\ref{tableSourceCounts} and \ref{tableExposureCounts} list their distribution between the various surveys and programs. We discuss below how targets were selected by DESI, how we identified objects for inclusion in our catalogue, and how to minimise bias when constructing samples from the catalogue.

\begin{table}
\centering
\caption{Source counts for white dwarf candidates observed by the various DESI surveys and programs. The same source may be observed by multiple
surveys or programs. The \texttt{main} row of the upper section corresponds to the primary sample.}
\label{tableSourceCounts}
\begin{tabular}{lrrrrr}
\hline
& \texttt{backup} & \texttt{bright} & \texttt{dark} & \texttt{other} & Total \\
\hline
\multicolumn{6}{l}{\textit{Sources targeted as white dwarfs:}}\\
\texttt{cmx} & 0 & 0 & 0 & 21 & 21 \\
\texttt{main} & 0 & 40152 & 8966 & 0 & 42839 \\
\texttt{special} & 0 & 380 & 12 & 0 & 386 \\
\texttt{sv1} & 587 & 551 & 1076 & 368 & 2406 \\
\texttt{sv2} & 0 & 263 & 285 & 0 & 317 \\
\texttt{sv3} & 0 & 842 & 681 & 0 & 868 \\
Total & 587 & 40951 & 10715 & 389 & 44579 \\
\\
\multicolumn{6}{l}{\textit{All sources:}}\\
\texttt{cmx} & 0 & 0 & 0 & 24 & 24 \\
\texttt{main} & 875 & 45169 & 22326 & 0 & 61178 \\
\texttt{special} & 8 & 396 & 43 & 18 & 458 \\
\texttt{sv1} & 597 & 614 & 1430 & 507 & 2964 \\
\texttt{sv2} & 2 & 271 & 427 & 0 & 461 \\
\texttt{sv3} & 83 & 1308 & 1361 & 0 & 1769 \\
Total & 1551 & 46390 & 24859 & 548 & 63968 \\
\hline
\end{tabular}
\end{table}

\begin{table}
\centering
\caption{Exposure counts for white dwarf candidates observed in the various DESI surveys and programs}
\label{tableExposureCounts}
\begin{tabular}{lrrrrr}
\hline
& \texttt{backup} & \texttt{bright} & \texttt{dark} & \texttt{other} & Total \\
\hline
\texttt{cmx} & 0 & 0 & 0 & 96 & 96 \\
\texttt{main} & 1512 & 53791 & 31655 & 0 & 86958 \\
\texttt{special} & 11 & 2557 & 109 & 57 & 2734 \\
\texttt{sv1} & 1217 & 6548 & 13454 & 5255 & 26474 \\
\texttt{sv2} & 2 & 986 & 653 & 0 & 1641 \\
\texttt{sv3} & 130 & 3982 & 4595 & 0 & 8707 \\
Total & 2872 & 67864 & 50466 & 5408 & 126610 \\
\hline
\end{tabular}
\end{table}

The selection function and observing strategy for white dwarfs in the Milky Way Survey \texttt{bright} program are described by \cite{Cooper2023}. The starting point is a list of sources identified in the DESI Legacy Survey, cross-matched (with proper-motion propagation) against \textit{Gaia}. Photometry is always drawn from \textit{Gaia}~DR2, while astrometry is usually taken from \textit{Gaia}~EDR3 \citep{Myers2023}. Selection criteria similar to but less restrictive than those used by GF19 are then applied to identify white dwarf candidates (see Section~4.4.1 of \citealt{Cooper2023} for details). \edit{They were observed with high priority in the \texttt{bright} program, but were also included at lower priority in the \texttt{dark} program.}

White dwarfs in DESI DR1 have also been \edit{used} as flux standards \edit{across all programs} \citep{Myers2023}\edit{, employing the same selection function as mentioned above. They have also been observed} as science targets under the \texttt{MWS backup} program, \edit{which has a separate} selection function \edit{that} captures stars of all types \citep{Dey2025}.

DESI DR1 contains a total of 44\,579 objects that were targeted as white dwarf candidates in any of the surveys. This information is encoded in their bitmasks, which are discussed extensively in the DESI online documentation\footnote{\url{https://desidatamodel.readthedocs.io/en/latest/bitmasks.html\#target-masks} and \cite{Myers2023, DESI2024EDR, DESI2025DR1}}. We extracted bitmask values from the calibrated exposure files\footnote{\url{https://desidatamodel.readthedocs.io/en/stable/DESI_SPECTRO_REDUX/SPECPROD/exposures/NIGHT/EXPID/cframe-CAMERA-EXPID.html}} (\texttt{cframe*.fits}, henceforth \textit{cframe files}) provided as part of DESI DR1, on which we base our work. We reproduce some of these header fields in supplementary Table~\ref{tableExposures} for convenience.

We also cross-matched all sources from all exposures in DESI DR1 against the GF19 and GF21 catalogues, using \textit{Gaia} designations or proper-motion-propagated coordinates within a 1.5-arcsec search radius. This yielded a further 13\,868 candidates not captured by the \texttt{MWS} white dwarf selection function. These objects were observed incidentally, having been targeted as e.g. \texttt{QSO} or main-sequence \texttt{MWS} objects, and thus have a different selection function to the white dwarf targets. Our catalogue is limited to objects with a \textit{Gaia} designation, as we did not attempt to identify white dwarfs that lack a \textit{Gaia} counterpart. To do so will likely require machine-learning tools, as DESI DR1 includes spectra for more than 18~million targets.

We found a further 5521 objects in the Milky Way Survey that were not targeted as white dwarfs, yet still appeared to be potential candidates. Specifically, we selected objects whose spectra were best fit by a white dwarf template according to the \texttt{MWS} stellar catalogue \citep{Koposov2026}, and which lay in the white dwarf region of the \textit{Gaia} HR~diagram (using equations 1 and 2 of GF21). More details are given in Appendix~\ref{appendixMWS}.

Our catalogue is inclusive, but heterogeneous. We therefore identify a subset with a straightforward and reproducible selection function, suitable for statistical analyses. It consists of the 42\,839 targets that have the \texttt{MWS\_WD} or \texttt{STD\_WD} bits set in their target masks and were observed under the \texttt{bright} or \texttt{dark} programs of the \texttt{main} survey. We refer to these as the \textit{primary sample}, and flag them in the \texttt{primary} column of our main catalogue (Table~\ref{tableMainCatalogue}).

\begin{figure}
\centering
\includegraphics[width=\columnwidth]{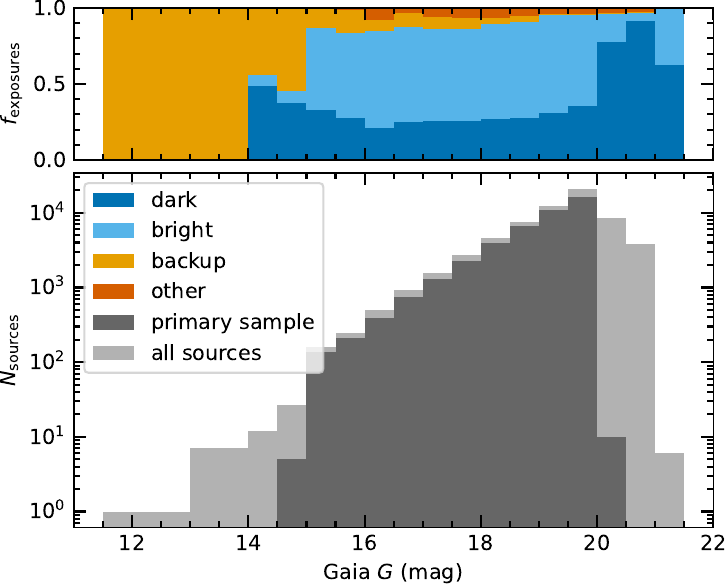}
\caption{\textit{Upper panel}: The fractional contributions of each program to exposures in each magnitude bin. \textit{Lower panel}: Magnitude distribution of white dwarf candidates in DESI DR1.}
\label{figureMagnitudes}
\end{figure}

\begin{figure}
\centering
\includegraphics[width=\columnwidth]{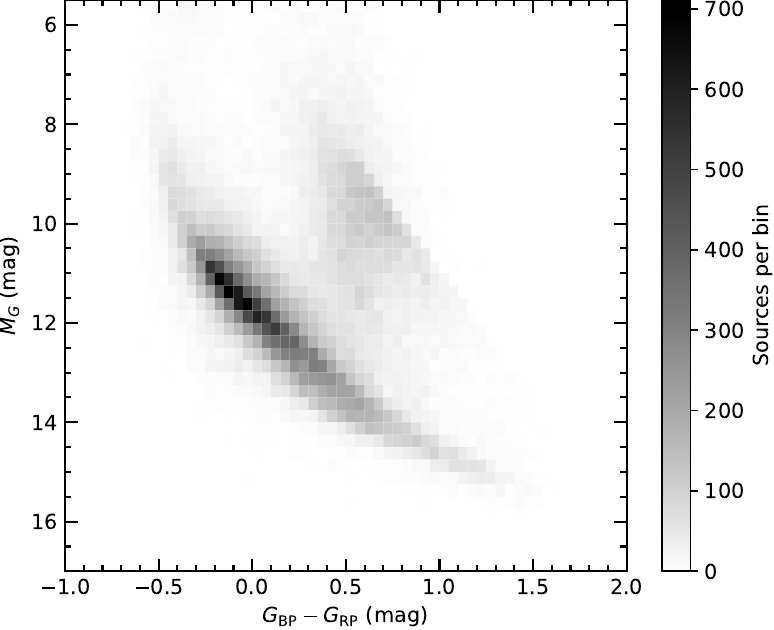}
\caption{Histogram showing density on the HR~diagram of white dwarf candidates in DESI DR1, using absolute \textit{Gaia G} magnitudes and \bprp colours.}
\label{figureHRDsourceDensity}
\end{figure}

Beyond the primary sample, almost all sources with $G\le15$ were observed solely by the \texttt{backup} program, and almost all with $G\geq20$ were observed solely by the \texttt{dark} program (Fig.~\ref{figureMagnitudes}). Not all of these candidates are genuine white dwarfs; for example, we classified the two brightest objects as a stellar contaminant (likely a subdwarf) and a white-dwarf--main-sequence binary (Section~\ref{sectionSpectralClassification}). Such objects are expected to lie above the main white dwarf track on the \textit{Gaia} HR~diagram, and indeed there is a population of objects observed in that region (Fig.~\ref{figureHRDsourceDensity}).

\begin{figure*}
\centering
\includegraphics[width=\textwidth]{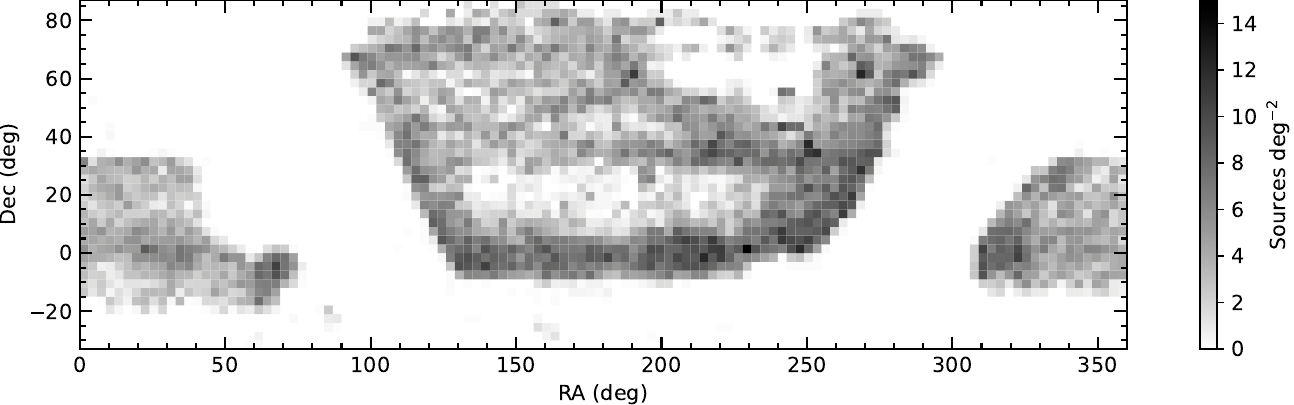}
\caption{Surface density of white dwarf candidates within DESI DR1. Bins are 3\,deg wide on each axis.}
\label{figureRAdecSourceDensity}
\end{figure*}

The sky surface density of observed white dwarf candidates (Fig.~\ref{figureRAdecSourceDensity}) broadly reflects the overall survey coverage at the DR1 cutoff date \citep{DESI2025DR1}. The relatively empty regions result from a combination of survey strategy and weather constraints, and will be filled in future data releases.

\begin{figure}
\centering
\includegraphics[width=\columnwidth]{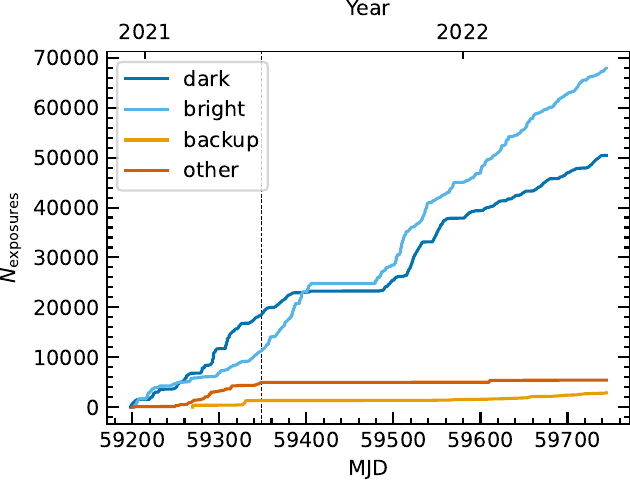}
\caption{Exposure accumulation for white dwarf candidates in DESI DR1, broken down by program. The vertical dashed line indicates the start of the main survey on 2021~May~14.}
\label{figureExposureTimeline}
\end{figure}

During commissioning and survey validation, white dwarfs were mostly observed under the \texttt{dark} program, but in the main survey they were observed more often by the \texttt{bright} program (Fig.~\ref{figureExposureTimeline}). The survey paused for hardware upgrades between 2021 July 10 and 2021 September 20.

\begin{figure}
\centering
\includegraphics[width=\columnwidth]{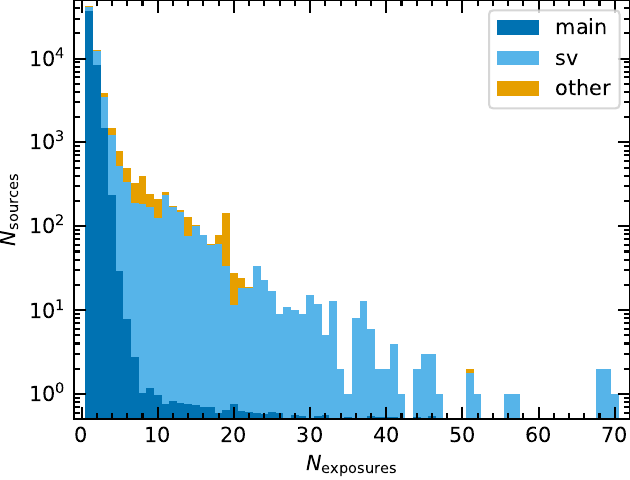}
\caption{Exposure counts for white dwarf candidates in DESI DR1. The total height of each bar gives the number of sources per bin on the logarithmic $y$-axis, but within each bar the coloured segments are divided \textit{linearly} to show the fractional contribution from each survey.}
\label{figureExposureCounts}
\end{figure}

The number of exposures per source, broken down by survey, is shown in Fig.~\ref{figureExposureCounts}. While DR1 contains more than 14 times as many white dwarf candidates as the EDR, \edit{it only has four times as many exposures}, and only half of the sources common to both catalogues have new exposures in DR1. The survey validation programs aimed for longer effective exposure times than the \texttt{dark} or \texttt{bright} programs. This means that objects that feature in the EDR white dwarf catalogue have a median of seven exposures with coadded $\SN=17$, whereas those that appear only in the DR1 catalogue have a median of one exposure with $\SN=8$.

\subsubsection{DESI spectra}
\label{subsectionDESIspectra}

DESI DR1 is a superset of the EDR, but all data have been reprocessed. DR1 uses the \texttt{Iron} version of the spectroscopic reduction pipeline \citep{Guy2023}, whose sky subtraction is superior to that in the \texttt{Fuji} version used for the EDR, amongst other improvements \citep{DESI2025DR1}. DESI spectra are supplied on a uniform grid of vacuum wavelengths in a Solar System barycentric frame.

DESI uses the `spectroperfectionism' algorithm \citep{Bolton2010} for spectral extraction, which yields a resolution matrix and a resolution-convolved spectrum. This technique decorrelates the pixel flux uncertainties, but at a cost: \textit{the matrix is not flux-conserving} \citep{Guy2023}. Therefore, model spectra should be convolved with the matrix before comparison with data. This is particularly relevant when dealing with sharp features near strong sky lines: if the resolution matrix is not applied before fitting a model, differences in sky subtraction residuals or resolution between epochs may be misinterpreted as astrophysical variability. Our tests suggest that it is acceptable to skip this step in our analysis (Section~\ref{subsubsectionResolutionMatrices}), as we only fit broad H and He features, but we can give no guarantees for other applications. We measured the distribution of resolutions in each arm (Table~\ref{tableResolutions}) by fitting Gaussian profiles to all resolution matrices for objects targeted as white dwarfs by DESI (around $10^5$ exposures). Resolutions are tightly distributed around their medians, with small low-resolution tails in the R and Z~arms.

\begin{table}
\centering
\caption{Distributions of resolutions (full width at half maximum $\Delta\uplambda$ in \AA) in each spectrograph arm for objects targeted as white dwarfs in DESI. The $\upsigma$~values correspond to the Gaussian-equivalent percentiles of the distribution. We adopt the median resolution during model fitting.}
\label{tableResolutions}
\begin{tabular}{lrrr}
\hline
 & B & R & Z \\
\hline
$-3\upsigma$ & 1.473 & 1.457 & 1.650 \\
$-2\upsigma$ & 1.504 & 1.483 & 1.690 \\
$-1\upsigma$ & 1.544 & 1.505 & 1.723 \\
median       & 1.597 & 1.530 & 1.762 \\
$+1\upsigma$ & 1.642 & 1.557 & 1.809 \\
$+2\upsigma$ & 1.693 & 1.593 & 1.894 \\
$+3\upsigma$ & 1.733 & 1.994 & 2.349 \\
\hline
\end{tabular}
\end{table}

\begin{figure*}
\centering
\includegraphics[width=\textwidth]{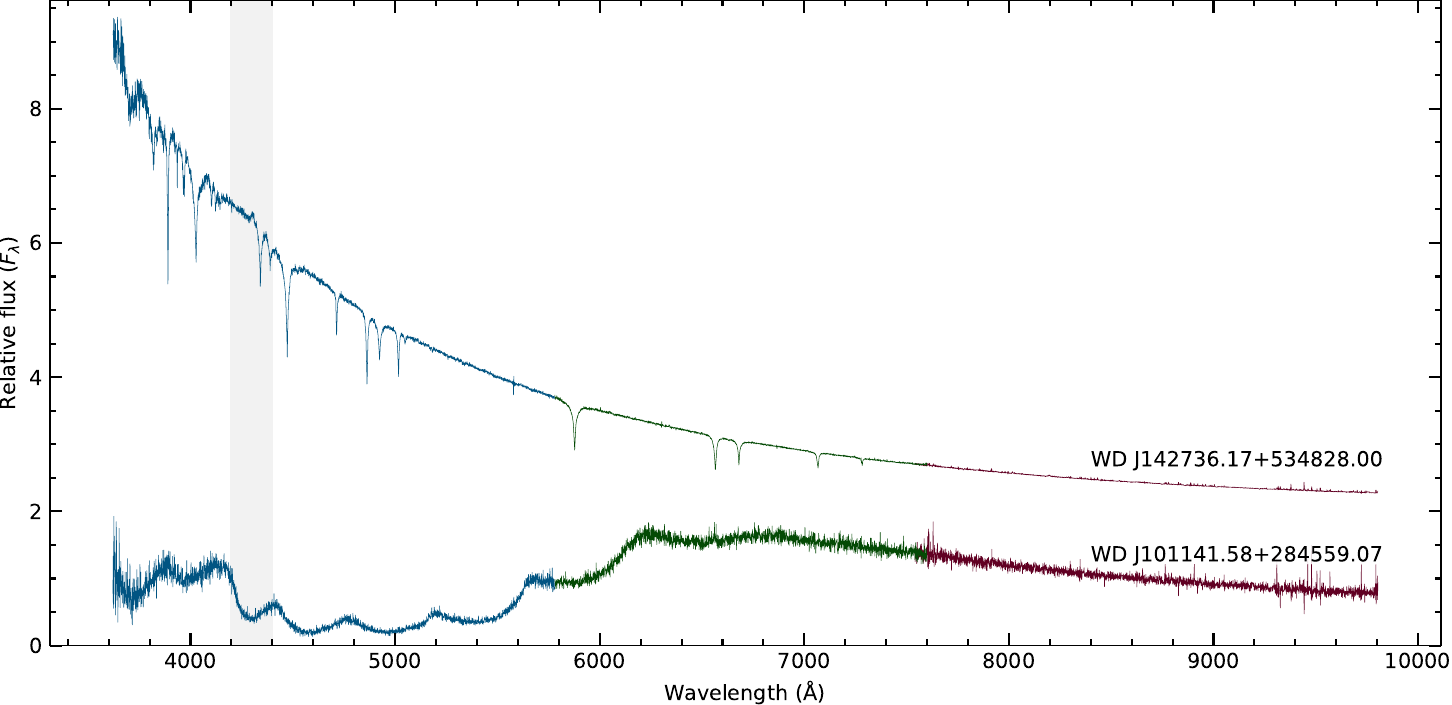}
\caption{Two examples of DESI spectra at different \SN levels. WD\,J142736.17+534828.00 is a DABZ white dwarf ($d=15$\,pc; \textit{Gaia} $G=15.0$\,mag) with 16 exposures in DESI DR1, which stack to create a coadd with $\SN>100$. WD\,J101141.58+284559.07 is a DQ white dwarf ($d=52$\,pc; $G=16.5$\,mag), which has only one exposure with $\SN=7$, in which we have masked a glitch at 7155\,\AA. The data are shown at full resolution, normalised and offset vertically. Grey shading indicates the region potentially affected by an instrumental artefact.}
\label{figureSpectraExamples}
\end{figure*}

DESI DR1 provides coadded spectra and resolution matrices for each source, but coadds are produced separately for exposures from different surveys and programs. We therefore created per-source coadds for each object (Section~\ref{subsubsectionDataPreparation}). Examples of DESI spectra are shown in Fig.~\ref{figureSpectraExamples}, one a high-\SN coadd created from 16 exposures, and the other a typical single-exposure spectrum with $\SN=7$. Both spectra demonstrate excellent agreement in relative flux calibration between the spectrograph arms.

\subsubsection{Data caveats}
\label{subsectionDataIssues}

We highlight a few issues that may affect the data presented here, some of which apply to DESI DR1 in general\footnote{\url{https://data.desi.lbl.gov/doc/releases/dr1/known-issues/}}, while others are specific to this catalogue.

At the DR1 level, the reported $\SN$ can be overestimated, and that effect grows stronger at higher $\SN$, such that spectra with reported $\SN=100$ may only have $\SN\approx20$. Instrumental defects impact spectra in certain wavelength regions \citep[fig.~24 of][]{Guy2023}, including a dip in collimator reflectivity at around 4200--4400\,\AA\ that overlaps with H$\upgamma$. Various quality flags in the bitmasks identify problematic pixels in reduced data.

Artefacts are occasionally encountered in DESI spectra, including spurious features common to multiple spectra from the same exposure, and discontinuities and offsets in flux calibration. Some examples are shown in Fig.~\ref{figureArtefacts}.

\begin{figure}
\centering
\includegraphics[width=\columnwidth]{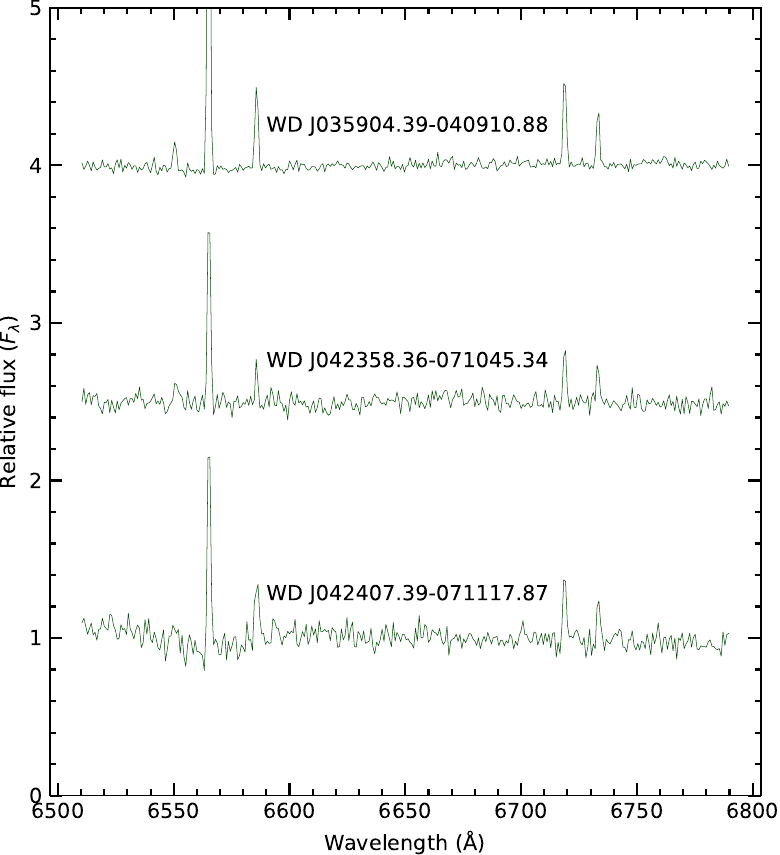}
\caption{Emission lines seen in spectra of white dwarfs in the direction of the Orion star-forming region. The data are shown at full resolution, normalised and offset vertically.}
\label{figureNebulaEmission}
\end{figure}

Considering white dwarfs specifically, Balmer emission appears to be present in hundreds of spectra, including otherwise-featureless DC stars. These are likely to be astrophysical contamination, or flux-calibration residuals caused by the use of main-sequence standard stars \citep{Guy2023}. Fig.~\ref{figureNebulaEmission} shows nebular emission lines contaminating the spectra of white dwarfs in the direction of the Orion star-formation region. Faint Ca\,H and K absorption is also seen in many spectra, most likely due to flux calibration residuals \citep[fig.~4]{RamirezPerez2024}.

There are four instances where two different WD\,J names map to the same fibre position (Table~\ref{tableDuplicateExposures}), all of which are known resolved binaries with separations of about 1.5\,arcsec.

Our catalogue contains 23 objects that appear under different WD\,J names in the GF19 catalogue and the reduced-proper-motion extension to the GF21 catalogue (Table~\ref{tableDuplicateWDJnames}). We use the names from the GF19 catalogue, on that basis that the WD\,J names from \textit{Gaia} DR2 should not be superseded by the astrometry in more recent \textit{Gaia} data releases.

\section{Supplementary data}
\label{subsectionSupplementaryData}

When fitting white dwarf models to DESI spectra (Section~\ref{sectionModelFitting}), we also used parallaxes, photometry and 3D extinction maps to constrain distances and reddening. These values are provided as supplementary data (see Table~\ref{tableExternalData} and the Data Availability Statement).

Where targets only had a \textit{Gaia} DR2 designation, we queried their DR3 designations from the \textsc{gaiadr3.dr2\_neighbourhood} table in the \textit{Gaia} archive. Only sources within 1\,arcsec were retained, to avoid spurious matches. We used parallaxes and photometry from DR3, falling back on DR2 where necessary. We applied a 0.029\,mas zero-point correction to DR2 parallaxes \citep{Lindegren2018}.

Additional photometry was obtained from SDSS and Pan-STARRS surveys as follows. First, we performed an initial cross-match between the \textit{Gaia} catalogues and the external surveys within a 45-arcsec radius to identify potential counterparts. We then propagated their coordinates to the observation epochs using \textit{Gaia} proper motions and selected the nearest source as the true counterpart, rejecting matches beyond 3\,arcsec.

We added a 1\,per cent error in quadrature to all photometric uncertainties to represent absolute calibration systematics. We discarded measurements where there is a risk of saturation, using thresholds fainter than the limits suggested by survey documentation, as saturation varies with conditions \citep{York2000SDSS, Magnier2013}. Specifically, we excluded SDSS magnitudes $u<13.5$, $g<14.5$, $r<14.5$, $i<14.5$, $z<13.5$, and Pan-STARRS magnitudes $g<14.0$, $r<14.0$, $i<14.0$, $z<13.5$, $y<12.5$. We also discarded SDSS or Pan-STARRS data with suspiciously large residuals in our model fits (Section~\ref{subsectionFittingProcedure}), or where there is a risk of flux contamination from nearby sources exceeding 1~per~cent (Section~\ref{subsectionCatalogueFluxContamination}). We also rejected any \textit{Gaia} $G_{\mathrm{BP}}$ or $G_{\mathrm{RP}}$ measurements where $\texttt{phot\_bp\_n\_contaminated\_transits} > 0$ or $\texttt{phot\_rp\_n\_contaminated\_transits} > 0$ in their \textit{Gaia} catalogue entries. We corrected SDSS photometry for small zero-point offsets \citep{Eisenstein2006}.

We used integrated extinction profiles from two independent maps \citep{Lallement2022, Vergely2022, Edenhofer2024} to construct priors as detailed in Section~\ref{subsubsectionFittingFramework}.

\edit{Light curves from the Zwicky Transient Facility \citep[ZTF;][]{Bellm2019} data release 24 were used to search for periodic variation in some of the objects of interest discussed in Section~\ref{sectionCatalogueContents}.}

\section{Spectral classification}
\label{sectionSpectralClassification}

Most catalogues present definitive classifications: an object either is or is not a white dwarf, and is identified with a single spectral type. That format is intuitive, but masks borderline cases and prevents uncertainty being propagated into statistical population studies. A probabilistic approach offers greater flexibility. A `probability of being a white dwarf', \pwd, was introduced by \citet{GentileFusillo2015photometric}, and became a key feature of the GF19 and GF21 catalogues, allowing users to select samples at arbitrary confidence thresholds and to control the trade-off between contamination and completeness. We adopt a similar approach here: each spectrum has been visually inspected, assigned a spectral type, and fitted with white dwarf atmosphere models, but both the classifications and atmosphere types have confidence scores, with the underlying data supplied so that users can tailor the catalogue to their own requirements. We also use the confidence scores to define our own \pwd metric (Section~\ref{subsectionCataloguePWD}). To avoid confusion, we distinguish the two schemes with superscripts: \pwdGaia means the scores published in the \textit{Gaia} white dwarf catalogues, while \pwdDESI means the scores assigned in this paper.

\begin{figure*}
\centering
\includegraphics[width=\textwidth]{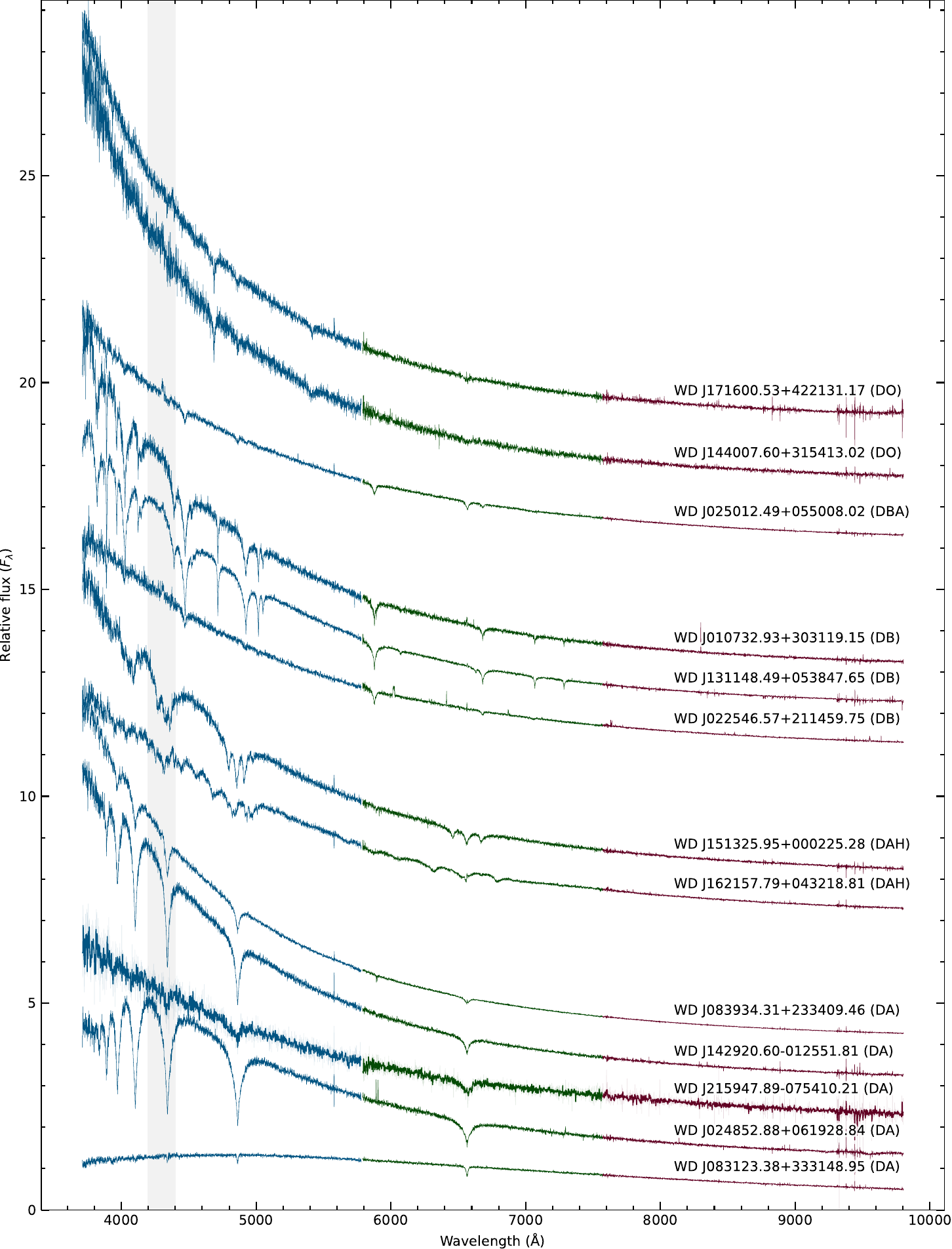}
\caption{Examples of spectral types found in the catalogue, smoothed with a 5-pixel boxcar median, normalised and offset vertically. Grey shading indicates the region potentially affected by an instrumental artefact.}
\label{figureSpectralAtlas}
\end{figure*}

Table~\ref{tableSpectralTypeDefinitions} presents our classification scheme, which is based on the established system for white dwarf spectral types \citep{Sion1983}, plus a few broad categories for non-white-dwarf objects or unclassifiable spectra. All white dwarf types begin with `D' (degenerate), and the second letter indicates the primary feature. Any subsequent letters indicate weaker features, though this is judged by eye, and depends on the wavelength region being considered. We avoid `O' types with `A' as a tertiary or later feature, as some Pickering lines are almost coincident with Balmer lines \citep{Manseau2016}. Examples of most types are shown in Fig.~\ref{figureSpectralAtlas}, including WD\,J215947.89$-$075410.21, a DA with a He-dominated atmosphere, which illustrates that classifications are based only on spectral features, not physical properties.

The non-white-dwarf objects in the catalogue are extragalactic sources (primarily quasars) and stars (primarily hot subdwarfs and early type main-sequence stars). We did not attempt to classify these contaminants beyond the broad types EXGAL and STAR.

\begin{table}
\centering
\caption{Spectral classification scheme. White dwarf types always begin with `D' (degenerate) and are followed by one or more feature letters, with optional suffixes.}
\label{tableSpectralTypeDefinitions}
\begin{tabular}{ll}
\hline
Identifier & Definition\\
\hline
\multicolumn{2}{l}{\textit{White dwarf features}} \\
A & H lines \\
B & \ion{He}{i} lines \\
C & Continuum-only (i.e. featureless) \\
O & \ion{He}{ii} lines \\
Q & Carbon lines or bands \\
Z & Metal lines \\[6pt]

\multicolumn{2}{l}{\textit{White dwarf suffixes}} \\
e & Emission \\
H & Zeeman splitting \\[6pt]

\multicolumn{2}{l}{\textit{Non-white-dwarf or binary types}} \\
CV & Cataclysmic variables \\
EXGAL & Extragalactic source \\
STAR & Main-sequence star or subdwarf \\
WD+MS & White-dwarf--main-sequence binary \\[6pt]

\multicolumn{2}{l}{\textit{Unclassifiable spectra}} \\
JUNK & Dominated by noise \\
UNCLASS & Unidentifiable \\
\hline
\end{tabular}
\end{table}

\begin{figure*}
\centering
\includegraphics[width=\textwidth]{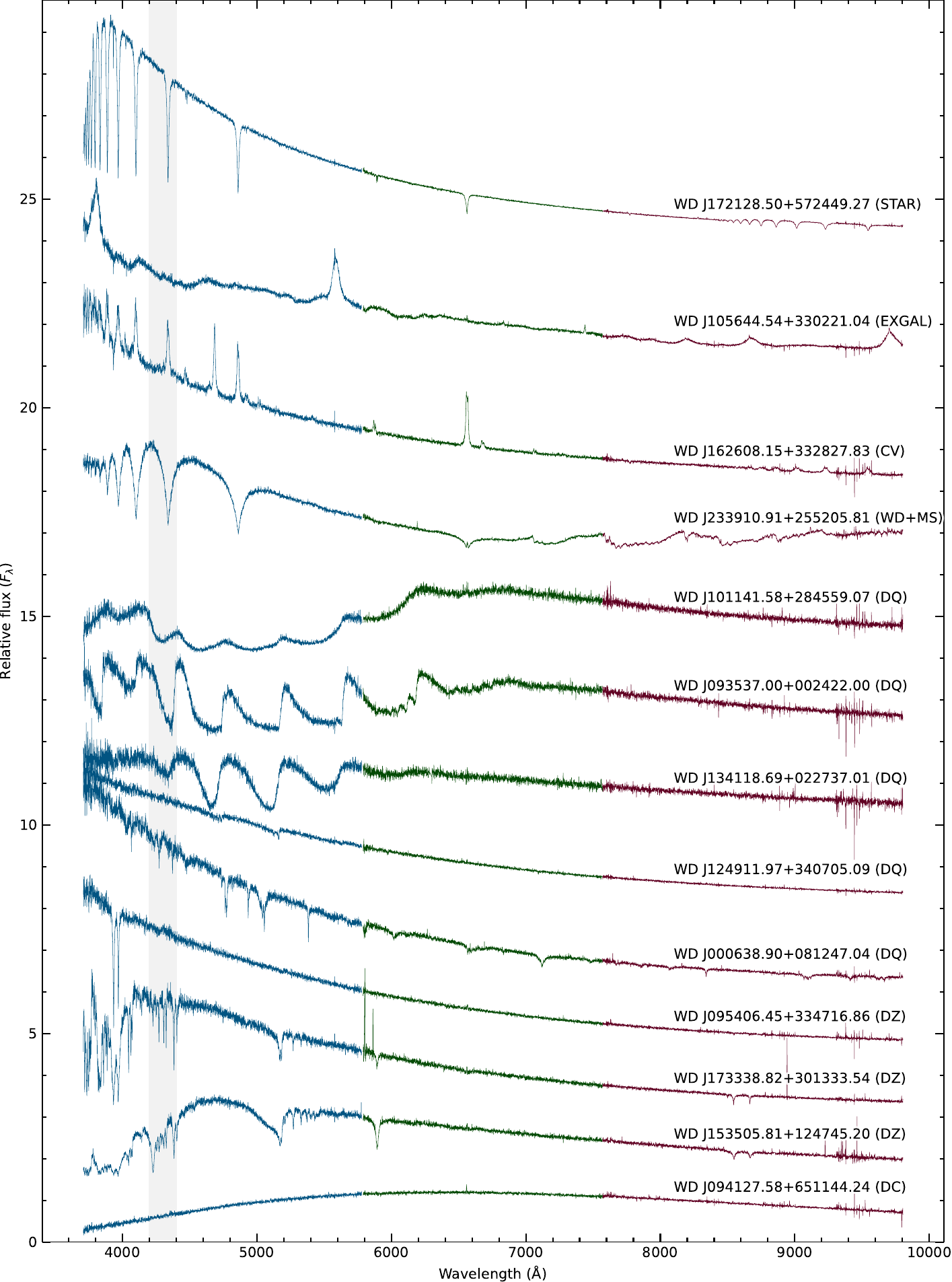}
\textbf{Figure~\ref{figureSpectralAtlas}} continued, but with spectra are shown at full resolution.
\end{figure*}

Four of the authors (henceforth `workers') inspected the spectra, with each seen by at least two people. One worker classified all spectra, and then others verified those results or provided independent classifications. Where differing opinions were recorded, further inspections were made\edit{, with no limit to the number of times each worker could review the same spectrum}. We also included classifications of DAHe stars made during a trawl of DESI spectra \citep{Manser2023DAHe}, as it was conducted by one of the workers in a similar fashion to the rest of the exercise.

Visual inspection was undertaken in two modes: classification and verification. When classifying, workers assigned spectral types, or recorded that they were unable to identify an object. Classifications could be flagged as uncertain. When verifying, workers were presented with spectra already labelled by another worker, and asked whether or not they agreed with that classification, or were unsure. Verification was performed on batches of objects originally classified as the same type. About 1 per cent of each batch was randomly sampled from objects classified as other spectral types, providing a control sample to gauge the random error rate of the workers.

After the initial round of classifications, opinions differed on classifications for around 10~per~cent of objects. Two workers each verified the disputed objects already seen by the other. Then three workers classified objects again where substantial disagreement remained, any spectra where features from metals or carbon were suspected, and some of the rarer types. Once only a few hundred contentious objects remained, mostly with low-S/N spectra, work stopped due to diminishing returns.

We used astrometry and photometry from \textit{Gaia} as an aid to classification. To help workers spot subtle differences between similar spectra, the HR~diagram was partitioned into Voronoi cells using the $k$-means algorithm, and spectra in each cell were classified in descending \SN order. When verifying, spectra were ordered by \bprp colour. Also, each spectrum was viewed alongside an HR~diagram showing the object being classified, the white dwarf track, and other populations such as white-dwarf--main-sequence binaries, cataclysmic variables and subdwarfs. For fainter objects with noisy parallaxes or unreliable colours, auxiliary data were potentially misleading if taken at face value, but all workers were aware of this and we judge that those risks were outweighed by the benefits of inspecting similar spectra together. An example classification image is shown in Fig.~\ref{figureClassificationImage}.

\edit{While there was unanimous agreement on the nature of most objects, there were differing opinions on others, so} we quantify our level of confidence in each classification for each object as follows. Each object is assigned one or more spectral types $T_1, T_2, \dots, T_n$ ($n\geq1$). For each type $T_i$ the number of positive, negative and uncertain classifications are $P_i$, $N_i$ and $U_i$, respectively. \edit{Each classification by each worker is a positive vote for that class. If a classification was flagged as uncertain, then an uncertain vote is also recorded alongside the positive vote. Each response by each worker during validation stages is treated as a single positive, negative, or uncertain vote.} \edit{We calculate the support $s_i$ for each spectral type by combining those votes with weighting factors and then applying the softplus transform:}

\begin{equation}
\label{equationSupport}
\edit{s_i=\ln{\left(1 + \exp{\left(P_i - \alpha U_i - \beta N_i\right)}\right)}}
\end{equation}

\edit{We set $\alpha=0.5$ and $\beta=1.0$ so that uncertainty erodes support and negative votes counter positive votes. Using softplus $[\ln(1+e^x)]$ instead of a ramp function $[\max(0,x)]$ ensures that all spectral types considered remain in contention, albeit at very low confidence if they were downvoted.}

\edit{Next, we normalised to find the fractional support for each type:}

\begin{equation}
\label{equationFractionalSupport}
\edit{f_i=\frac{s_i}{\sum_j{s_j}}}
\end{equation}

\edit{We then calculated the confidence $c_i$ for each spectral type $T_i$ by applying a global penalty for any uncertainty:}

\begin{equation}
\label{equationConfidence}
\edit{c_i = f_i \left(1-\gamma\frac{U}{V}-\delta\frac{N}{V}\right)}
\end{equation}

\noindent
\edit{where}

\begin{equation*}
\label{equationConfidence}
\edit{P=\sum_i{P_i},\quad U=\sum_i{U_i},\quad N=\sum_i{N_i},\quad V=P+U+N}
\end{equation*}

\edit{Confidence values can lie in the range $0<c\leq1$, where $\sum_k c_k \le 1$ for any given object. Unanimous positive votes for a single type result in $c=1$. We set $\gamma=0.5$ and $\delta=0.25$ so that uncertainty and disagreement both erode the total confidence across all spectral types. This ensures that even if only one type was considered, but it received a negative vote during validation, it will score $c<1$.}

White dwarf types that are permutations of each other (e.g. DBA and DAB) were treated as identical when calculating confidence scores, and the permutation that received the most votes became the reported type. That avoided low confidence scores being assigned in cases where the only disagreement was over the relative strength of features, which is somewhat subjective.

The output of this process was one or more spectral types per object, with associated confidence scores. The highest-scoring type for each object is presented in the main catalogue, and summary statistics are given in Table~\ref{tableSpecTypes}. Alternative classifications are given in the supplementary data (Table~\ref{tableClassifications}), as are the underlying data from which our results can be reconstructed (Table~\ref{tableClassificationLogs}).

\begin{table}
\centering
\caption{Statistics per spectral type (based on the classifications presented in the main catalogue): the number of objects \textit{N} of each type in total and in the primary sample; the mean confidence $\bar{c}$ of their classifications; and mean \pwd score in DESI DR1 and the \textit{Gaia} GF19 (G2) and GF21 (G3) white dwarf catalogues. \pwdDESI is defined in Section~\ref{subsectionCataloguePWD} and the \pwd metrics are discussed and compared in Section~\ref{subsubsectionContentsPWD}.}
\label{tableSpecTypes}
\begin{tabular}{lrrrrrr}
\hline
 & $N$ & $N_{\text{primary}}$ & $\bar{c}$ & & $\bar{p}_{\mathrm{WD}}$ & \\
 &  & &  &  DESI & G2 &  G3 \\
\hline
CV & 232 & 0.96 & 0.99 & 0.66 & 0.50 \\
DA & 32892 & 0.98 & 0.96 & 0.97 & 0.97 \\
DAB & 41 & 0.82 & 0.79 & 0.99 & 0.99 \\
DABO & 1 & 0.41 & 0.99 & 0.99 & 1.00 \\
DABZ & 15 & 0.78 & 1.00 & 1.00 & 1.00 \\
DABZE & 1 & 0.47 & 0.99 & 1.00 & 1.00 \\
DAE & 3 & 0.46 & 1.00 & 0.66 & 0.97 \\
DAH & 522 & 0.88 & 0.93 & 0.99 & 0.99 \\
DAHE & 32 & 0.57 & 0.86 & 0.99 & 0.99 \\
DAO & 58 & 0.64 & 0.74 & 0.81 & 0.91 \\
DAQ & 20 & 0.69 & 0.89 & 0.98 & 0.98 \\
DAZ & 609 & 0.82 & 0.93 & 0.96 & 0.97 \\
DAZB & 7 & 0.67 & 1.00 & 0.99 & 0.99 \\
DAZH & 2 & 0.39 & 1.00 & 0.99 & 0.98 \\
DB & 2416 & 0.96 & 1.00 & 0.98 & 0.99 \\
DBA & 342 & 0.90 & 0.99 & 0.99 & 0.99 \\
DBAZ & 50 & 0.70 & 1.00 & 0.99 & 1.00 \\
DBH & 7 & 0.63 & 1.00 & 0.87 & 0.89 \\
DBO & 17 & 0.58 & 0.89 & 0.95 & 0.96 \\
DBZ & 89 & 0.80 & 1.00 & 1.00 & 1.00 \\
DBZA & 27 & 0.79 & 1.00 & 0.96 & 0.98 \\
DC & 4614 & 0.93 & 0.99 & 0.96 & 0.98 \\
DH & 118 & 0.57 & 0.90 & 0.99 & 1.00 \\
DO & 47 & 0.61 & 0.89 & 0.83 & 0.90 \\
DOA & 98 & 0.72 & 0.90 & 0.90 & 0.94 \\
DOB & 24 & 0.49 & 0.91 & 0.91 & 0.96 \\
DQ & 806 & 0.90 & 1.00 & 0.98 & 0.98 \\
DQA & 18 & 0.52 & 0.98 & 0.98 & 0.98 \\
DQAE & 1 & 0.30 & 1.00 & 1.00 & 0.99 \\
DQB & 1 & 0.32 & 1.00 & NaN & 0.99 \\
DQH & 1 & 0.51 & 1.00 & 1.00 & 0.99 \\
DQZ & 3 & 0.76 & 1.00 & 1.00 & 0.99 \\
DZ & 842 & 0.94 & 1.00 & 0.98 & 0.99 \\
DZA & 97 & 0.88 & 0.99 & 0.99 & 0.99 \\
DZAB & 13 & 0.72 & 1.00 & 1.00 & 0.98 \\
DZB & 26 & 0.87 & 1.00 & 0.99 & 1.00 \\
DZBA & 37 & 0.82 & 1.00 & 0.97 & 0.99 \\
DZH & 6 & 0.70 & 1.00 & 0.97 & 0.99 \\
DZQ & 1 & 0.85 & 1.00 & 1.00 & 0.99 \\
EXGAL & 10225 & 1.00 & 0.00 & 0.03 & 0.47 \\
JUNK & 47 & 0.70 & 0.11 & 0.86 & 0.63 \\
STAR & 7958 & 0.96 & 0.01 & 0.22 & 0.18 \\
UNCLASS & 36 & 0.55 & 0.50 & 0.87 & 0.82 \\
WD+MS & 1290 & 0.96 & 0.99 & 0.65 & 0.56 \\
\hline
\end{tabular}
\end{table}

\section{Model fitting}
\label{sectionModelFitting}

Our methods are described in detail below, but are similar to those used for the EDR white dwarf catalogue \citep{Manser2024DESIEDR}. Improvements and changes include the use of a new sampler, the ability to perform model comparison, the inclusion of reddening as a free parameter, updates to white dwarf models and reddening maps, and a modified distance prior.

We fitted model spectra and synthetic photometry to every object in the catalogue, in a Bayesian framework described in Section~\ref{subsectionFittingProcedure}. Three model grids were used: pure H, pure He, and mixed He+H (Section~\ref{subsectionWhiteDwarfModels}). Free parameters common to all models are effective temperature ($\Teff$), surface gravity ($\log{g}$), distance ($d$) and reddening ($E_{B-V}$). Hydrogen abundance ($\logh$) was left free for the mixed He+H model. This procedure was completely independent of the spectral classification exercise, so we fitted each object with all three models and calculated the Bayesian evidence to enable model comparison.

Photometric and spectroscopic fits for the same star can be inconsistent \citep{Izquierdo2023}, so we compromise by fitting them simultaneously to generate the primary catalogue results. This joint fitting approach allows the parallax and photometry to constrain the surface gravity. In turn, that weakens the degeneracy between `hot' and `cold' solutions that can arise in pure spectroscopic fits due to Balmer line equivalent widths reaching a maximum near 13\,000\,K \citep[e.g.][]{RebassaMansergas2007}. He-atmosphere stars show a similar degeneracy at $22\,000\lesssim\Teff\lesssim30\,000$\,K \citep{Bergeron2011}.

In addition to the white dwarf fits, we also used a hydrogen model grid with a wider surface gravity range ($4.5\leq\log{g}\leq9.5$) to help identify subdwarfs or main-sequence stars, as their spectra can be challenging to distinguish visually from white dwarfs (Section~\ref{subsectionLowLogg}).

\subsection{White dwarf atmosphere models}
\label{subsectionWhiteDwarfModels}

We computed synthetic spectra using the latest version of the \cite{Koester2010} code, where most of the recent updates are related to the treatment of non-ideal effects in cool atmospheres. We use the H-line profiles of \citet{Tremblay2009}.

For pure H-atmospheres, a grid of models was generated spanning $3000\leq\Teff\mathrm{(K)}\leq80\,000$ and $7.00\leq\log{g}\leq9.75$. Models were spaced by 0.25\,dex in $\log{g}$ across the grid, and in steps of 250\,K between 3000\,K and 20\,000\,K, then in steps of 1000\,K up to 30\,000\,K, then in steps of 2000\,K up to 30\,000\,K, and then in steps of 5000\,K up to 80\,000\,K.

Pure He-atmospheres used an almost identical grid, except that models hotter than 60\,000\,K were spaced by 10\,000\,K rather than 5000\,K.

The grid of mixed He+H atmospheres used the same $\Teff$ and $\log{g}$ points as the He grid, but extended only to 20\,000\,K and 9.5\,dex at the upper bounds. The $\logh$ abundance was varied from $-6.00$ to $-2.00$ in steps of 0.25\,dex, and then in steps of 1\,dex up to 0.

The models were generated on a wavelength grid whose spacing was dynamically adjusted to the local level of detail in the spectrum, so that all features are finely resolved. These models were then convolved with a Gaussian kernel, whose width in each spectrograph arm was set to the median resolution that we measured from the resolution matrices for all white dwarf candidates targeted by DESI (Table~\ref{tableResolutions}). For computational efficiency, we did not apply the resolution matrices themselves, having verified that approximating them with Gaussian kernels is sufficient for our purposes (Section~\ref{subsubsectionResolutionMatrices}).

\subsection{Fitting procedure}
\label{subsectionFittingProcedure}

Several steps were involved in generating the final results, with some iteration. Exposures were coadded before fitting, for computational convenience. An initial photometry-only fit was performed, and residuals inspected to detect and remove problematic outliers. Spectroscopy was then included, radial velocities measured, and the fit repeated with the data and model in the same velocity frame. These steps are described in detail below.

\subsubsection{Data preparation}
\label{subsubsectionDataPreparation}
Data points flagged as unusable via the \textsc{specmask} bitmasks in the DESI cframe files were discarded before processing. In some cases all points are flagged, so we did not fit those spectra. We also excluded potentially problematic photometric points, where we suspect flux contamination from neighbouring sources (Section~\ref{subsectionCatalogueFluxContamination}), or where isolated outliers were detected during the first pass of the fitting process (detailed below).

Where multiple exposures were available, we created co-added spectra for fitting by taking means weighted by the inverse flux variances reported in the cframe file. The DESI spectroscopic pipeline reduces spectra onto a standard wavelength grid, so averaging exposures does not require interpolation, nor does it introduce covariances, meaning that fitting the combined spectrum is statistically equivalent to jointly fitting the individual exposures.

\subsubsection{Fitting framework}
\label{subsubsectionFittingFramework}
We sampled parameter posterior distributions and estimated the Bayesian evidence using \texttt{pocoMC}, a preconditioned Monte Carlo sampler that can cope with non-linear and multimodal problems \citep{Karamanis2022pocoMC, Karamanis2022PMC}. We used it with default settings, starting with 1024 effective particles and 512 active particles, and requiring at least 4096 samples for the posterior, and at least 2048 for the evidence.

We adopted uniform priors for $\Teff$, $\log{g}$ and $\logh$, whose intervals were defined by the model grid boundaries (Section~\ref{subsectionWhiteDwarfModels}).

Inferring distances from \textit{Gaia} parallaxes requires careful treatment to avoid introducing potentially strong biases \cite[e.g.][]{Luri2018}. We therefore used an exponentially-decreasing distance prior following the prescription of \cite{BailerJones2018} but calibrated against sources with $\log{g}>7$ from the mock Gaia EDR3 catalogue \citep{Rybizki2020}.

Reddening varies with both sky position and distance, so we used three-dimensional extinction maps to construct priors. We queried the integrated extinction profile towards each object in our catalogue at all distances sampled by two independent all-sky maps \citep{Lallement2022, Vergely2022, Edenhofer2024}, and converted to reddening by assuming $A_V = 3.1\times E_{B-V}$. During sampling, values were interpolated from each profile at the proposed distance, and then used to define the prior: a Student's~$t$ distribution with five degrees of freedom, whose centre is the mean of the interpolated values, and whose width is their difference plus a systematic error. That additional error was estimated by fitting a quadratic in distance to the fractional difference between the two maps along all sightlines. The Student's~$t$ distribution was chosen as it is more forgiving of outliers than a Gaussian, thus accommodating local variations in extinction not captured at the resolution of the maps.

We generated models by linear interpolation over the grids described in Section~\ref{subsectionWhiteDwarfModels}, but using $\log{\Teff}$ as the temperature dimension. We evaluated the models against the data using a Gaussian likelihood function.

\subsubsection{Photometric fits}
\label{subsubsectionPhotometricFits}

To generate synthetic photometry, we convolved our model spectra with the various photometric bandpasses. We then scaled by distance and solid angle, calculated by interpolating the white dwarf radius from evolutionary models \citep{Althaus2013,Camisassa2016,Camisassa2017,Camisassa2019}.

Photometric surveys can contain problematic data, for example where diffraction spikes or other artefacts affect the images. For objects classified as DA, DB or DBA, we performed an initial fit using H, He and He+H models, respectively, in an attempt to identify and remove problematic data. In this first pass, we fitted data for each survey (\textit{Gaia}, Pan-STARRS, SDSS) independently, and inspected their residuals to detect potentially problematic data as follows.

For each data point, we compared the observed value $y$ against the expected value $\hat{y}$, and divided by the uncertainty $\sigma$ to obtain {$r=|y-\hat{y}|/\sigma$}, the \textit{normalised residual}. We then calculated the median normalised residual $\tilde{r}_s$ for each survey $s$. To identify cases where all data from a survey are suspect, we compared the per-survey median normalised residuals for each object: where a per-survey median was much larger than the minimum per-survey median for the same object, i.e. $\tilde{r}_s - \min{\tilde{r}_s\prime} > 3$, we excluded the data from that survey. To deal with single outliers among otherwise usable data, we excluded points where $r>3\tilde{r}_s$ and $\tilde{r}_s<2$. Outlier detection was performed only once, and not iterated.

Having masked the problematic data identified above, we fitted the full photometric data for each object against all three models. For completeness, we also fitted against the data for each survey independently, and we record the results without further analysis.

\subsubsection{Spectroscopic fits}
\label{subsubsectionSpectroscopicFitting}

Both data and model were normalised to their respective means within {200-\AA}-wide windows, to mitigate against any flux calibration errors, following the method used for the EDR white dwarf catalogue \citep{Manser2024DESIEDR}. This normalisation means we could not fit for distance, so we kept reddening fixed at the value inferred during the photometric fit. We show some example fits in Appendix~\ref{appendixFits}, where we also fit SDSS spectra and compare results against previous work.

We also measured radial velocities, as detailed in Section~\ref{subsubsectionRVs}, alternating between parameter and velocity fits. In each iteration, we used the result of the parameter fit as a template to fit the velocity shift of the B and R~arms of the coadded spectrum (there are no suitable lines in the Z~arm). Where a fit returned $|v_r|<750$\,\kms, we shifted the model into the same frame as the data and re-fitted \edit{(larger values were treated as failed fits and we did not not shift the model)}. We iterated up to three times, though convergence was usually reached within two iterations. After the final pass, we measured radial velocities for individual exposures.

\subsubsection{Radial velocities}
\label{subsubsectionRVs}

Velocity shifts are challenging to measure for white dwarfs, as they typically have only a few broad lines, and the sharp core of H$\upalpha$ is not resolved in DESI spectra. Furthermore, the data are often noisy, especially individual exposures. We therefore restricted the radial-velocity fit to the strongest features. For H-atmosphere models, we used {40-\AA} windows centred on the H$\upalpha$, H$\upbeta$ and H$\updelta$ lines (H$\upgamma$ was not used as it is coincident with the collimator reflectivity artefact). For He-atmosphere models, we used {32-\AA} windows centred on 23 \ion{He}{i} lines in the range 3600--7300\,\AA. For mixed-atmosphere models we combined both sets of windows. The window sizes were chosen by experimentation, where we selected values that minimised scatter between successive iterations during fitting.

We expect radial velocity to be uncorrelated with the other fit parameters \citep[e.g.][fig.~15]{BadenasAgusti2024}, so we did not solve for it in the Bayesian framework. Instead, we fitted the velocity offset between our model spectrum and the data via maximum likelihood estimation. We used the Huber loss function (equation~\ref{equationHuberLoss}, Appendix~\ref{appendixRVs}) to mitigate against outliers, and a Gaussian prior of width {1000\,\kms} centred at {30\,\kms} (the gravitational redshift of a typical white dwarf of mass 0.6\,\Msun). The weak prior improves interpretability of the results: a fit will converge to the prior where the data are minimally constraining, while even greater uncertainties suggest bad data or a mismatched model. The likelihood was evaluated at {10-\kms} intervals from $-5000$ to {5000\,\kms}, and then at {1-\kms} resolution in the high-likelihood region. We then refined the solution using \texttt{scipy.optimize.minimize\_scalar}, bounded by the {2-\kms} interval centred on the grid maximum. The uncertainty was estimated as half the 16th--84th-percentile range of the cumulative likelihood distribution. Velocities are quoted as measured; we do not attempt to correct for gravitational redshifts.

\edit{If we detected radial velocity variation between exposures (see Section~\ref{subsectionRVvariation}) then we repeated the analysis for that star, jointly fitting the individual exposures instead of our coadd.}

\subsubsection{Joint fits}
\begin{figure*}
\centering
\begin{minipage}{0.64\textwidth}
    \includegraphics[width=\linewidth]{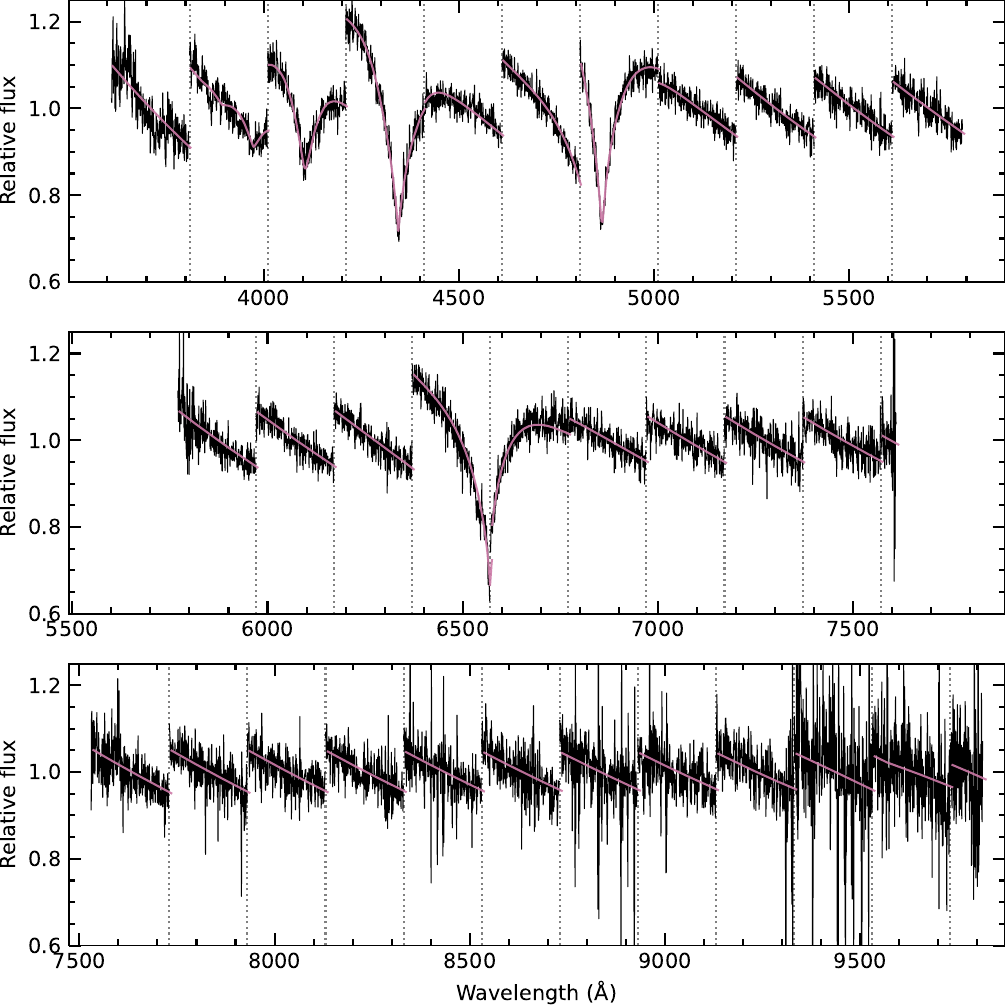}
\end{minipage}
\hfill
\begin{minipage}{0.33\textwidth}
    \includegraphics[width=\linewidth]{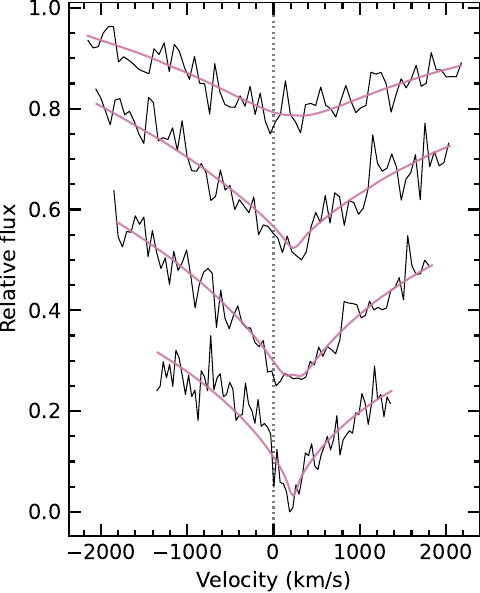}
    
    \vspace{0.5em}
    
    \includegraphics[width=\linewidth]{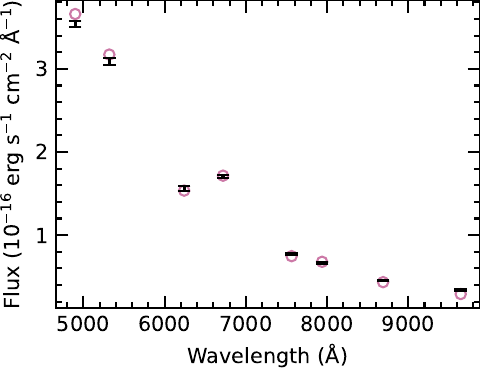}
\end{minipage}
\caption{Joint fit of the pure-H atmosphere model to WD\,J103812.22+825135.62, a DA white dwarf with $\Teff=30\,300\pm100$\,K. \textit{Left}: data (black) and model (red) normalised within {200-\AA}-wide windows, whose boundaries are indicated by dotted lines. The unmodified spectrum is shown in Fig.~\ref{figureClassificationImage}. \textit{Upper right}: The central {60\,\AA} of Balmer lines H$\upalpha$--$\updelta$, offset vertically, which clearly show the radial velocity shift of $230\pm30$\,\kms from the rest frame (the dotted vertical line). \textit{Lower right}: Spectral energy distribution showing photometric data (black) and model (red).}
\label{figureSpectralFitsH}
\end{figure*}

Finally, we made joint fits to the spectra, photometry and parallax for each object. For computational efficiency, we only allowed reddening to affect the photometry, as changes in reddening have minimal effect on the normalised spectra, which we reddened by the amount inferred from the photometric fit. We used the same iterative fitting procedure as for spectroscopic fits (Section~\ref{subsubsectionSpectroscopicFitting}) to perform the parameter and radial velocity fits. The outputs of this stage are our primary results. Fig.~\ref{figureSpectralFitsH} shows an example of a fit using the pure-H model, and further examples for the pure-He and mixed He+H models are shown in Figures~\ref{figureSpectralFitsHe} and \ref{figureSpectralFitsHe+H}.

\subsubsection{Outputs}
Summary statistics were generated from the posterior samples. Results for the joint fits are given in the main catalogue (Table~\ref{tableMainCatalogue}), and results from the purely photometric and spectroscopic fits are made available in the supplementary data (Table~\ref{tableFullFitResults}). Fig.~\ref{figureTEFFlogg} shows the outcomes for all three types of fits, where the degeneracies inherent to pure spectroscopic fits are particularly obvious.

\subsubsection{Resolution matrices}
\label{subsubsectionResolutionMatrices}

We tested the impact of neglecting the resolution matrix when fitting, using a sample of about 350 confidently-identified DB stars, which have narrower lines than DA stars and are thus more likely to be affected by resolution issues. We repeated our fits against their individual exposure spectra, convolving our models with their resolution matrices. The median difference in $\Teff$ between the two fitting methods is 0.15~per cent across the sample. We thus conclude that our method of fitting coadded spectra with models convolved to the median resolution of each arm is sufficient for our purposes.

\subsection{Radial velocity variation}
\label{subsectionRVvariation}

We tested for velocity variation in objects with $N>1$ exposures, using the inverse-variance-weighted mean of the velocities from all exposures as the reference point. Under the null hypothesis of constant velocity, normalised residuals around the mean should follow a standard Gaussian distribution, and the sum of their squares should follow a $\chi^2$ distribution with $N-1$ degrees of freedom. We calculated $\chi^2$ against the mean and report the corresponding $p$-values, where lower values indicate a higher probability of variation. However, low-$\SN$ data and failed fits can introduce spurious outliers, so we also report $\chi^2$ values calculated using robust loss functions: the Huber loss (our primary metric), and Tukey's biweight. The 3-$\upsigma$ thresholds are $\ln{p}<-5.91$ for the standard analysis, $-3.34$ for the Huber loss, and $-3.05$ for Tukey's biweight. Values below those thresholds are good candidates for velocity variation. \edit{We selected the Huber loss as our primary metric as Tukey's biweight proved too aggressive in its suppression of outliers and failed to detect some known binaries.} Appendix~\ref{appendixRVs} gives more detail on the robust loss functions and how their 3-$\upsigma$ thresholds were calibrated.

\begin{figure}
\centering
\includegraphics[width=\columnwidth]{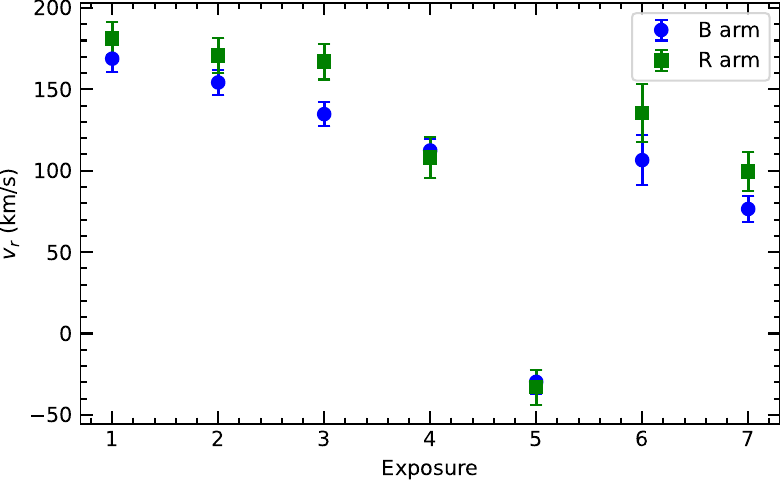}
\caption{Radial velocities measured from individual exposures of WD\,J101606.87$-$011917.14, a known single-lined WD--WD binary, which scores highly in our velocity variability test.}
\label{figureRVexample}
\end{figure}

Fig.~\ref{figureRVexample} shows the radial velocity data for a high-confidence variable system that we identified (WD\,J101606.87$-$011917.14; $\ln{p_{\text{Gaussian}}}=-345$; $\ln{p_{\text{Huber}}}=-32$; $\ln{p_{\text{Tukey}}}=-0.1$). The first three DESI exposures were taken within about 1\,hr of each other, and show decreasing velocities in both the B and R~arms. We measured velocities spanning ranges of $198\pm11$ and {$213\pm12$\,\kms} in the B and R~arms, respectively, but did not attempt to fit an orbital model. These values appear consistent with results from the SPY survey, which identified it as a binary with a period of 0.44\,d and a velocity semi-amplitude of {122\,\kms} \citep{Nelemans2005}, highlighting the potential for discovering supernova progenitors and other binaries from repeat observations by DESI.

\subsection{Identifying subdwarfs and main-sequence stars}
\label{subsectionLowLogg}

Distinguishing visually between hot white dwarfs, subdwarfs and main-sequence stars (especially types B, A and F) can be difficult. Therefore, we fitted all objects with an H-atmosphere grid covering a wide range of $\log{g}$ values. These fits are not a science product: they were only used as an additional diagnostic for identifying white dwarfs (Section~\ref{subsectionCataloguePWD}). There is a dedicated catalogue for main-sequence stars in DESI DR1 \citep{Koposov2026}.

The H-atmosphere model grid spans $4000\leq\Teff\text(\,K)\leq130\,000$ and $4.5\leq\log{g}\leq9.5$. Models were spaced by 0.1\,dex in $\log{g}$ across the grid, and in steps of 500\,K between 4000\,K and 18\,000\,K, then in steps of 1000\,K up to 30\,000\,K, then in steps of 2500\,K up to 65\,000\,K, then in steps of 5000\,K up to 100\,000\,K, and then in steps of 10\,000\,K up to 130\,000\,K. Between 55\,000\,K and 65\,000\,K the lower $\log{g}$ bound increases due to the Eddington limit, reaching $\log{g}=6.0$ above that temperature range.

We performed only a spectroscopic fit with this grid, as we do not have a mass--radius relation covering the whole parameter space, precluding the use of photometry or parallaxes.

\section{Catalogue content}
\label{sectionCatalogueConstruction}

The main catalogue contains spectral classifications, stellar parameters, radial velocities, and data quality metrics for the 63\,968 white dwarf candidates observed as part of DESI DR1. Table \ref{tableMainCatalogue} lists the column headings used in the data file, with brief descriptions. We describe below in more detail how it was assembled.

\begin{table*}
\centering
\caption{Main catalogue: column names and descriptions. Provided as \texttt{catalogue.csv}.}
\label{tableMainCatalogue}
\begin{tabular}{ll}
\hline
Column & Description \\
\hline
WDJname & White dwarf name as defined by GF19 \\
designation & \textit{Gaia} designation \\
specType & Spectral type \\
specType\_confidence & Confidence in spectral type \\
model & Atmosphere type used in fit \\
model\_confidence & Confidence in atmosphere type \\
TEFF & Effective temperature (K) \\
TEFF\_err & Error on effective temperature \\
LOGG & Surface gravity (log [\cms]) \\
LOGG\_err & Uncertainty on surface gravity \\
H & Numerical H abundance as log(H/He) \\
H\_err & Uncertainty on numerical H abundance \\
distance & Distance (pc) \\
distance\_err & Uncertainty on distance \\
EBminusV & Reddening as $E_{B-V}$ \\
EBminusV\_err & Uncertainty on reddening \\
\edit{mass} & \edit{Mass calculated using mass--radius relation used during fitting (Section~\ref{subsubsectionPhotometricFits})} \\
SN\_B & Median $\SN$ calculated from flux and uncertainty in blue arm \\
SN\_R & Median $\SN$ calculated from flux and uncertainty in red arm \\
SN\_Z & Median $\SN$ calculated from flux and uncertainty in near-infrared arm \\
SNmeasured\_B & $\SN$ measured statistically in blue arm \\
SNmeasured\_R & $\SN$ measured statistically in red arm \\
SNmeasured\_Z & $\SN$ measured statistically in infrared arm \\
fractionMasked\_B & Fraction of pixels masked in blue arm \\
fractionMasked\_R & Fraction of pixels masked in red arm \\
fractionMasked\_Z & Fraction of pixels masked in near-infrared arm \\
pwd\_DESI & Probability of being a white dwarf \pwdDESI as determined here \\
pwd\_GF19 & Probability of being a white dwarf \pwdGaia as listed in the GF19 catalogue \\
pwd\_GF21 & Probability of being a white dwarf \pwdGaia as listed in the GF21 catalogue \\
fluxContamination\_DESI & Expected fractional fibre contamination from field stars based on \textit{Gaia} data \\
nNeighbours\_DESI & Number of field stars within 60\,arcsec \\
nExposures & Number of exposures \\
primary & Member of the primary sample: objects targeted as white dwarfs in the \texttt{dark} or \texttt{bright} programs of the \texttt{main} survey \\
logP\_Gaussian & RV variability: $\ln{p}$ for $\chi^2$ test against constant-velocity null hypothesis (using Gaussian residuals) \\
logP\_Huber & RV variability: $\ln{p}$ for $\chi^2$ test against constant-velocity null hypothesis (using Huber loss function) \\
logP\_Tukey & RV variability: $\ln{p}$ for $\chi^2$ test against constant-velocity null hypothesis (using Tukey biweight) \\
pHalo & Probability of halo membership (indicative only; see Section~\ref{subsectionContentsKinematics} for method used) \\
RV & Radial velocity (\kms) \\
RV\_error & Uncertainty on radial velocity \\
\hline
\end{tabular}
\end{table*}

\subsection{Spectral classification}
\label{subsectionCatalogueSpectralClassification}

This is a purely visual exercise, performed entirely separately from model fitting. We report the highest-confidence spectral type assigned by the human classifiers, together with its confidence score, as determined in Section~\ref{sectionSpectralClassification}.

\subsection{Probability of being a white dwarf}
\label{subsectionCataloguePWD}

White dwarf models were fitted to all spectra, but they are not appropriate for non-white-dwarf objects, where Bayesian model comparison merely identifies the least-bad fit, and the inferred parameters are not physically meaningful. We therefore estimate the probability of being a white dwarf \pwdDESI for each object \edit{by using the fit results and the fractional support $f$ for white dwarf spectral types as follows.}

Almost all objects that we classified as stellar contaminants have $\log{g}<7$ in the fits to the extended $\log{g}$ grid, as expected. However, hundreds of objects that we classified as DA types also fall in this range, but white dwarfs formed by single-star evolution within a Hubble time have $\log{g}\gtrsim7.5$, illustrating that visual classification can be challenging for objects \edit{near the subdwarf--white dwarf boundary}. Therefore, we \edit{introduce a} penalty function $\rho$ for DA types $T_i$ based on their $\log{g}$ values from the extended grid fit. This function is defined as a Gaussian cumulative distribution $\Phi$, centred at $\mu_0=7.0$ with width $\sigma_0=0.2$, convolved with the fit uncertainty $\sigma_{\log{g}}$. This ensures that any object whose fit is inconsistent with \edit{being a white dwarf} $(\log{g}\lesssim7.5)$ will be assigned $\rho\approx0$. The full expression is:

\begin{equation}
\rho_i=\begin{cases}
    0 & \text{for } T_i\in{\{\mathrm{EXGAL, STAR}\}}\\
    \Phi\left(\frac{\log{g}-\mu_0}{\sqrt{\sigma_{\log{g}}^2+\sigma_0^2}}\right) & \text{for } T_i\in{\{\mathrm{DA, DAH, DAZ,} \dots\}}\\
    1 & \text{otherwise}
  \end{cases}
\end{equation}

\edit{We modified Equation~\ref{equationFractionalSupport} to exclude the uninformative classes UNCLASS and JUNK, as they are agnostic about whether or not an object is a white dwarf:}

\begin{equation}
\label{equationModifiedFractionalSupport}
\edit{\tilde{f}_i=\frac{s_i}{\displaystyle\sum\limits_{j
}{s_j}}\quad\text{for}\quad T_i\not\in\mathcal{U}\quad\text{where}\quad\mathcal{U}=\{\text{UNCLASS, JUNK}\}}
\end{equation}

\edit{We then combined these quantities to estimate the probability of being a white dwarf:}

\begin{equation}
\pwd=\sum_i{\rho_i\edit{\tilde{f}_i\quad\text{for}\quad T_i\not\in\mathcal{U}}}
\end{equation}

\edit{We did not calculate \pwd for the few objects where no other classes were recorded besides UNCLASS or JUNK.}

The inclusion of $\log{g}$ in the definition of \pwdDESI limits its usefulness for some objects, such as extremely-low-mass (ELM) white dwarfs \citep[e.g.][]{Brown2010ELM,Pelisoli2019sDA} and their precursors \citep[e.g.][]{Maxted2011,Maxted2013}. Such objects overlap with the sDA stars \citep{Brown2017sDA,Pelisoli2018sDA}, whose subdwarf-like spectra appear hydrogen-rich and suggest surface gravities below the threshold where we assign low \pwdDESI scores. The \texttt{MWS} survey did not specifically target ELMs, so more may await discovery among spectra from other DESI surveys.

Our $\pwdDESI$ metric is based primarily on spectral classifications, and thus supersedes the HR-diagram-based metrics (\pwdGaia) of GF19 and GF21. However, we still include those values in our catalogue for comparison (see Section~\ref{subsectionContentsClassifications}) and as they serve to identify input catalogue membership.

\subsection{Model type and  stellar parameters}
\label{subsectionCatalogueModelTypeStellarParameters}

Fitting every object with models $\mathcal{M}_j$, where $\mathcal{M}_j\in\{\text{H, He, He+H}\}$, yielded three sets of inferred parameter values with associated Bayesian evidences $\mathcal{Z}_j$. We select the best-performing model, as described below, and report that in the main catalogue (Table~\ref{tableMainCatalogue}) with an associated confidence.

For objects with $\pwdDESI\geq0.5$ we also report stellar parameters, distance and extinction for the best-performing model. We use the median and half the 16th--84th percentile range of the marginal posterior distribution as the value and uncertainty for each parameter. We emphasise that for white dwarfs not well-described by our models (e.g. carbon-rich DQs), results should be treated with caution.

The best-performing model returns the largest Bayesian evidence $\mathcal{Z}$. When comparing the relative performance of two models A and B, a Bayes factor of $\ln{\mathcal{Z}_A}-\ln{\mathcal{Z}_B}\gtrsim5$ is typically considered decisive support for model~A. We use the evidence to select the best model, but with an adjustment for mixed-atmosphere models.

As discussed in Section~\ref{subsectionContentsStellarParameters}, values of \logh near the upper bound of the model grid are likely not physical, as they are not expected theoretically and would have DA spectra almost indistinguishable from those of H-dominated stars. In hindsight a physically-motivated prior on \logh would have been a better choice than the uniform prior we adopted, so we made a post hoc adjustment as follows.

We defined a survival function $s(\theta)$ for $\theta=\logh$, using a threshold value of $\theta_0=-2$, softening factor $\sigma_0=0.3$, and the Gaussian cumulative probability function $\Phi$:

\begin{equation}
s(\theta)_j=\begin{cases}
    1 & \text{for } \mathcal{M}_j\in{\{\mathrm{H, He}\}}\\
    \Phi\left(\frac{\edit{\theta_0-\theta}}{\sqrt{\sigma_\theta^2+\sigma_0^2}}\right) & \text{for } \mathcal{M}_j={\mathrm{He+H}}
  \end{cases}
\label{equationEvidencePenalty}
\end{equation}

We then combined the survival function with the Bayesian evidences from the model fits to yield a pseudo-posterior probability for each atmosphere type. Using evidences $\mathcal{Z}_j$ for the fits to the three models $\mathcal{M}_j$, we calculated a relative weight for each model:

\begin{equation}
\label{equationAtmosphereTypeRelative}
\tilde{w}_j=s(\theta)_j\mathcal{Z}_j
\end{equation}

Normalising so that $\sum_j w_j=1$ for each object, we report the weight $w_j$ as the model confidence:

\begin{equation}
\label{equationAtmosphereTypeNormalised}
w_j=\frac{\tilde{w}_j}{\sum_{k}{\tilde{w}_k}}
\end{equation}

The highest-confidence model (as defined by Eqn.~\ref{equationAtmosphereTypeNormalised}) is listed in the main catalogue.

\subsection{Radial velocity}
\label{subsectionCatalogueRV}

We report a single radial velocity in the main catalogue for objects where a reasonable fit using Balmer or \ion{He}{i} lines might be expected, specifically those whose spectral type includes an `A' or `B', but not an `H' (which indicates Zeeman-split lines). We take an inverse-variance-weighted average of the values measured independently from the B and R spectrograph arms (Section~\ref{subsubsectionRVs}). The values given are the line velocities as measured, i.e. they include both space velocities and gravitational redshifts. Only average velocities that are consistent with bound Galactic orbits ($|v_r|<750$\,\kms) are included in the main catalogue, with the rest provided in supplemental Table~\ref{tableRVs}.

We also report $p$-values for our radial velocity variability test for all three loss functions we considered (Section~\ref{subsectionRVvariation}).

\subsection{Signal-to-noise ratio}
\label{subsectionCatalogueSN}

We calculate \SN using short continuum regions in each spectrograph arm: 4550--4650\,\AA, 6000--{6100\,\AA} and 8150--8250\,\AA. Within each region, we divide fluxes by their uncertainties and treat the median as the nominal \SN. However, as mentioned in Section~\ref{subsectionDataIssues}, uncertainties can be underestimated by the DESI reduction pipeline, so we compute an alternative \SN estimate from the data by fitting a quadratic function to each region, calculating the residuals, and dividing their median by half the 16--84th percentile range. Throughout the paper, we use the nominal values for the blue arm when quoting \SN. Comparison against our measurements confirms that nominal {\SN} is systematically overestimated when $\SN\gtrsim20$ (Fig.~\ref{figureSN}).

\subsection{Flux contamination}
\label{subsectionCatalogueFluxContamination}

DESI observations of white dwarfs can be contaminated by flux from nearby stars, leading to misidentification of spectral type, or incorrect estimates of stellar parameters, luminosity and distance. For spectroscopic surveys like DESI, contamination can occur both spatially and spectrally. Spatially, the point-spread functions (PSFs) of nearby stars may overlap with the fibre aperture at the focal plane, and we attempt to quantify this effect as detailed below. Spectrally, light exiting adjacent fibres at the spectrograph end may scatter onto the region of the CCD where the target spectrum is measured, but we do not investigate this further.

We queried the \textit{Gaia} DR2 or DR3 catalogue in a 60-arcsec radius around each object in our catalogue. While even the brightest sources contribute negligibly to fibre contamination when 20\,arcsec away, a wider search radius captures high-proper-motion objects. We estimated the flux entering the fibre from each source, using their proper-motion-propagated positions in each exposure. We assumed that fibres are circular with radius 0.75\,arcsec, although in reality their projected apertures vary across the focal plane \citep{DESI2022}. We also assumed that the \textit{Gaia G}~band approximates the DESI spectral response, and that source PSFs are Gaussians with a full-width-at-half-maximum defined by the seeing. Where there are multiple exposures we sum their flux contributions based on their effective exposure times. We report the fractional contamination, i.e. the contribution to flux entering the aperture from field stars relative to the target.

We repeated the calculation for the photometric surveys whose data we use, to identify and exclude potentially problematic measurements before fitting. We propagated source positions to the average epoch for each survey (2003 for SDSS, 2012 for Pan-STARRS, 2015.5 for \textit{Gaia} DR2, and 2016.0 for {Gaia} DR3). Instead of a fibre radius, we used an aperture that captures 99~per~cent of the encircled energy. We assumed Gaussian PSFs with FWHM set to the median plus $1\upsigma$ uncertainty of the band with the largest FWHM for SDSS and Pan-STARRS (1.52 and 1.41\,arcsec, respectively), and adopt 0.25\,arcsec for \textit{Gaia G}. We roughly approximated the \textit{Gaia} $G_\mathrm{BP}$ and $G_\mathrm{RP}$ photometry by using a circular aperture and Gaussian PSF both of radius 1.77\,arcsec. However, in reality they are slitless prism spectra integrated over a rectangular $3.5\times2.1$\,arcsec window \citep{Carrasco2021}, the reported photometry combines numerous measurements made at different dispersion angles, and the \textit{Gaia} data reduction pipeline attempts to deblend overlapping sources. Photometric flux contamination ratios are reported in Table~\ref{tableExternalData}.

\begin{figure*}
\centering
\includegraphics[width=0.49\textwidth]{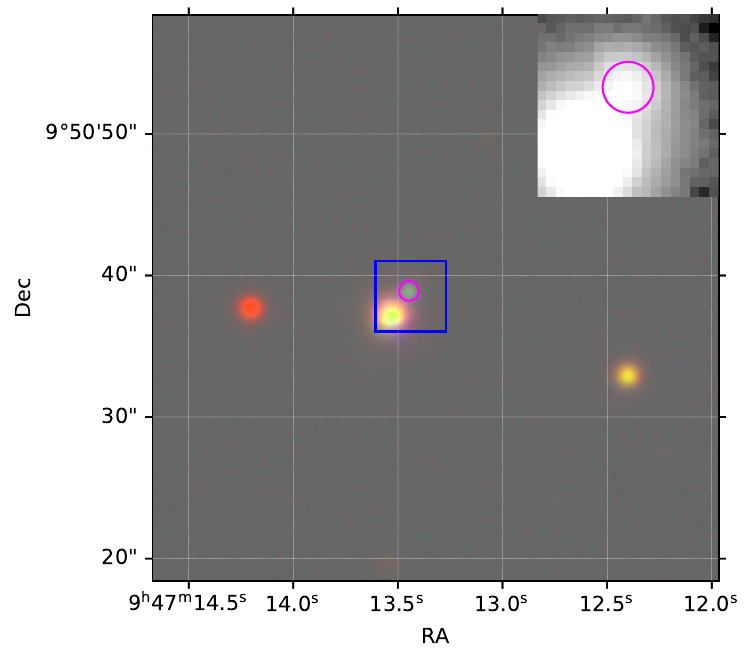}
\hfill
\includegraphics[width=0.49\textwidth]{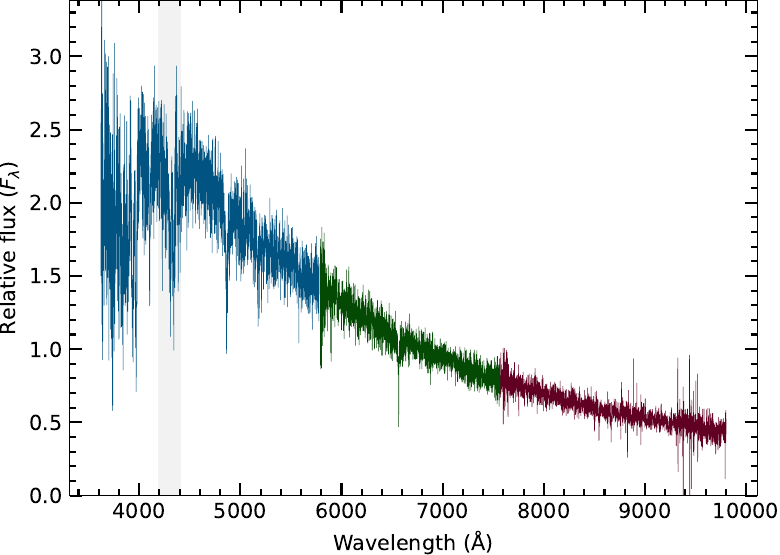}
\caption{White dwarf WD\,J094713.44+095038.94, whose DAZ spectral type likely results from a main-sequence neighbour within the DESI fibre radius.
\textit{Left panel}: target and field stars as seen by the DESI Legacy Imaging Survey, with the fibre aperture (radius 0.7\,arcsec) shown at the target position. The $g$, $r$ and $z$ band data are assigned to the blue, green and red channels, respectively, and rendered with an asinh stretch (softening factor 7). The inset shows the $g$-band data in a 5\,\arcsec square around the target, shown with a logarithmic stretch to emphasise the contamination from the neighbouring star, which is over 20 times brighter.
\textit{Right panel}: The spectrum, shown at full resolution and normalised. Grey shading indicates the region potentially affected by an instrumental artefact.}
\label{figureContamination}
\end{figure*}

Fig.~\ref{figureContamination} shows a white dwarf where we estimate fractional flux contamination of 0.76 by the neighbouring main-sequence star. Its DAZ spectrum suggests a white dwarf spectacularly enriched in metals, which would typically be caused by planetesimal accretion, but we suspect those features originate in the neighbouring star instead.

The catalogue contains 532 sources where we estimate spectral contamination at 1 per cent or higher, the threshold above which we rejected contaminated photometry when fitting stellar parameters. The class with the highest rate of contamination at that level is WD+MS, where 10 per cent of sources are affected. The DBAZ class has a 2 per cent contamination rate, further illustrating the need for caution when interpreting spectral types.

Our method uses a Gaussian approximation to the PSF, and so does not account for second-order effects like diffraction spikes or the wings of the Airy disc. For example, WD\,J073717.69+395858.25 drew our attention during the classification exercise as it presents a DC spectrum in the B~arm, while the R and Z~arms show subtle, unusual features, but they are likely associated with the naked-eye star lying 2 arcmin away, which is too distant to be flagged by our metric.

\subsection{Supplementary information}
\label{subsectionCatalogueSupplementaryInfo}

The main catalogue presents our best estimates of the types and parameters of each object, within certain limitations, and should be sufficient for most purposes. However, users may wish to understand how a particular entry in our catalogue came to be, or to assemble their own samples or catalogue. Therefore, intermediate data products are also supplied, without any of the quality cuts that we applied to the main catalogue, as detailed in Appendix~\ref{appendixSupplementaryDataFiles}. They include: selected metadata for the individual exposures analysed here, reproduced for convenience from the DESI cframe file headers (Table~\ref{tableExposures}); full records of the classification process (Table~\ref{tableClassificationLogs}); and radial velocity measurements for all co-added spectra and individual exposures using all model templates (Table~\ref{tableRVs}).

We emphasise that intermediate products are exactly that: \textit{they are not curated, science-ready outputs}, and should be treated with the same caution as e.g. uncalibrated CCD images.

\section{Properties of the DESI DR1 white dwarf sample}
\label{sectionCatalogueContents}

Our catalogue contains 63\,968 objects, including 44\,409 white dwarfs ($\pwdDESI\ge0.5$), of which 38\,904 form the primary sample, and 5312 had not been targeted as white dwarfs but were observed incidentally for other reasons. We now explore the catalogue in detail, assessing its completeness, showing the distribution of various types on the HR~diagram, comparing our classifications against other white dwarf catalogues, assessing the results of the model fitting, examining radial velocities and kinematics, and discussing some rare white dwarf classes.

\subsection{Completeness and biases}
\label{subsectionCompleteness}

We characterise the completeness of the DESI DR1 white dwarf sample, and contrast it with SDSS, the survey most comparable in scale and spectral resolution. As \textit{Gaia} white dwarf candidates were used exclusively in the DESI target selection, our results inherit any biases inherent to the input sample. However, such biases should be minimal: the \textit{Gaia} white dwarf sample is estimated to be 97 per cent complete within 40\,pc \cite{OBrien2024_40pc}, and up to 93 per cent complete for objects with $G\le20$\,mag and $\pwdGaia\ge0.75$ (GF21).

We define a baseline sample by selecting sources with $G\le20$\,mag and $\pwdGaia\ge0.75$ from the GF21 white dwarf catalogue. We then calculate the fraction of those objects that were actually observed as a function of various quantities, detailed below. We rely on the SDSS~DR16 spectral classifications in the \textit{Gaia} white dwarf catalogue to identify SDSS observations.

\begin{figure*}
\centering
\includegraphics[width=0.9\textwidth]{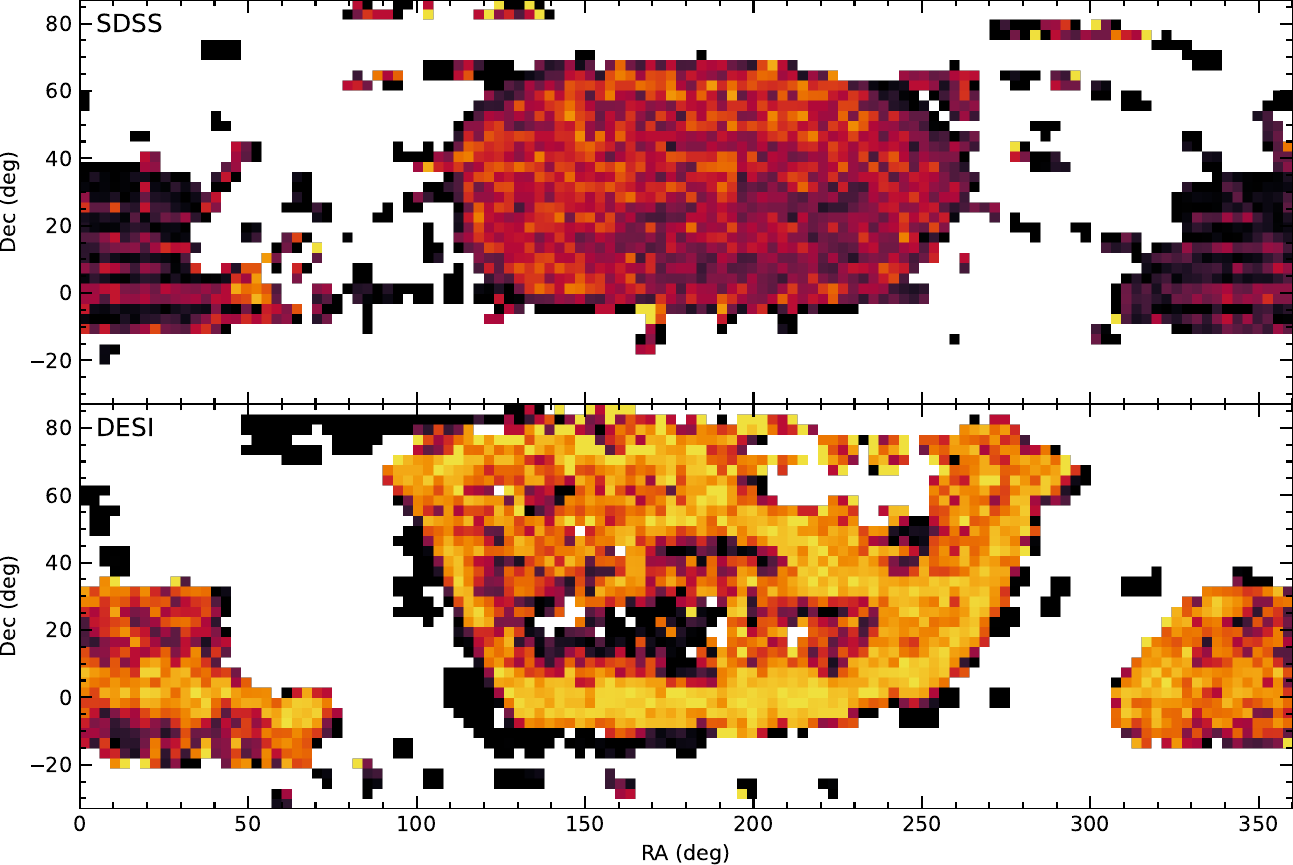}
\includegraphics[width=\textwidth]{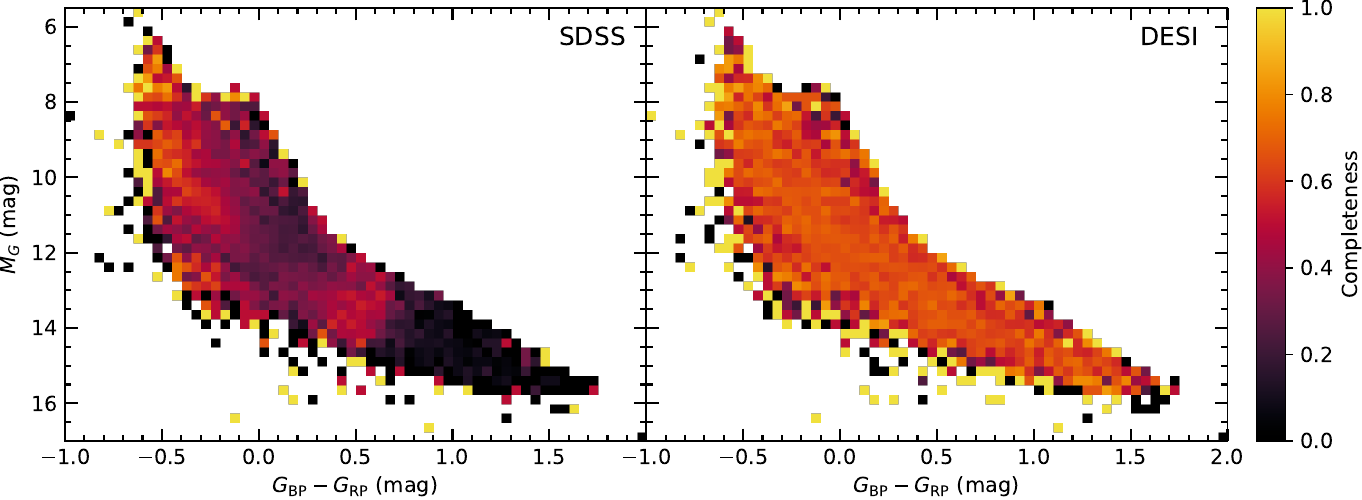}
\includegraphics[width=\textwidth]{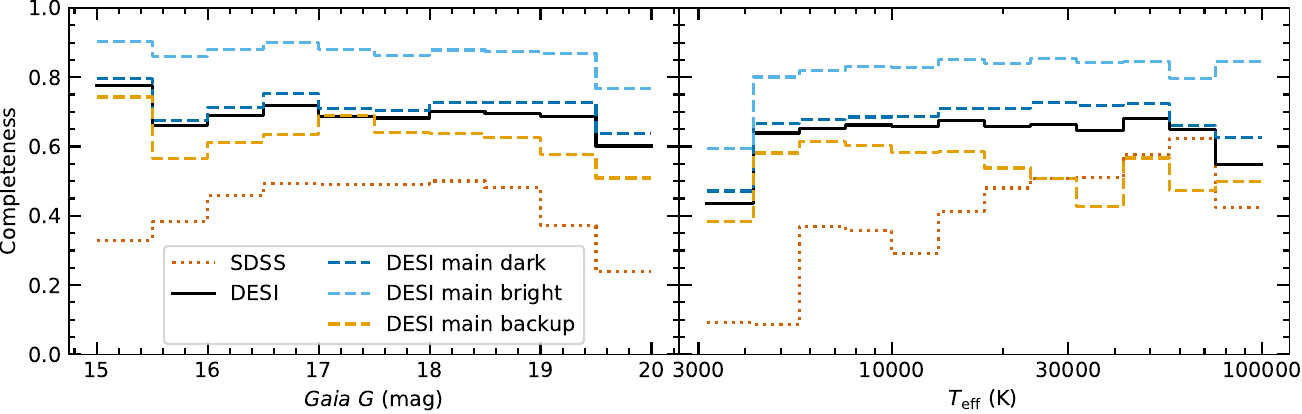}
\caption{SDSS vs DESI comparison of observational completeness of white dwarf candidates ($\pwdGaia\ge0.75$; $G\le20$\,mag) as a function of sky position (\textit{top panel}), HR~diagram position (\textit{middle panels}), and \textit{Gaia G} magnitude and effective temperature (\textit{bottom panel}). The colour bar is common to all 2D histograms. The SDSS~DR16 sample is based on spectral classifications in the GF21 dwarf catalogue.}
\label{figureCompleteness}
\end{figure*}

To calculate on-sky completeness, we use sources that lie within 1.6\,deg of the centre of a DESI tile, i.e. within the field of view. For SDSS, we use sources within 1.49\,deg of a plate centre\footnote{\url{https://data.sdss.org/sas/dr16/casload/spCSV/plates/sqlPlateX.csv.bz2}}. While not all regions had been covered by the DR1 cut-off date, DESI still achieves higher completeness than SDSS across large areas of the sky (Fig.~\ref{figureCompleteness}, upper panels). The low-completeness patches at the borders of and beyond the main body of DESI observations are tiles observed only by the \texttt{backup} program, whose brighter magnitude limit leads to fewer sources being targeted.

DESI outperforms SDSS in both completeness and uniformity, as measured on the \textit{Gaia} HR~diagram (Fig.~\ref{figureCompleteness}, middle panels). This is unsurprising: DESI deliberately targeted white dwarfs, with an inclusive selection function, whereas the majority of white dwarfs in SDSS were observed incidentally, sometimes by programs that specifically sought to exclude them \citep[e.g.][]{Richards2002}. As a result, SDSS is biased towards hotter white dwarfs, and there is a noticeable dip in its completeness near $\bprp\approx0.2$ (corresponding to $\Teff\approx10\,000$\,K), and a sharp drop beyond $\bprp\gtrsim0.7$ ($\Teff\lesssim6000$\,K).

Similarly, DESI is largely unbiased across \textit{Gaia G} magnitude and effective temperature, except at extreme values where far fewer sources are available (Fig.~\ref{figureCompleteness}, bottom panels). However, investigating completeness for subsets of DESI DR1 reveals that the the \texttt{backup} program is biased by its more complex selection function.

Overall, DESI DR1 achieves good uniformity across the entire white dwarf sequence, validating its utility for statistical samples. Gaps in sky coverage within the survey footprint will be filled by future data releases. The main limitation of the DESI white dwarf sample is that it is magnitude-limited. Nearly-complete volume-limited samples have been assembled within 13\,pc \citep{Holberg2002,OBrien2026_13pc}, 20\,pc \citep{Giammichele2012,Hollands2018_20pc}, 25\,pc \citep{Holberg2016} and 40\,pc \citep{OBrien2024_40pc}, providing clean local benchmarks, but they are orders of magnitude smaller than the DESI sample.

\subsection{HR diagrams}
\label{subsectionHRDiagrams}

\begin{figure*}
\centering
\includegraphics[width=\textwidth]{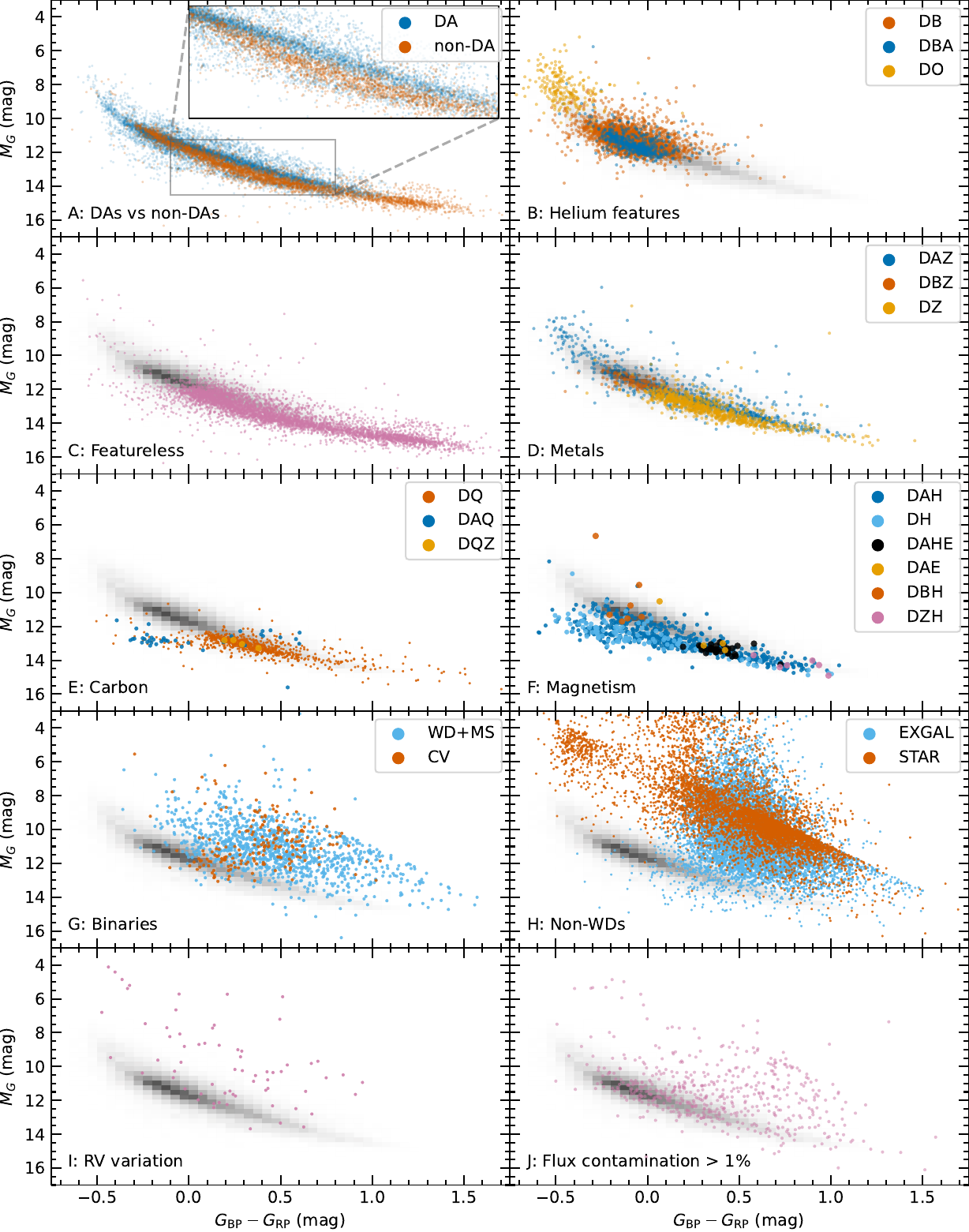}
\caption{HR diagrams of various subsets from the catalogue, described in detail at the beginning of Section~\ref{sectionCatalogueContents}. Symbol sizes are varied only for clarity, and have no other meaning.}
\label{figureHRdiagrams}
\end{figure*}

Fig.~\ref{figureHRdiagrams} shows HR~diagrams highlighting various features from the catalogue. Labels in the legends are inclusive, covering all subtypes within the class (e.g. `DA' represents both DA and DAZ types).

Most panels include a 2D histogram of white dwarfs observed by DESI (selected using $\pwdDESI\ge0.8$), shaded on a logarithmic scale. Absolute magnitudes are calculated from fitted distances when creating the 2D histogram, to reduce scatter, but all over-plotted points use absolute magnitudes calculated using inverse-parallax distances.

Panel~A separates DAs~--~the most common white dwarf spectral type~--~from the various non-DA classes, many of which feature in the other panels. The bifurcation in the white dwarf sequence first identified in the \textit{Gaia}~DR2 HR~diagram \citep{Babusiaux2018} is clearly visible, and our classifications clearly show DA types populating the upper track, with non-DA types on the lower track. The upper branch follows the cooling track for H-atmosphere stars of $0.6\,\Msun$, where the white dwarfs mass distribution peaks \citep[e.g.][]{Bergeron1992,Kepler2007,Falcon2010,Genest-Beaulieu2019DB}. The lower branch is less well understood, and could represent a secondary peak in the mass distribution at $0.8\,\Msun$ \citep{ElBadry2018,Kilic2018}, spectral evolution \citep{Ourique2020}, or He-atmosphere stars with traces of H \citep{Bergeron2019} or C \citep{Blouin2023carbonTheory}. There is increasing evidence for the latter explanation, where convective mixing in He atmospheres dredges up C from the interior in quantities sufficient to modify the spectral continuum, but insufficient to produce detectable features at optical wavelengths \citep{Blouin2023carbonObservation, Camisassa2023}. 

Panel B shows the transition from DO to DB types (with \ion{He}{ii} and \ion{He}{i} features, respectively) towards the warmer end of the cooling track, reflecting the change in ionisation balance with temperature. Below that point, the picture is more complex: He-atmosphere white dwarfs can undergo spectral evolution, changing their appearance as they cool: H can diffuse upwards from deep in the envelope, or a surface layer of H can be mixed into the atmosphere as convection zones develop \citep{Bedard2024}, and H can be delivered to the surface by accretion of planetary material \citep[][Izquierdo et al. submitted]{GentileFusillo2017}. The lower end of the cooling track is unpopulated in this panel, where \ion{He}{i} transitions fade as stars cool below around 10\,000\,K and the relevant excited states depopulate.

Panel C shows the white dwarfs with spectral type DC (featureless continuum). Most DC stars have He-dominated atmospheres, and thus the distribution in this panel is approximately the complement of that shown in panel B. Below the main track, a number of DCs reside on or near the Q branch (an overdensity at $G_{\text{abs}}\approx13$ and $\bprp\lesssim0.2$, most obviously seen in panel~E) possibly as a result of high magnetic fields altering spectral features beyond recognition \citep{Berdyugin2023}. A few stars populate a separate region further below the cooling track where ultra-blue white dwarfs reside \citep{Kilic2020, Blouin2024CIA}, whose colour is a result of collisionally-induced absorption in a mixed He+H atmosphere.

In Panel D, metal features are seen across the HR~diagram, including below the main white dwarf track where high-mass stars reside, and at the red end where the oldest stars are found. In most white dwarfs the metals are photospheric, having been accreted from remnant planetary systems \citep{Farihi2010rockyPlanetesimals, Koester2015}, but there are other origins to consider. Stars are brighter at the hot end of the white dwarf cooling track, and thus observable to greater distances, increasing the likelihood of interstellar material being responsible for absorption features. Stars above the main track may have low masses, may have inaccurate parallaxes or colours, or may be in binaries where metals could be a signature of wind accretion \citep{Debes2006}. We do not examine stars with metals further here, as they will be the focus of separate papers.

Panel E shows some populations of stars with carbon features in their spectra. At the blue end, the Q~branch below the main white dwarf track on the HR~diagram has a high concentration of hot D\edit{(A)}Qs. That branch reflects a pile-up of high-mass stars experiencing a cooling delay \citep{Tremblay2019, Bedard2024BuoyantCrystals}, which could be merger products \citep{Shen2023}, whose atmospheres may be carbon-dominated with H and He present only in trace quantities \citep{Koester2020atmospheres, Hollands2020, Sahu2025}. Those on the main white dwarf cooling track are the classical DQs, which result from convective dredge-up of carbon from the core into the He envelope \citep{Pelletier1986}. These may represent the observable tail of a population that is otherwise hidden among the featureless DC stars shown in panel~C and forms the lower branch of the bifurcation in panel~A that was discussed above. \edit{Finally, the cooler DAQ stars appear to form a separate population from those on the Q branch, mostly lying above the main cooling track, which would be consistent with them being double-degenerates \citep{Vennes2012,Manser2024DESIEDR}.}

The magnetic objects in Panel F also trace the Q branch, supporting a merger origin for magnetic fields in massive white dwarfs \citep{Bagnulo2022}. The increased frequency of magnetism toward the older, cooler end of the white dwarf track is consistent with the delayed emergence of magnetic fields in white dwarfs undergoing single-star evolution, where high field strengths are mostly observed after the onset of core crystallisation \citep{Bagnulo2022}. The DAHe subclass show Zeeman-split Balmer lines in emission, of uncertain origin, and cluster together on the HR~diagram \citep{Gansicke2020DAHe}. There are a couple of outliers, although one is faint enough that its \textit{Gaia} colour is unreliable.

Binary types are mostly offset from the white dwarf track in Panel~G, due to additional flux from a companion star or accretion disc. They include cataclysmic variables, where a white dwarf is accreting material from a companion \citep{Warner1995,Inight2023}, and white-dwarf--main-sequence binaries \citep{RebassaMansergas2010,RebassaMansergas2025} where both stars contribute noticeably in the optical wavelength range.

Panel H shows that the distributions of non-white-dwarf objects overlap with the white dwarf sequence, emphasising the necessity of spectroscopic confirmation of white dwarf candidates (compare also Fig.~\ref{figureHRDsourceDensity}). Only 18~per cent of the objects shown in this panel were targeted as white dwarfs; the rest were observed incidentally by other programs and picked up by our cross-match against the entirety of DESI DR1 (see Table~\ref{tableSpecTypes} and the discussion in Section~\ref{subsubsectionContentsPWD}).

Panel I shows that more radial velocity variables lie above the white dwarf track more often than would be expected if they were randomly distributed throughout the white dwarf sample, consistent with velocity variations being caused by a companion star. They are discussed further in Section~\ref{subsectionContentsRVs} and tallied by spectral type in Table~\ref{tableRVvariation}.

Panel J shows that objects with a risk of fibre contamination from nearby sources are scattered across the HR~diagram, most likely because the photometry that determines their coordinates is also contaminated.

\subsection{Classifications}
\label{subsectionContentsClassifications}

We now assess the reliability of our \edit{visual} classifications by exploring how confidence scores vary between spectral types, quantifying the random error rate of human classifiers, and comparing our results against other catalogues. We also assess the performance of the \pwdGaia metric from the GF21 white dwarf catalogue.

\begin{figure*}
\centering
\includegraphics[width=\textwidth]{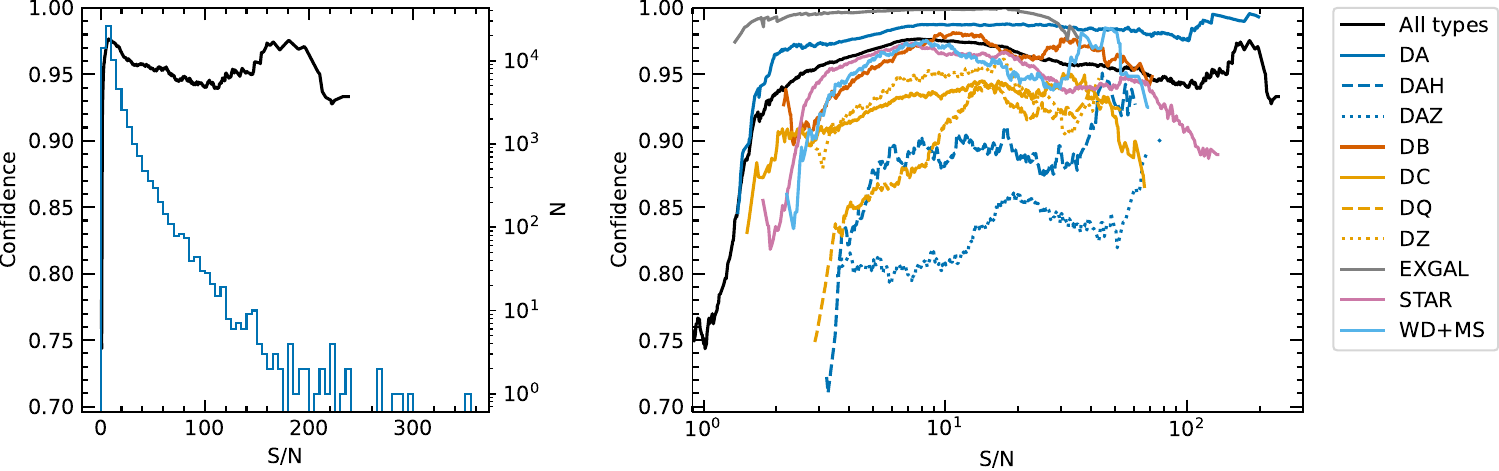}
\caption{\textit{Left:} Moving average of per-source spectral type confidence scores vs nominal signal-to-noise ratio of coadded spectra (black). The secondary $y$-axis shows a logarithmic histogram of $\SN$ for coadded spectra (blue). \textit{Right:} Moving averages of confidence for common spectral types. The $x$-axis is logarithmic, to highlight details at low $\SN$.}
\label{figureConfidenceVsSN}
\end{figure*}

The left panel of Fig.~\ref{figureConfidenceVsSN} shows how spectral type confidence varies with $\SN$. At $\SN\lesssim4$, confidence rapidly falls off, as spectral features become obscured by noise. The slight downward trend towards $\SN\approx120$ reflects the spectral similarity between hot white dwarfs and subdwarfs. The right panel of Fig.~\ref{figureConfidenceVsSN} shows how confidence depends on spectral class. The broad emission features of quasars are easy to identify at almost any \SN, so they maintain an average confidence close to 1.0. Metal lines are identified less confidently than most other features, as they are often weak or intrinsically narrower than H or He lines, and can be confused with residuals from the main-sequence standard stars used in the data reduction. Magnetic splitting and carbon features are also less confidently recognised, as they are often subtle.

While spectral types were unanimously assigned to most of the stars in the catalogue, around 5000 objects received multiple classifications. The most common disagreements involving white dwarfs were between DA/DAZ, DA/DC, DA/DAH, and DQ/DC types, and stellar contaminants are also confused with several white dwarf types. These are all common classes, so it is no surprise that they are well represented among disputed objects. However, the fact that there are thousands of disagreements highlights the need for a more nuanced approach than attempting to reach consensus on every object. Our confidence-based approach retains potentially useful information, such as the suspicion of metal lines that would signpost a planetary system, which would otherwise be lost in a consensus catalogue.

As our classification procedure is necessarily subjective, we attempted to gauge its reliability. During verification, workers processed batches of objects classified with the same spectral type. Some batches were injected with objects of other types selected at random, providing a control sample of 553 `misfits'. The workers rejected 89~per cent of these, which at first glance seems a surprisingly low success rate. However, most of the misfits that were not rejected were of spectral types similar to the primary type of their batch (e.g. a DAZ in a DA batch), and only half of those were accepted, with the remainder flagged as uncertain. When considering only obviously dissimilar types (e.g. DA vs DB, DA vs DQ), 1.8~per cent of the misfits were accepted, and we take this as the baseline error rate for human workers processing the DESI sample. If classification errors occurred randomly at that rate, then the expected number of random errors in the catalogue is less than one, as each spectrum received at least three independent classifications, and disagreements prompted re-inspection.

A further test of reliability is enabled by DAHe stars, whose Zeeman-split Balmer emission lines can be subtle and challenging to identify. A focused search of DESI spectra specifically targeting this rare class has been undertaken \citep{Manser2023DAHe}. That study used all DESI spectra that had been taken at the time, of which the majority are included in DR1, totalling 19 secure identifications and 13 uncertain candidates. As the search was conducted on DESI spectra by one of the workers who also performed visual classification for this catalogue, we included their DAHe identifications in the consensus exercise described in Section~\ref{sectionSpectralClassification}. However, re-generating the catalogue \textit{without} those votes reveals that we would otherwise have classified three of the 19 secure identifications as DC stars. While 16 out of 19 (84~per cent) is a respectable recovery rate for a classification exercise where spectra were typically viewed for only a few seconds in total, it highlights the risk that subtle features may escape detection.

\subsubsection{Performance of \pwd metrics}
\label{subsubsectionContentsPWD}
\begin{figure*}
\centering
\includegraphics[width=0.475\textwidth]{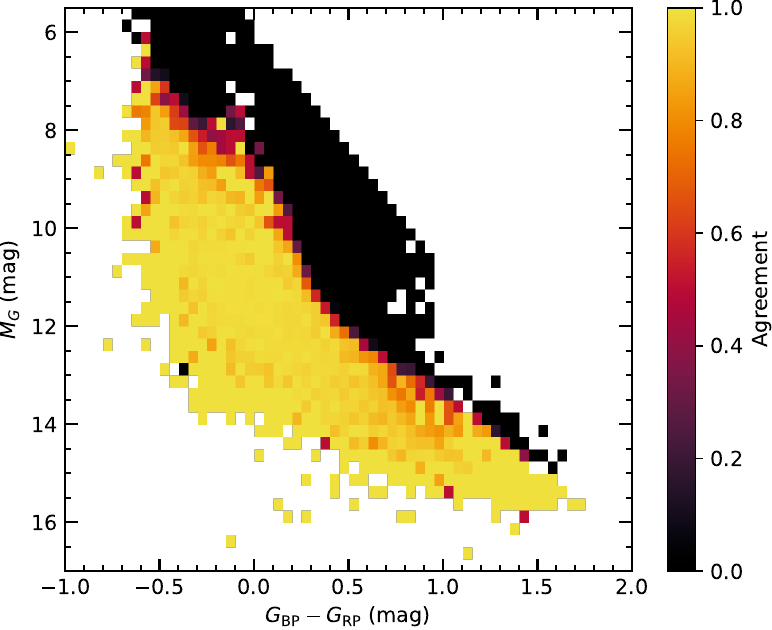}
\hfill
\includegraphics[width=0.475\textwidth]{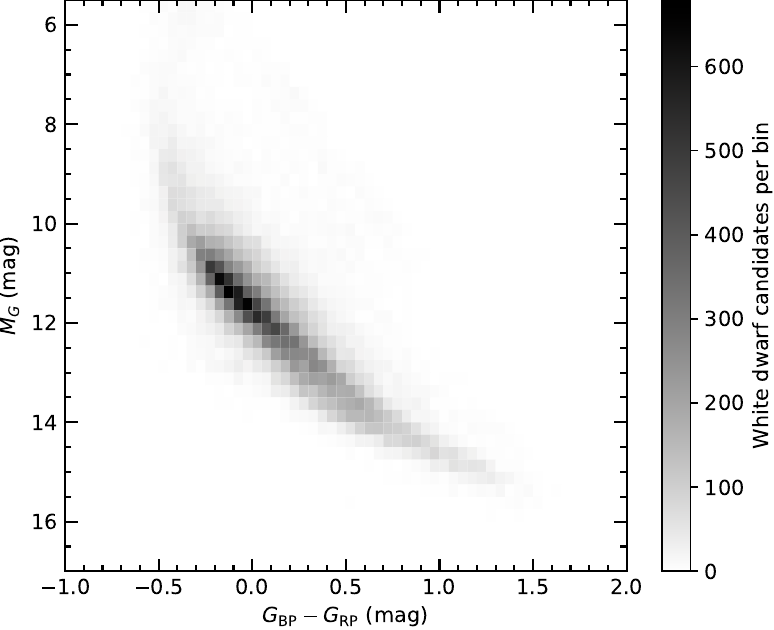}
\caption{Performance of the $\pwdGaia>0.75$ cut often used to select a clean sample of white dwarfs from the GF21 white dwarf catalogue, measured using the primary sample. \textit{Left panel}: The histogram shows the fraction of objects in each cell that score $\pwdDESI>0.5$. \textit{Right panel}: Source counts in the sample. Note that the low-agreement region contains very few objects.}
\label{figurePwdComparison}
\end{figure*}

The mean values of \pwdGaia and \pwdDESI (Table~\ref{tableSpecTypes}) are generally high and in good agreement for white dwarf classes, which is unsurprising as the \textit{Gaia} metrics were calibrated against a spectroscopic sample from SDSS. The recommended $\pwdGaia>0.75$ cut to select a high-confidence photometric sample of white dwarfs performs well when tested against our catalogue: 97~percent of such objects are assigned $\pwdDESI>0.5$, both overall and when restricted to the primary sample. Fig.~\ref{figurePwdComparison} shows how the $\pwdGaia>0.75$ threshold performs across the HR~diagram for the primary sample. Disagreements are concentrated above the main white dwarf track.

Objects classified as DAB have a lower mean $\pwdDESI$ score than other classes associated with mixed He+H atmospheres (Table~\ref{tableSpecTypes}). Most low-\pwdDESI DABs also have fitted $\log{g}<7.2$ and $30\,000\le\Teff\le40\,000$\,K, suggesting that human classifiers struggle to distinguish white dwarfs from stellar contaminants in this region of parameter space. Similarly, DAHe stars have a lower mean \pwdDESI score than their mean \pwdGaia, owing to the number of `unidentifiable' votes they received during classification. Conversely, non-white-dwarf and binary classes are confidently identified by our $\pwdDESI$ metric, as their spectra are often straightforward to classify.

Notably, \textit{Gaia} DR3 white dwarf candidates that we identified as extragalactic objects from their DESI DR1 spectra have a mean $\pbarwdGaia=0.45$. While surprisingly high at first glance, it does not imply that the GF21 white dwarf catalogue is heavily contaminated by extragalactic objects, merely that those observed by DESI represent a sample that can be selected from the \textit{Gaia} catalogue in a similar way to white dwarfs. Indeed, only 18 of the 10\,222 extragalactic objects were targeted as white dwarfs in the primary sample (Table~\ref{tableSpecTypes}), i.e. almost all were observed under other programs, and happen to lie in the white dwarf region of the \textit{Gaia} HR~diagram. Further, the proper-motion selection criteria were relaxed between the GF19 and GF21 white dwarf catalogues, making it easier for quasars to be photometrically selected as white dwarf candidates. The stark difference between the \pwdDESI and \pwdGaia statistics for extragalactic sources demonstrates the necessity of spectroscopic confirmation of photometrically-selected white dwarf candidates.

\subsubsection{Comparison against automated methods}

As data volumes begin to exceed the practical limits of human inspection, there has been a timely surge of interest in automatic spectral classification via machine learning \citep[e.g.][]{Byrne2024,Kao2024,PerezCouto2025,Zhang2025}. Here, we compare our human classifications of DESI spectra against two studies that used machine-learning methods to classify \textit{Gaia XP} spectra. One applied a gradient-boosting classifier to about 100\,000 white dwarf candidates, with a training set whose labels were assigned based on SDSS spectra \citep{Vincent2024}. The other applied the random forest algorithm to about 80\,000 white dwarf candidates within 500\,pc, trained on labels from the MWDD \citep{GarciaZamora2025}. Both studies assign one of six spectral types (DA, DB, DC, DO, DQ, DZ) to each object, and can thus be compared with our classifications.

\begin{figure*}
\centering
\includegraphics[width=0.49\textwidth]{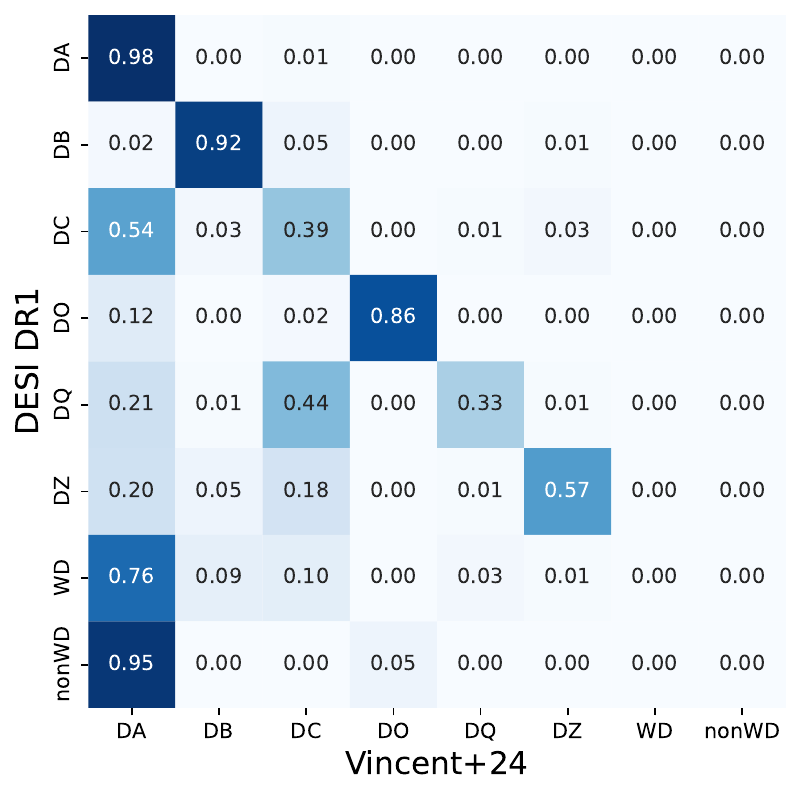}
\hfill
\includegraphics[width=0.49\textwidth]{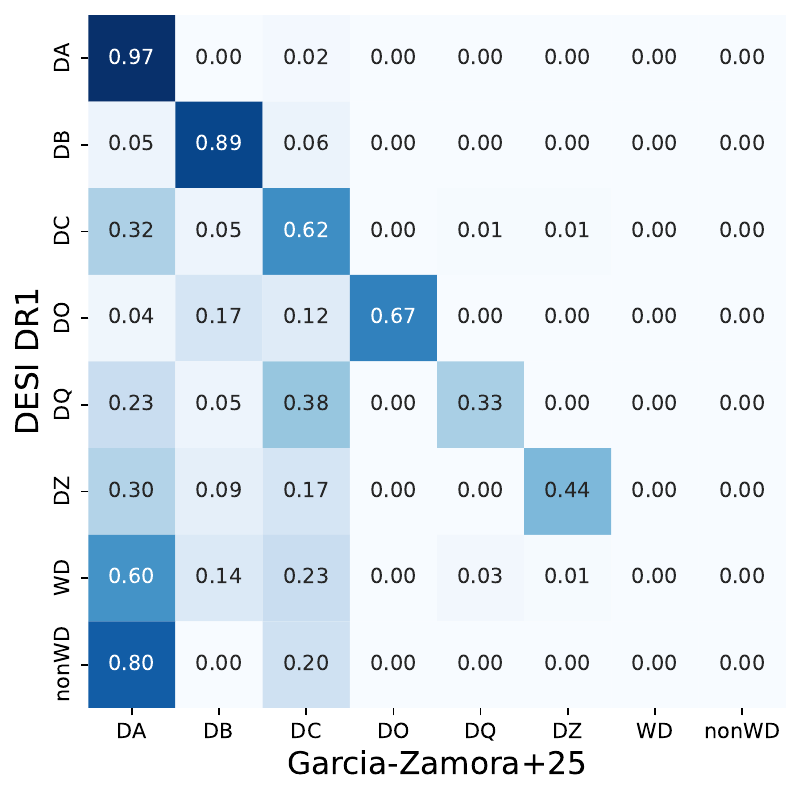}
\caption{Confusion matrices comparing automated classifications based on \textit{Gaia} spectrophotometry \citep{Vincent2024,GarciaZamora2025} against our human classifications based on DESI spectra. Each row of the matrix represents the objects assigned to a given type based on their DESI spectra, and the values in each cell show the fractions of those objects assigned the various labels considered in the \textit{Gaia}-based studies, so that values sum to unity along rows. The WD and non-WD classes encompass all types in the DESI catalogue that were not used by the \textit{Gaia} studies.}
\label{figureConfusion}
\end{figure*}

About 19\,000 white dwarf \edit{candidates} are common to the DESI DR1 catalogue and the \textit{Gaia} machine-learning studies. We group subtypes in the DESI catalogue together into their primary classes to make the datasets compatible. The remaining peculiar and binary types (e.g. CV, DH, WD+MS) are grouped under the `WD' label, while non-white-dwarf types are grouped together as `non-WD'. We investigated various methods for calculating confusion matrices, such as using all potential classifications for each object weighted by their confidences, but they all gave similar results, so we prefer the simplest method where only the highest-confidence spectral type for each object is used. The results are shown in Fig.~\ref{figureConfusion}.

The overall agreement is respectable, especially for DAs and DBs, and only a handful of non-white-dwarf objects passed the strict selection criteria of the machine-learning studies. However, there is significant disagreement for over 1000 objects that we classified as DC, DQ, or DZ, presumably because the higher-resolution DESI spectra make it easier to identify the often-subtle features or their absence. Interestingly, the gradient-boosting method of \cite{Vincent2024} is better at classifying DOs and DZs, while the random forest method of \cite{GarciaZamora2025} more reliably identifies DCs. When encountering spectral types absent from the training samples (WD and non-WD), the methods in both \textit{Gaia} studies tend to label objects as DAs. \edit{For example, the non-WD types labelled as DA turn out either to be classified as STAR or UNCLASS in our catalogue. The former have Balmer lines in their spectra, so the confusion is unsurprising, but the latter are likely an example of} the majority-class bias that can arise from an imbalanced training set.

\subsection{Stellar parameters and model types}
\label{subsectionContentsStellarParameters}

\begin{figure*}
\centering
\includegraphics[width=\textwidth]{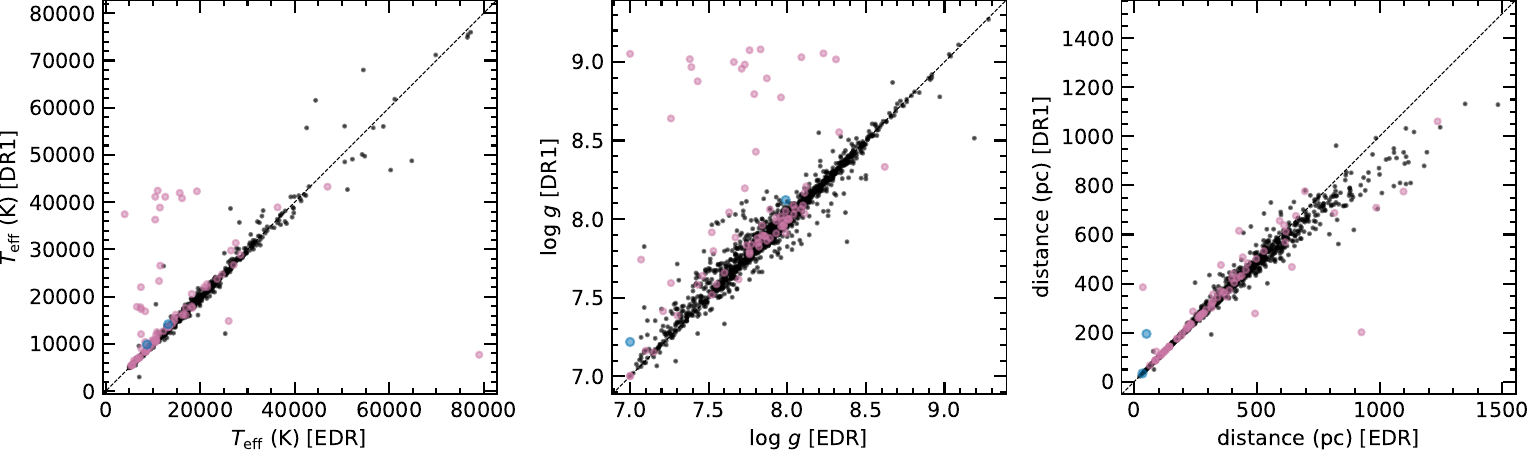}
\caption{Photometric parameters for DA- and DB-type WDs fitted common to both EDR and DR1. Stars with contaminated or saturated photometry are shown as red or blue points, respectively.}
\label{figureEDRvsDR1}
\end{figure*}

Fig.~\ref{figureEDRvsDR1} compares our photometric fit results against those we obtained for the DESI EDR white dwarf catalogue \citep{Manser2024DESIEDR}. The evolution of our methods between the two catalogues accounts for some of the outliers, particularly our rejection of contaminated and saturated photometry. The slight systematic reduction in larger distances from EDR to DR1 reflects the change to a simpler distance prior. The median fractional differences between the two catalogues are 0.003 in $\Teff$ and distance, and 0.0007 in $\log{g}$, half that found between the EDR results and those from independent fits to \textit{Gaia} photometry.  Given the good agreement between the EDR and DR1 results, we do not repeat the detailed comparisons against external results undertaken for the EDR.

Spectral classification is an unavoidably subjective visual task, and does not necessarily reveal the physical nature of an object. Model fitting is more objective, but can be influenced by systematics and artefacts that human classifiers would disregard. Nevertheless, there is strong agreement between results from the two methods: 98 per cent of objects that we classified as DAs were best fit by the pure-H model, and over 99 per cent of DB(A)s used the He or He+H models.

Discrepant results have several potential explanations, astrophysical and otherwise. For white dwarfs not well-described by the models (e.g. DQs with strong molecular bands or heavily metal-enriched DZs), the selected model is the least-bad fit, and may not reflect the dominant atmospheric element. Similarly, bad fits may indicate a double-degenerate where absorption lines are diluted by flux from a companion, but we only model single stars.

There is no reason to expect a H-dominated atmosphere to show He lines, yet the H-atmosphere solution \edit{was the best fit for one DB (WD\,J151024.78+734909.08). The fit returned $\log{g}\approx7$, at the edge of the model grid, suggesting a problem, and indeed the photometry turned out to be unreliable, skewing the fit.} While we tried to detect and exclude such data, inevitably some remains, and in such cases the spectroscopic fits (Table~\ref{tableFullFitResults}) are likely to be more reliable.

\begin{figure}
\centering
\includegraphics[width=\columnwidth]{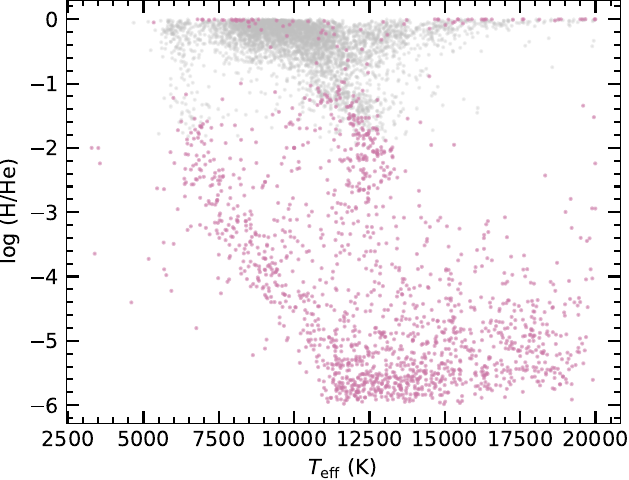}
\caption{Stellar parameters for stars with an `A' or `B' in their spectral types where the mixed He+H model provided the best fit. Grey points show results that were suppressed, i.e. those where the mixed He+H solution was disfavoured after application of the evidence (Equation~\ref{equationEvidencePenalty}). Coloured points show results that survived and are presented in the catalogue.}
\label{figureLogH}
\end{figure}

Conversely, He-dominated atmospheres can present a DA spectrum when sufficient quantities of H are mixed in. However, the similarity between the spectra of stars with pure-H and H-rich He atmospheres makes definitive identification a challenge. The DA types that were best fit by mixed He+H models include He-dominated stars with trace H (e.g. WD\,J215947.89$-$075410.21; Fig.~\ref{figureSpectralAtlas}). However, we also found an unexpected cluster of results near the model grid boundary with $\logh\approx0$ (Fig.~\ref{figureLogH}). Stars with such high H abundances would present spectra that are almost indistinguishable from H-dominated atmospheres \citep{Bergeron1991}, but are not known to be common \citep{Tremblay2010}, and are not theoretically expected from stellar evolution \citep{Bedard2024}. Moreover, fits that converge at the edge of a model grid are often problematic.

Inspection of the fits suggests that the high $\logh$ results are unreliable. Often, the difference between the high $\logh$ model and a pure-H model is subtle, but the spectroscopic data nevertheless overwhelm the photometry and parallax. That may be due to incorrectly-estimated uncertainties, but an investigation into systematics is beyond the scope of this paper, so as a compromise we retrospectively applied a penalty to the Bayesian evidence to discourage high-$\logh$ results, as described in Section~\ref{subsectionCatalogueModelTypeStellarParameters}. Fig.~\ref{figureLogH} shows the \logh results and the effect of applying the evidence penalty. While effective, it is a post hoc solution adopted after inspection of the results. We therefore recommend that future work explore a physically-motivated prior on \logh, considering both stellar evolution \citep{Rolland2018, Rolland2020} and external sources of H \citep{GentileFusillo2017}.

\begin{figure}
\centering
\includegraphics[width=\columnwidth]{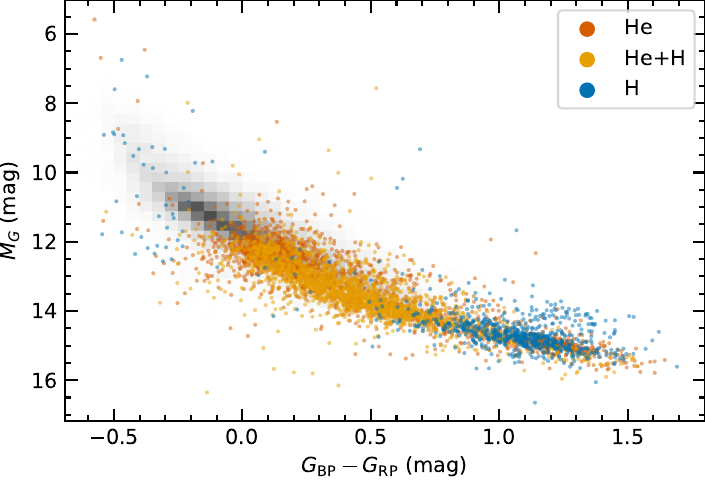}
\caption{HR diagram showing which atmosphere type gave the best model fit to DC stars.}
\label{figureDCs}
\end{figure}

Fig.~\ref{figureDCs} reproduces panel~C of Fig.~\ref{figureHRdiagrams}, but with the points coloured according to the best-fitting model, revealing a clear preference for mixed-atmosphere models for warmer stars. Determining the composition of DC stars is challenging, requiring well-calibrated data \citep[e.g.][]{Blouin2019cool,Caron2023}, so we refrain from interpreting these results.

\subsection{Distances and reddening}
\label{subsectionContentsDistancesReddening}

These two quantities are fitted primarily to avoid underestimating uncertainties on $\Teff$ and $\log{g}$, which can be strongly correlated with distance (as is reddening). However, they are not purely nuisance parameters, and we now examine our results for stars with reliable fits (specifically, high-confidence DA and DB type stars with model fits consistent with their spectral type).

\begin{figure*}
\centering
\begin{minipage}[t]{0.3\textwidth}
  \includegraphics[width=\linewidth]{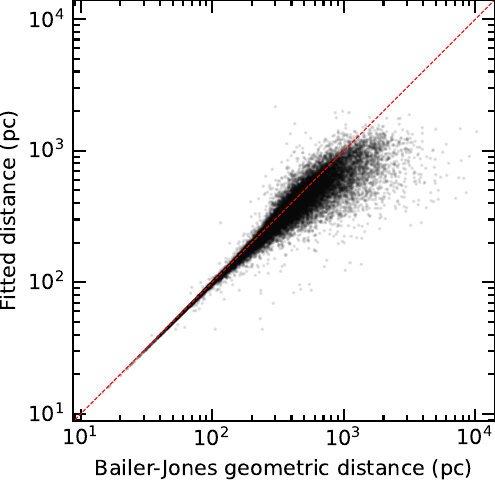}
  \captionof{figure}{Comparison between distances inferred for this catalogue and those estimated from their \textit{Gaia} data \citep{BailerJones2021}, showing that the latter are systematically larger.}
  \label{figureDistance}
\end{minipage}
\hfill
\begin{minipage}[t]{0.3\textwidth}
  \includegraphics[width=\linewidth]{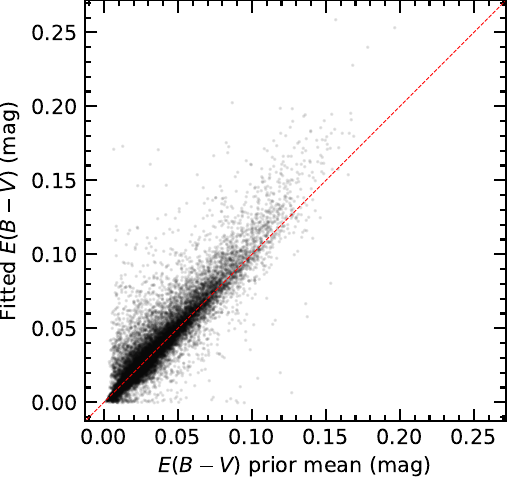}
  \captionof{figure}{Comparison between the reddening we infer and the prior we constructed from 3D extinction maps.}
  \label{figureReddening}
\end{minipage}
\hfill
\begin{minipage}[t]{0.3\textwidth}
  \includegraphics[width=\linewidth]{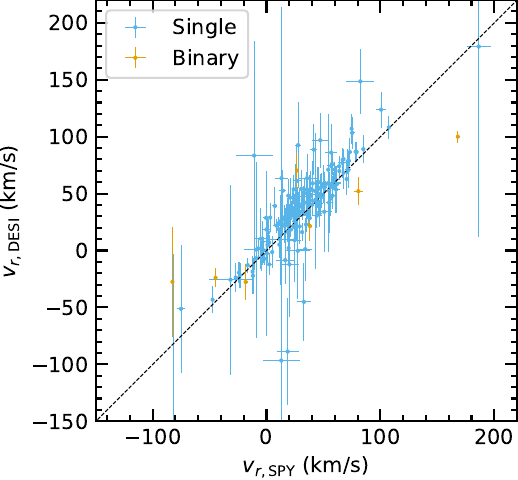}
  \captionof{figure}{Comparison between radial velocities measured here and by the SPY survey, for stars common to both.}
  \label{figureSPY}
\end{minipage}
\end{figure*}

Fig.~\ref{figureDistance} compares our fitted distances against those inferred under a popular geometric prior \citep{BailerJones2021} available from the \textit{Gaia} archive (\texttt{external.gaiaedr3\_distance.r\_med\_geo}). Both distance inferences rely on the \textit{Gaia} parallax, so agree well for nearby stars with small parallax uncertainties, where the prior has a negligible influence. However, the standard Bailer--Jones estimates are larger than ours for more distant stars whose noisy parallaxes are only weakly constraining, exceeding 10\,kpc in extreme cases. This is expected: while our prior follows the Bailer--Jones prescription, it is calibrated for a white dwarf population, whereas the original Bailer--Jones distances are calibrated for a population dominated by main sequence and giant stars. A substantial fraction of their prior probability distribution covers distances of several kpc or more, and they are therefore inappropriate for white dwarfs, which typically have $G_{\mathrm{abs}}>10$\,mag and thus must lie within 1.6\,kpc to be observable by \textit{Gaia}, whose faint limit is $G\lesssim21$\,mag. Indeed, for white dwarfs in the DESI sample, Bailer--Jones distances are on average larger than those obtained by inverting parallaxes. They might be appropriate for edge cases, such as cataclysmic variables that only become observable by \textit{Gaia} during outburst, but otherwise should not be used for white dwarfs.

Fig.~\ref{figureReddening} compares the fitted reddening against our prior (the average of two reddening maps) at the inverse-parallax distance. Our values generally agree well with those from the maps, albeit with some scatter. The maps are generated from millions of observations across the sky, so on average we would expect our results to be in good agreement with the maps. However, our Student's~$t$ prior is not strongly constraining, so the scatter may reflect local variations in interstellar material not captured at the maps' resolution, or imperfections in the models.

\subsection{Radial velocities}
\label{subsectionContentsRVs}

\begin{figure}
\centering
\includegraphics[width=\columnwidth]{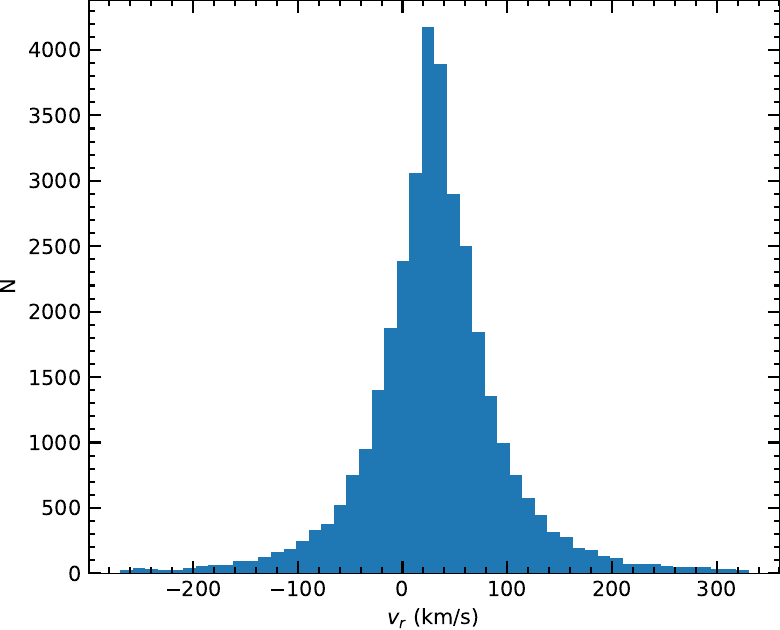}
\caption{Distribution of radial velocities.}
\label{figureRVs}
\end{figure}

Despite the often {low-\SN} spectra and usually having only three Balmer lines to work with, 28\,754 objects in the main catalogue have radial velocity measurements with $\upsigma_{\text{RV}} < 100$\,\kms. Comparison with velocities measured by the SPY survey shows a strong correlation (Fig.~\ref{figureSPY}), confirming that our measurements are reliable, albeit noisy. Fig.~\ref{figureRVs} shows the distribution of velocities presented in the main catalogue. The median radial velocity of all white dwarfs in DR1 is {31\,\kms}, consistent with the gravitational redshift of a typical white dwarf, which is the expected value when averaging measurements over many stars across a large region of sky \citep{Falcon2010, Chandra2020redshifts}. Velocities measured from the B~arm are systematically lower by a median of {8\,\kms} than those measured from the R~arm, but that effect is small compared to the median velocity uncertainty of {45\,\kms} (calculated excluding uncertainties exceeding 500\,\kms). As a sanity check, we compare our velocities against those in the DESI DR1 main-sequence star catalogue, where white dwarfs are \edit{typically fitted} as rapidly-rotating stars with $\log{g}=6$ \citep{Koposov2026}. Despite that imperfect approximation, for stars with radial velocity uncertainties below 50\,\kms, the offset between the two catalogues is $0\pm21$\,\kms.

Of 9869 objects that we test for variability, 153 show 3-$\upsigma$ velocity variation based on Gaussian uncertainties, 85 based on the Huber loss, and just eight based on Tukey's biweight. All of the variables identified in the Tukey version exceed the threshold under the Huber version, confirming that its conservative treatment of outliers avoids false positives, but its utility is diminished by its excessive rejection rate. By contrast, the Gaussian version produces more candidates, but their reliability is undermined by its acceptance of outliers. We therefore use the Huber version to detect variation, as it provides an acceptable compromise. Table~\ref{tableRVvariation} gives the number of 3-$\upsigma$ variables by spectral type.

\begin{table}
\centering
\caption{Numbers of radial-velocity variables identified by the Huber metric (Section~\ref{subsectionRVvariation}), broken down by spectral type. See also panel~I of Fig.~\ref{figureHRdiagrams}.}
\label{tableRVvariation}
\begin{tabular}{lr}
\hline
Spectral type & $N_{\text{var}}$ \\
\hline
DA & 73 \\
DAB & 1 \\
DAZ & 1 \\
DB & 6 \\
DBA & 4 \\
\hline
\end{tabular}
\end{table}

We validate the variability test using stars targeted by binary surveys \citep{Brown2020ELM, Napiwotzki2020, Munday2024} that have velocities measured from multiple exposures in DESI DR1. There are 96 such stars, of which 11 are identified as binaries by those surveys. The Gaussian test flags four (36\,\%) of those binaries as variable, but also two (2.4\,\%) of the single stars. The Huber method finds three (27\,\%) of the binaries, and flags two (2.4\,\%) of the single stars, while the Tukey test does not flag any of the stars. The binaries flagged by the Huber method have or exceed the median number of DESI exposures (three) within the validation sample, suggesting that the variability test performs better when more measurements are available.

As noted in Section~\ref{subsectionHRDiagrams}, the distribution of radial-velocity variables on the HR~diagram (Fig.~\ref{figureHRdiagrams}, panel I) is weighted towards objects that lie above the white dwarf cooling track, as expected if unresolved companions are contributing excess flux. However, we did not attempt to fit two-star models to spectra or photometry, nor did we flag composite spectral types other than WD+MS, as double degenerates are often challenging to identify by eye.

The influence on \textit{Gaia} astrometry of photocentre motion induced by a companion allows identification of binary candidates via the RUWE statistic \citep{Belokurov2020}. However, there is no correlation between RUWE and the radial velocity variability metrics explored here. The RUWE statistic is most sensitive to orbital periods of months to years, whereas the large velocity uncertainties here bias detections towards high-amplitude variation associated with orbital periods of a few weeks or less.

DESI DR1 is a challenging dataset from which to measure velocity variation, as in most cases there are few exposures to work with (Fig.~\ref{figureExposureCounts}). Nevertheless, it is a useful starting point for identifying variable candidates. The observing cadence covers a wide range of timescales, with 44~per cent of repeat exposures being less than 1\,hr after the previous one. Among the remainder, the median time between exposures is 9\,d, with an inter-quartile range of 33\,d and a maximum of 439\,d.

Care is required in interpreting the results of a variability test, as demonstrated by the widely divergent candidate counts and false negative rates that we find using our three methods. For users contemplating a follow-up observing campaign and wishing to avoid non-detections, we recommend a more nuanced approach to target selection than simply using the candidates flagged in our main catalogue. For example, our methods in Appendix~\ref{appendixRVs} can be adapted to select a sample with a defined false discovery rate.

Radial velocity fits can be compromised in several ways. Stars may have noisy or masked data within the narrow windows used to measure velocities, or imperfect wavelength calibration in one or more spectrograph arms. Further, the model template used to measure velocity may be a poor match to the spectral type, especially if there are emission features (e.g. from a companion), Zeeman splitting, or unmodelled lines from carbon or metals. Potential improvements to our method include model templates better tailored to the targets, which would increase performance for types such as DQs or DZs. Alternatively, line-by-line fitting using an analytic model (e.g. a polynomial continuum with Lorentzian line profiles) could replace templates. Line widths and relatives strengths are temperature-dependent, so varying the width of the wavelength windows accordingly may improve the fit. The literature on radial velocity fitting is extensive, but some recent work with SDSS data is comparable in terms of sample size and spectral resolution \citep{Chandra2020redshifts, Arseneau2024, Crumpler2025}.

The potential for problematic fits means that the largest radial velocities listed in the catalogue are likely unreliable, but inspection shows that there are a handful of stars with apparently genuine velocities above 200\,\kms. For example, WD\,J023809.44+273329.79 ($v_r=-228\pm16$\,\kms), which is likely a double-degenerate binary as we identify velocity variation among its individual exposures, and WD\,J083446.91+304959.37 ($v_r=340\pm60$\,\kms), which has only one exposure in DESI DR1 but is a known double degenerate \citep{Kilic2017}. There are higher-velocity stars in the DR1 sample, such as WD\,J110907.96+000132.92 (Section~\ref{subsectionContentsExotica}; \citealt{ElBadry2023}), but identifying and measuring them will require an alternative method.

\begin{figure*}
\centering
\includegraphics[width=0.79\textwidth]{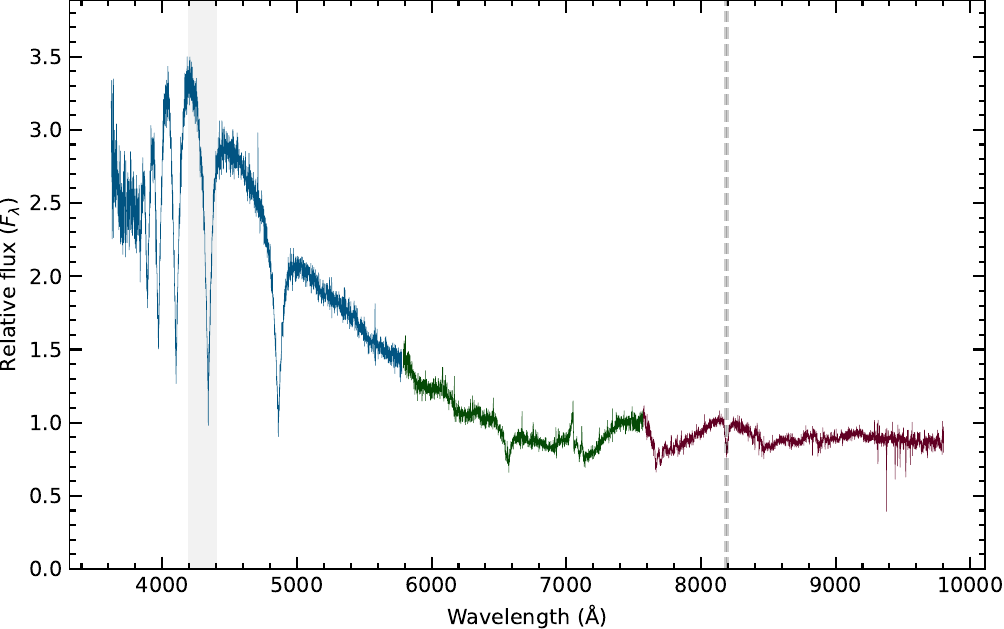}
\includegraphics[width=0.20\textwidth]{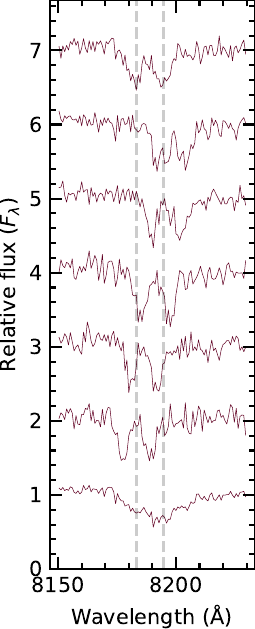}
\caption{WD+MS system WD\,J083845.86+191416.78. \textit{Left}: coadded spectrum, normalised. \textit{Right:} Zoom on the \ion{Na}{i} doublet (rest-frame wavelengths shown with vertical dashed lines) showing radial velocity variation between individual exposures, with their coadd shown at the bottom of the panel. Spectra are shown at full resolution, normalised and offset vertically. Grey shading indicates the region potentially affected by an instrumental artefact.}
\label{figureWDJ0838}
\end{figure*}

We did not separately analyse \edit{velocities for} WD+MS systems, but DESI is likely to be a rich source of information on them. Fig.~\ref{figureWDJ0838} shows spectra from individual exposures and our coadd for WD\,J083845.86+191416.78 to demonstrate the potential of the multi-epoch spectroscopy. The infrared \ion{Na}{i} doublet is smeared out in the coadd, but clearly resolved in the individual exposures, where velocity shifts caused by the binary orbit are evident. Our radial velocity fits to Balmer lines in the blue arm also show clear variation with an amplitude of at least $150\pm30$\,\kms, consistent with results from a dedicated study \citep{Parsons2017}. \edit{We also recover its 3.1-hr orbital period from ZTF photometry (Section~\ref{subsubsectionWDMS} and Table~\ref{tableZTFperiods}).}

\subsection{Kinematic populations}
\label{subsectionContentsKinematics}

We combined our distance and radial velocity measurements with \textit{Gaia} proper motions to calculate stellar velocities in a Galactocentric frame, and then inferred whether stars belong to the ancient halo population or the younger disc\edit{. Previous studies of large samples of white dwarfs have achieved this using random-forest classification of their \textit{Gaia} data \citep{Torres2019}, or by modelling their orbits in a Galactic potential \citep{Pauli2006, Raddi2022}. Here, we rely on the relation between age and velocity dispersion for stellar populations \citep[e.g.][]{Wielen1977}. We used Gaussian mixture models to estimate the probabilities of disc or halo membership for each star, a method that has been used both for white-dwarf and main-sequence populations \citep{Anguiano2017,Nikakhtar2021}}. Our result is only indicative, given the simplicity of the method, but it demonstrates the potential of a large catalogue with 3D velocity information.

We calculated Galactocentric $UVW$ velocities with \texttt{astropy} routines, using the default settings of its \texttt{Galactocentric} frame. Sky coordinates (RA, dec) were taken from the \textit{Gaia} catalogues. We drew proper motion samples from a multivariate normal distribution specified by the \textit{Gaia} astrometry (including correlations), conditioned on parallaxes derived by inverting the distance samples from our model fits (Section~\ref{subsectionFittingProcedure}). We drew radial velocity samples from a normal distribution defined by our fitted velocities and their errors (Section~\ref{subsubsectionRVs}), and then subtracted gravitational redshifts derived from our $\Teff$ and $\log{g}$ samples and the mass--radius relation used during fitting. Putting these together, we calculated samples of the $UVW$ velocities that reflect the correlations in the input variables.

A Gaussian mixture model was fitted to the means of the $UVW$ samples for DA white dwarfs with radial velocity uncertainties less than 30\,\kms. \edit{The aim was to capture the halo population with one of the Gaussian components, on the basis that its velocities should be isotropically distributed, and well separated from the rotationally-supported disc.} We employed \texttt{BayesianGaussianMixture} from the \texttt{scikit-learn} package, setting its \texttt{weight\_concentration\_prior} parameter to $10^{-4}$ to encourage a low number of components. \edit{The model determines during fitting how many components are necessary, subject to a maximum that we set to 10 to ensure any kinematic structures in the disc were captured}, but only six were used. Experimentation showed that the outcome was qualitatively the same regardless of the number of components used: one component is an obvious outlier, having a lower mean $V$ velocity and larger velocity dispersions than the remainder. We interpret these components as representing the halo and disc, respectively.

\begin{figure*}
\centering
\includegraphics[width=\textwidth]{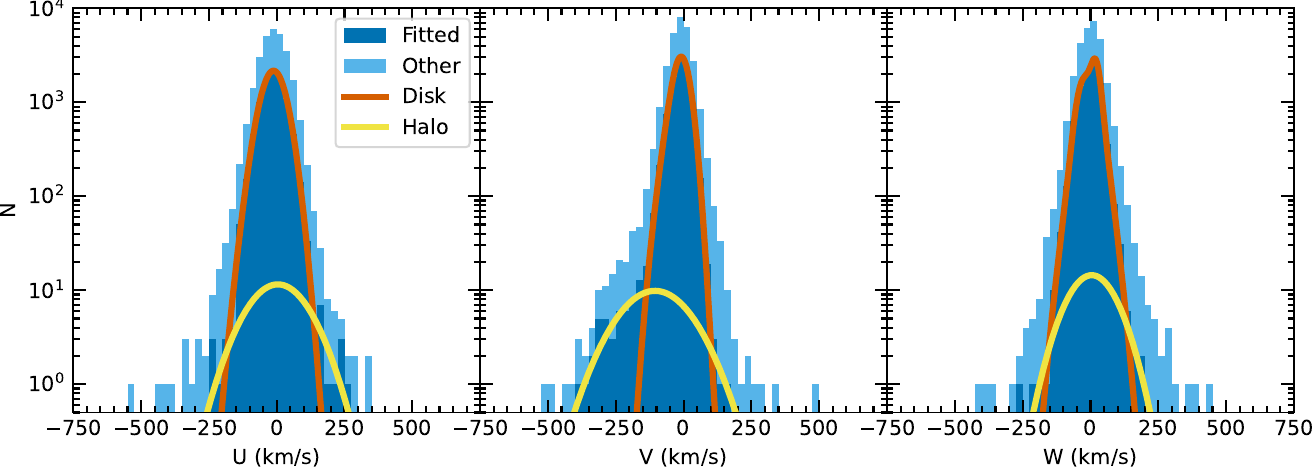}
\caption{Galactocentric $UVW$ velocity histograms for white dwarfs with 3D velocity information, overlaid with the marginal probability distributions of our Gaussian mixture model components that we interpret as representing the Galactic disc and halo. The mixture model was fitted only to DA white dwarfs with radial velocity uncertainties of {30\,\kms} or lower, so the histograms are shaded to identify that sample, with the remainder being white dwarfs with radial velocity uncertainties 30--200\,\kms.}
\label{figureUVW}
\end{figure*}

\begin{figure}
\centering
\includegraphics[width=\columnwidth]{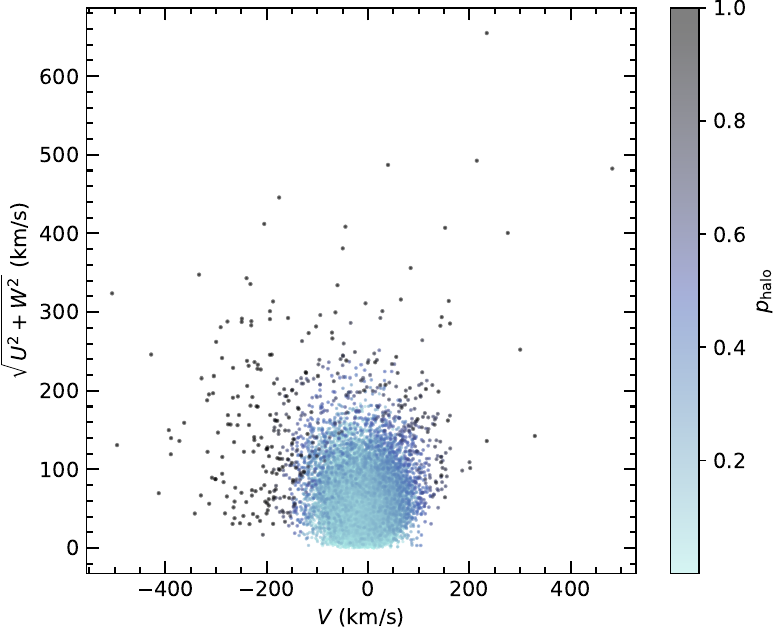}
\caption{Toomre diagram, shaded by probability of halo membership, as estimated in Section~\ref{subsectionContentsKinematics}. Only stars with radial velocity uncertainties below {200\,\kms} are plotted, using the median of their $UVW$ distributions.}
\label{figureToomre}
\end{figure}

The $UVW$ distributions and the marginal distributions of the mixture model fit are shown in Fig.~\ref{figureUVW}. We evaluated the model against all samples for all objects with 3D data to obtain probabilities of membership of the halo component. We report the means of the probability samples in the main catalogue (Table~\ref{tableMainCatalogue}) for stars whose radial velocity errors are 50\,\kms or less. One per cent of those stars are assigned halo probabilities above 0.5, consistent with the halo fraction of 0.02 or less found by previous studies \citep{Pauli2006, Torres2019, Raddi2022}. Fig.~\ref{figureToomre} shows a Toomre diagram based on the median $UVW$ velocity samples for each star.

We emphasise that this is only an indicative result from a toy model intended to demonstrate the utility of the catalogue. Stars will be assigned high halo probabilities if their radial velocity and astrometric uncertainties cause their $UVW$ velocity distributions to extend across large swathes of the Toomre diagram. We therefore only provide the full results from our model as supplementary information in Table~\ref{tableKinematics}, and urge caution in the interpretation of halo probabilities that were not included in the main catalogue. A more physically-motivated and statistically-robust approach might consider orbits in a Galactic potential and use a hierarchical model, but such work is beyond the scope of this catalogue.

\subsection{Exotica}
\label{subsectionContentsExotica}

\subsubsection{Ultra-high-excitation lines}
\label{subsubsectionUHE}
\begin{figure}
\centering
\includegraphics[width=\columnwidth]{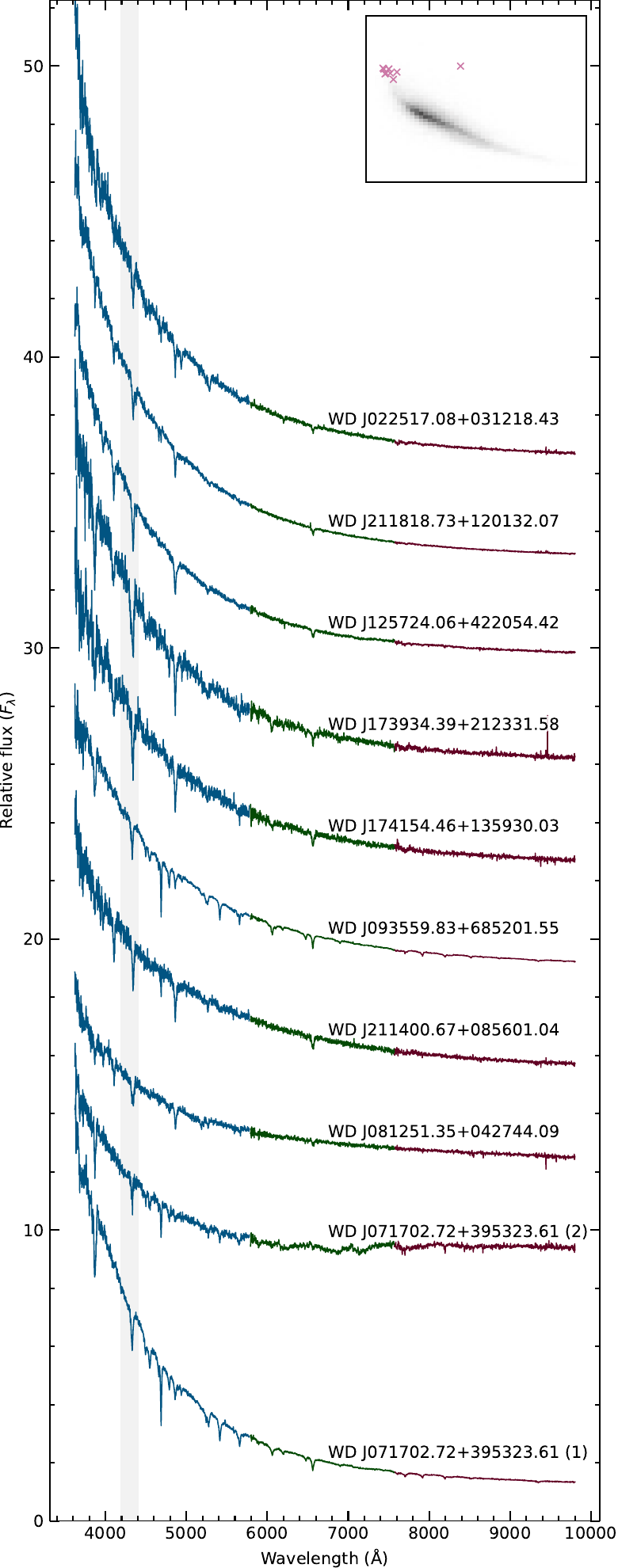}
\caption{Ultra-high-excitation line objects. Spectrum~(1) of WD\,J071702.72+395323.61 targets the white dwarf and spectrum~(2) targets its M-dwarf companion 1\,arcsec away. The inset shows their positions relative to the white dwarf cooling track on the HR~diagram, using the same axes, bins and shading as Fig.~\ref{figureHRdiagrams}. Spectra are shown smoothed with a 5-pixel boxcar median, normalised and offset vertically. Grey shading indicates the region potentially affected by an instrumental artefact.}
\label{figureUHE}
\end{figure}

Some hot white dwarfs were found to have absorption lines from ultra-high-excitation (UHE) metals (e.g. 5243\,\AA, 5280\,\AA). The species involved are only partially identified, but include O and N. These unusual features are poorly understood, but may arise from shocks in magnetically-confined winds \citep{Reindl2021}, and are often photometrically variable \citep{Reindl2023}. We show the examples we noted in Fig.~\ref{figureUHE}, some of which are newly identified.

\begin{figure}
\centering
\includegraphics[width=\columnwidth]{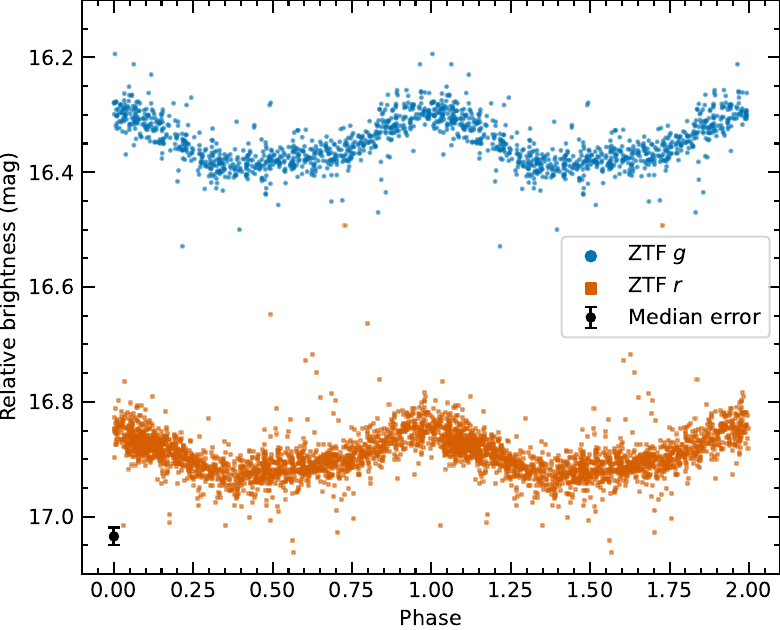}\\
\vspace{6pt}
\includegraphics[width=\columnwidth]{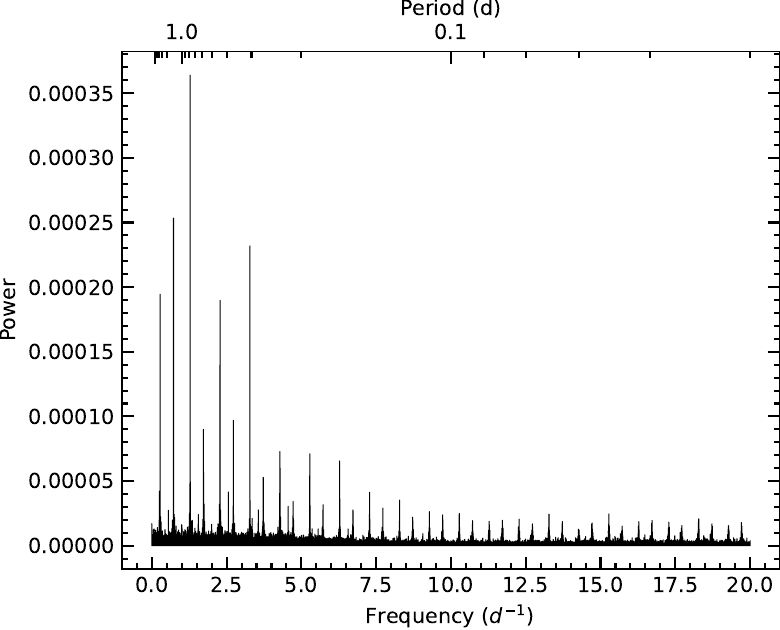}
\caption{\textit{Top:} Phase-folded light curve for the UHE star WD\,J071702.72+395323.61 in the ZTF \textit{g} and \textit{r} bands, folded to a period of $0.7823890\pm0.0000033$\,d. \textit{Bottom}: Discrete Fourier transform of the combined light curve.}
\label{figureUHEZTF}
\end{figure}

Fig.~\ref{figureUHE} includes one of the class prototypes \citep[WD\,J071702.72+395323.61][]{Werner1995UHE}\edit{, which has an M~dwarf companion at 1\,arcsec \citep{Werner2018}, and another common-proper-motion main-sequence companion at 17\,arcsec based on its \textit{Gaia} astrometry}. Three of the four DESI exposures appear similar to each other, and their coadd is labelled (1) in Fig.~\ref{figureUHE}. However, the exposure labelled (2) \edit{targeted the M~dwarf companion rather than the white dwarf, but was picked up by our cross-match due to its close proximity, creating the illusion of spectral variation in the white dwarf, whose flux strongly contaminates the M~dwarf spectrum. The white dwarf is flagged as photometrically variable in \textit{Gaia} DR3, and we measure a period of $0.7823890\pm0.0000033$\,d from ZTF light curves}, in line with previous work \citep{Reindl2021}. However, we note that the phase-folded lightcurve shows non-sinusoidal morphology and a phase offset between the \textit{g} and \textit{r} bands (Fig.~\ref{figureUHEZTF}). Similar variability has been seen in another member of this class, whose equivalent-width and photometric variation follows the 0.24-d rotational period of that star \citep{Reindl2019}. We include WD\,J081251.35+042744.09 in Fig.~\ref{figureUHE} as a potential UHE star, as it shows features near 5250\,\AA\ that are common to most of the class, and has a ZTF photometric period ($2.45759\pm0.00021$\,d; Fig.~\ref{figureUHEcandidate}). However, we note that it is an outlier on the HR~diagram and that its spectrum has \ion{Ca}{ii} emission, both of which may indicate the presence of a companion.

\subsubsection{Cataclysmic variables}
These are compact binaries containing a white dwarf that accretes material from a low-mass Roche-lobe-filling companion \citep{Warner1995} with typical orbital periods approximately in the range of $80$\,min to $12$\,hr \citep{Gansicke2009CVs, Thorstensen2010, Peters2005}. Cataclysmic variables exhibit a very rich observational morphology in terms of their spectroscopic appearance, reflecting a wide range in their underlying physical parameters, which include the white dwarf temperature (largely set by the mass transfer rate, \citealt{Townsley2009}), the magnetic field strength of the white dwarf (which determines the geometry of the accretion flow between the two stars), and the companion star radius and temperature. Correspondingly, cataclysmic variables occupy a large range in the HR~diagram in between the main sequence and the white dwarf cooling track \citep{Abril2020, Inight2023}. The target selection of white dwarfs within DESI \citep{Cooper2023} follows closely the cuts on the \textit{Gaia} data defined by GF19, thus favouring intrinsically faint cataclysmic variables in which the companion star contributes relatively little to the optical flux, i.e. systems with low mass transfer rates and short orbital periods. 

\begin{figure*}
\centering
\includegraphics[width=\textwidth]{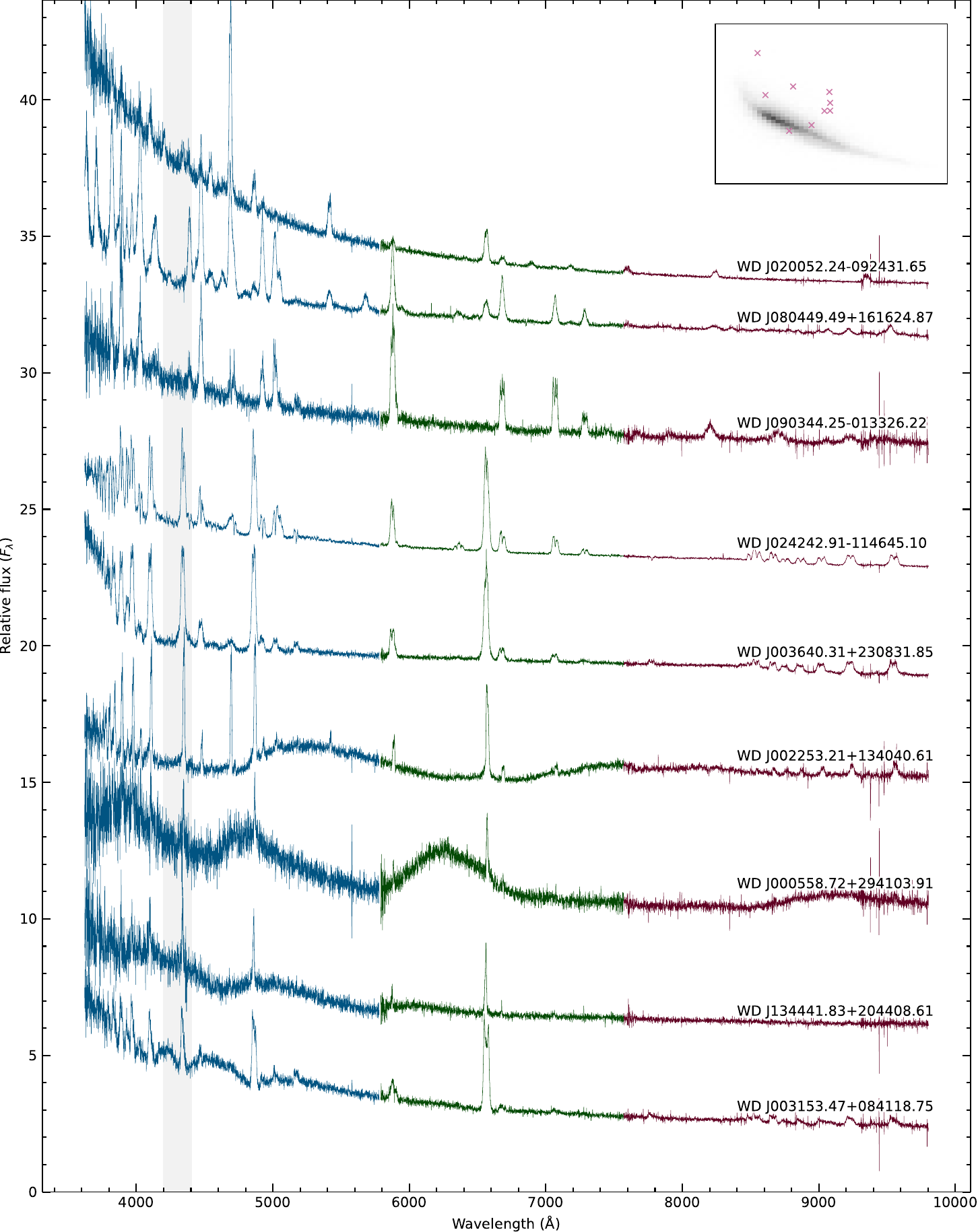}
\caption{Examples of cataclysmic variables. The inset shows their positions relative to the white dwarf cooling track on the HR~diagram, using the same axes, bins and shading as Fig.~\ref{figureHRdiagrams}. Spectra are shown at full resolution, normalised and offset vertically. Grey shading indicates the region potentially affected by an instrumental artefact.}
\label{figureCVs}
\end{figure*}

We have identified 232 cataclysmic variables in this catalogue, and a representative sample of their spectra are shown in Fig.~\ref{figureCVs}. WD\,J020052.24$-$092431.65, WD\,J090344.25$-$013326.22 and WD\,J080449.49+161624.87 are AM\,CVn systems \citep{Warner2002, Inight2023, Roelofs2009}, in which the white dwarf accretes from a hydrogen-deficient companion; WD\,J024242.91$-$114645.10 is a system that may have a stripped donor star, revealing the helium-enriched composition of its nuclear evolved donor \citep{Wils2011, McAllister2015}; WD\,J003640.31+230831.85 is a typical short-period cataclysmic variable with a non-magnetic white dwarf, harbouring an accretion disc that quasi-regularly undergoes thermal instabilities \citep{Thorstensen2016}; WD\,J002253.21+134040.61 and WD\,J000558.72+294103.91 contain white dwarfs that are sufficiently strongly magnetic to suppress the formation of an accretion disc and funnel the accreting material towards the magnetic poles \citep{Hou2026CVs,Inight2026}.

More systematic searches of the DESI DR1 and DR2 spectroscopy resulted in the identification of 412 \citep{Hou2026CVs} and 1020 \citep{Inight2026} cataclysmic variables, spanning the full range of their diverse population. 

\edit{Accreting white dwarfs with high magnetic fields ($B\gtrsim10$\,MG) can emit cyclotron radiation detectable in the optical \citep{Visvanathan1979, Ferrario2015}. Fig.~\ref{figureCVs} includes candidates that we identified in DESI DR1. Notably, the spectrum of WD\,J134441.83+204408.61 has markedly changed since its SDSS observation in 2009 \citep{Szkody2011}, with the cyclotron humps peaking at different wavelengths and the emission lines becoming more prominent in the DESI spectrum.}

\subsubsection{\edit{Low-accretion-rate polars}}
\edit{Besides the 234 CVs, we found seven systems containing a magnetic white dwarf that display evidence for low-level accretion, with the mass transfer driven either by Roche-lobe overflow or a stellar wind, either in the form of cyclotron emission or weak H$\upalpha$ emission. Six of these seven systems were found as part of a final visual check of the classification of our white dwarf catalogue, highlighting the challenges in distinguishing the subtle features among morphologically similar spectra. } 

\begin{figure*}
\centering
\includegraphics[width=\textwidth]{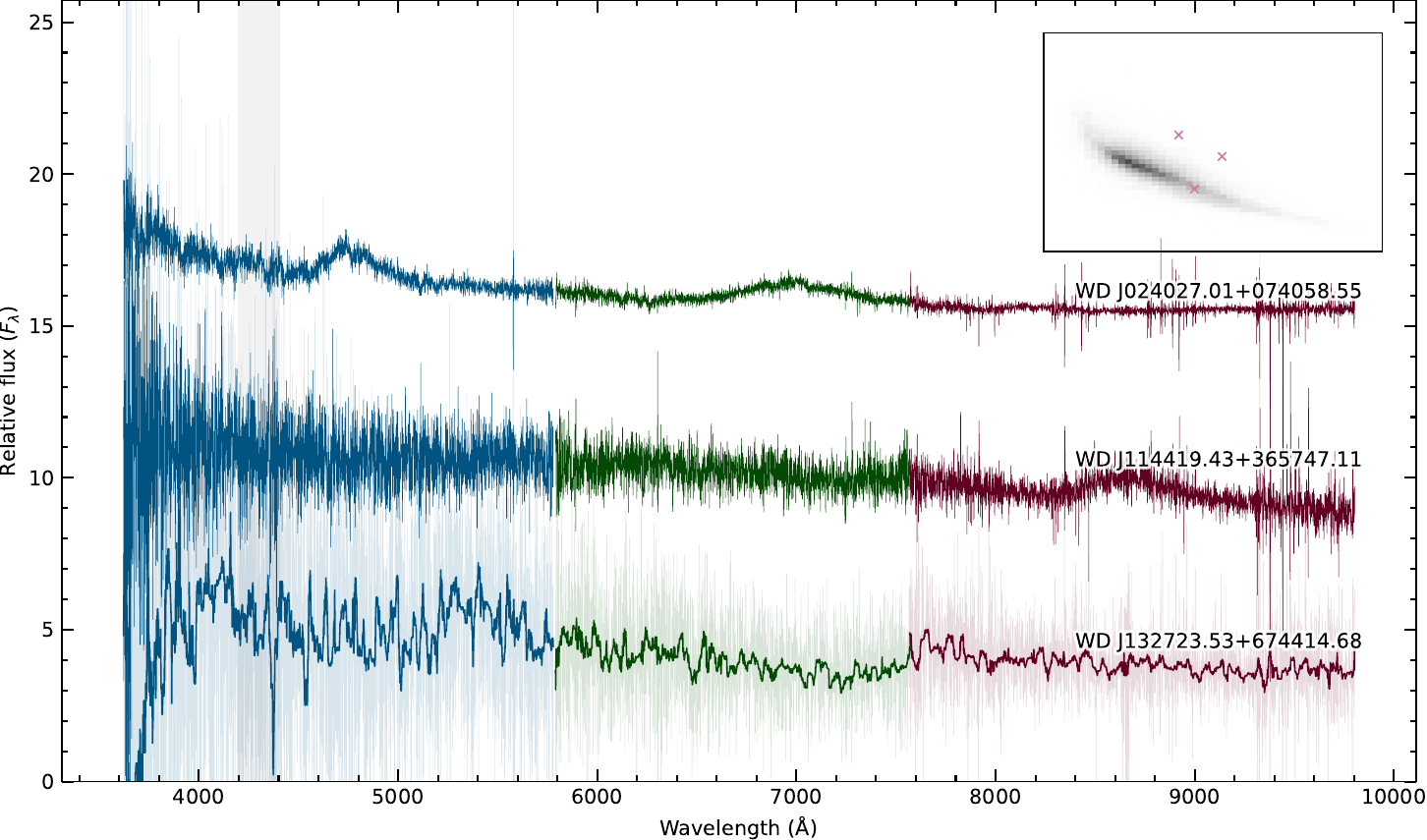}
\caption{\edit{Examples of low-accretion-rate polars with cyclotron emission. The inset shows their positions relative to the white dwarf cooling track on the HR~diagram, using the same axes, bins and shading as Fig.~\ref{figureHRdiagrams}. Spectra are shown either at full resolution or smoothed with a 30-pixel boxcar median, normalised and offset vertically, and Vertical scaling is exaggerated in one case to highlight the features. Grey shading indicates the region potentially affected by an instrumental artefact.}}
\label{figureCyclotron}
\end{figure*}

\edit{Systems with prominent cyclotron features are shown in Fig.~\ref{figureCyclotron}, namely WD\,J024027.01+074058.55, WD\,J114419.43+365747.11 and WD\,J132723.53+674414.68 (0240+0740, 1144+3657 and 1327+6744 hereafter). The latter two were among around 50 objects with significant proper motions and parallaxes whose spectra we had originally classified as extragalactic, demonstrating that \textit{Gaia} DR3 astrometry alone is not fully sufficient to distinguish between galactic and extragalactic objects. Analysing the ZTF light curves we found short-period variability, confirming both systems as binaries. 1144+3657 was previously identified as a cyclotron-emitting white dwarf binary by \citet{vanRoestel2025} on the basis of its ZTF light curve. We applied analysis-of-variance (AoV, using the algorithm of \citealt{SchwarzenbergCzerny1996}) to the ZTF photometry of these three systems, revealing strong periodic signals in all three of them. Assigning orbital periods based on the AoV analysis requires an assumption on whether the orbital phase-folded light curve exhibits one \citep[e.g. VV\,Pup][]{Szkody1983} or two maxima per cycle \citep[e.g. SDSS\,J085414.02+390537.3][]{Dillon2008}, which in turn depends on the geometry of the accretion column \citep{Shani1990, Gansicke2001}. Adopting two cyclotron humps per orbital cycle, we measure orbital periods of $0.124696669\pm0.000000071$\,d, $0.0757301695\pm0.0000000094$\,d and $0.11418566\pm0.00000039$\,d for 0240+0740, 1144+3657 and 1327+6744, respectively. Confirmation of these periods would ideally require time-series spectroscopy of polarimetry. In the case of 1144+3657, the period determined by \citet{vanRoestel2025} confirms our value. The general consensus is that the majority of white dwarf binaries emitting almost pure cyclotron emission, with little or no sign of Balmer or He emission lines, are in a detached state, with some ongoing discussion around whether they are prior to or past the onset of mass transfer \citep{Vogel2007, Schreiber2021}. We note that 1327+6744 exhibited a state change in its ZTF light curve at around MJD~58\,660, fading from $g\simeq20$\,mag to near the ZTF detection limit at $g\simeq21.5$\,mag, suggesting that this system may be a semi-detached polar that entered a prolonged and deep low state.}

\begin{figure*}
\centering
\includegraphics[width=\textwidth]{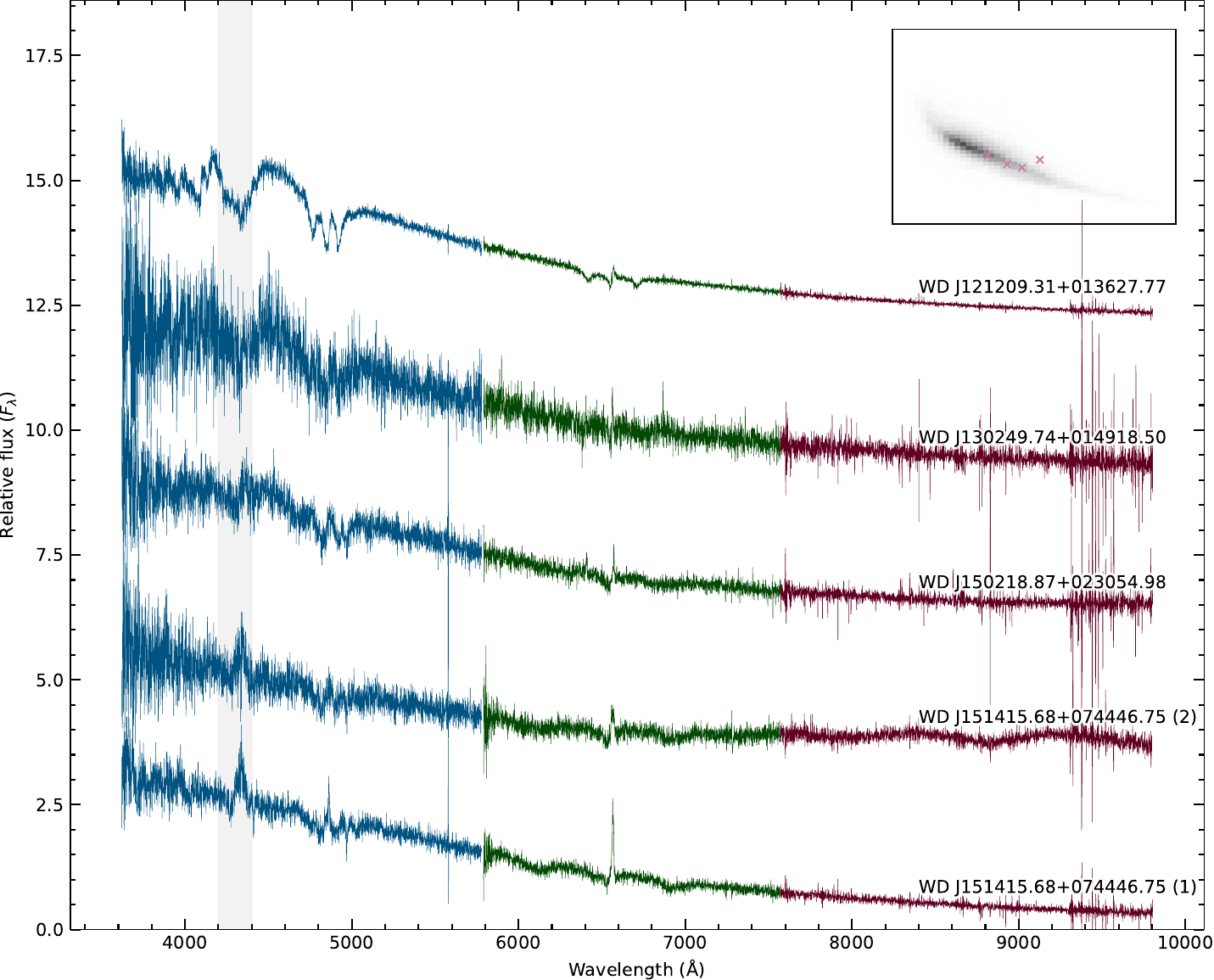}
\caption{\edit{Examples of low-accretion-rate polars. Individual exposures taken about 2\,d apart are shown for WD\,J151415.68+074446.75 to illustrate its variable nature. The inset shows their positions relative to the white dwarf cooling track on the HR~diagram, using the same axes, bins and shading as Fig.~\ref{figureHRdiagrams}. Spectra are shown at full resolution, normalised and offset vertically. Grey shading indicates the region potentially affected by an instrumental artefact.}}
\label{figureLARPs}
\end{figure*}

\edit{WD\,J121209.31+013627.77, WD\,J130249.74+014918.50, WD\,J150218.87+023054.98, and WD\,J151415.68+074446.75 (1212+0136, 1302+0149, 1502+0230, and 1514+0744 hereafter) form a group of apparently single DAH white dwarfs, but on closer inspection they feature weak, radial-velocity-variable H$\alpha$ emission identifying them as close binaries (Fig.~\ref{figureLARPs}). 1212+0136 and 1514+0744 are well-studied systems \citep{Schmidt2005, Burleigh2006, Farihi2008, Breedt2012}, and 1302+0149 and 1502+0230 were recently discussed as a CV candidates by \citet{HernandezDiaz2026} and \citet{Inight2026}. Commonalities between these stars are white dwarf temperatures of around 10\,000\,K, very low-mass companions (in most cases brown dwarfs), and very low accretion rates (see Table\,3 in \citealt{Cunningham2025} for an overview of the current roster). As with the cyclotron systems above, there remains some ongoing discussion around whether these systems are detached or semi-detached CVs that simply have very low accretion rates, with the latter option being favoured \citep{MunozGiraldo2023}. Analysing the ZTF light curves of these four systems, we recover the published values for 1212+0136 ($0.061406934\pm0.000000092$\,d, \citealt{Koen2006}), 1502+0230 \citep[$0.062594439\pm0.000000044$\,d][]{Inight2026}, and 1514+0744 ($0.061605473\pm0.000000031$\,d, \citealt{Breedt2012}) and we report the first period measurement for 1302+0149 ($0.065046125\pm0.000000048$\,d), confirming that it also is located near the minimum period of CVs. The case of 1514+0744 is of particular interest, as the two DESI spectra taken two days apart show relatively broad H$\upalpha$ emission as well as variability of the spectral slope, contrasting the weak H$\upalpha$ reported by \citet{Breedt2012}. The ZTF light curve spanning the epoch of the DESI spectroscopy (MJD~59\,731) shows an increase in the amplitude of the orbital modulation. Together, these findings corroborate the X-ray study of \citet{MunozGiraldo2023} that the systems in this small class are indeed cataclysmic variables in extended dormant states. }

\subsubsection{Detached white dwarf plus M-dwarf binaries}
\label{subsubsectionWDMS}
\edit{In addition to the semi-detached white dwarf binaries discussed in the two previous sections, DESI is a rich source for detached white dwarf plus main-sequence binaries (WD+MS), with  systems identified in DR1. Among those, systems containing a low-mass (typically M-dwarf spectral type) companion are easy to recognise as both stars contribute similar amounts of flux at optical wavelengths. The white dwarfs dominate in the blue part of the DESI spectrum, in most cases showing Balmer absorption lines, and the M-dwarfs dominate the red part of the spectrum and often exhibit strong molecular bands of TiO. It is important to bear in mind that these systems comprise both post-common envelope binaries with typical periods of hours to days, as well as wider (tens of au) separation binaries that did not interact during their evolution, but remain spatially unresolved from the ground \citep[see e.g.][]{Willems2004, RebassaMansergas2010}.} 

\edit{We downloaded the ZTF \citep{Bellm2019} light curves for the WD+MS binaries (where available) and computed AoV and discrete Fourier transform periodograms spanning 0--20\,$\mathrm{d}^{-1}$. The AoV technique is particularly suited to identify sharp eclipses in the light curves, and the Fourier transform is ideal for the detection of quasi-sinusoidal modulations. Inspection of these periodograms resulted in 73 systems with a clear detection of a periodic signal (Table~\ref{tableZTFperiods}), including 24 eclipsing ones. For the non-eclipsing systems, we assumed that the photometric variability is due to the ``reflection effect'', i.e. the variation in the visibility of the inner hemisphere of the companion heated by the white dwarf, resulting in one maximum per orbital cycle. This assumption is most likely correct for the majority of the systems analysed here, given the moderately warm white dwarfs in these systems. However, a caveat to the results in Table~\ref{tableZTFperiods} is that the true orbital period could be twice the reported value if the photometric modulation is caused by ellipsoidal modulation, i.e. the variation in the projected area of the unheated companion, resulting in two maxima per orbital cycle. Ultimately, a small number of radial velocity measurements spanning the putative period would unambiguously distinguish between the two cases \citep{GomezMoran2011}.} 

\begin{figure*}
\centering
\includegraphics[width=\textwidth]{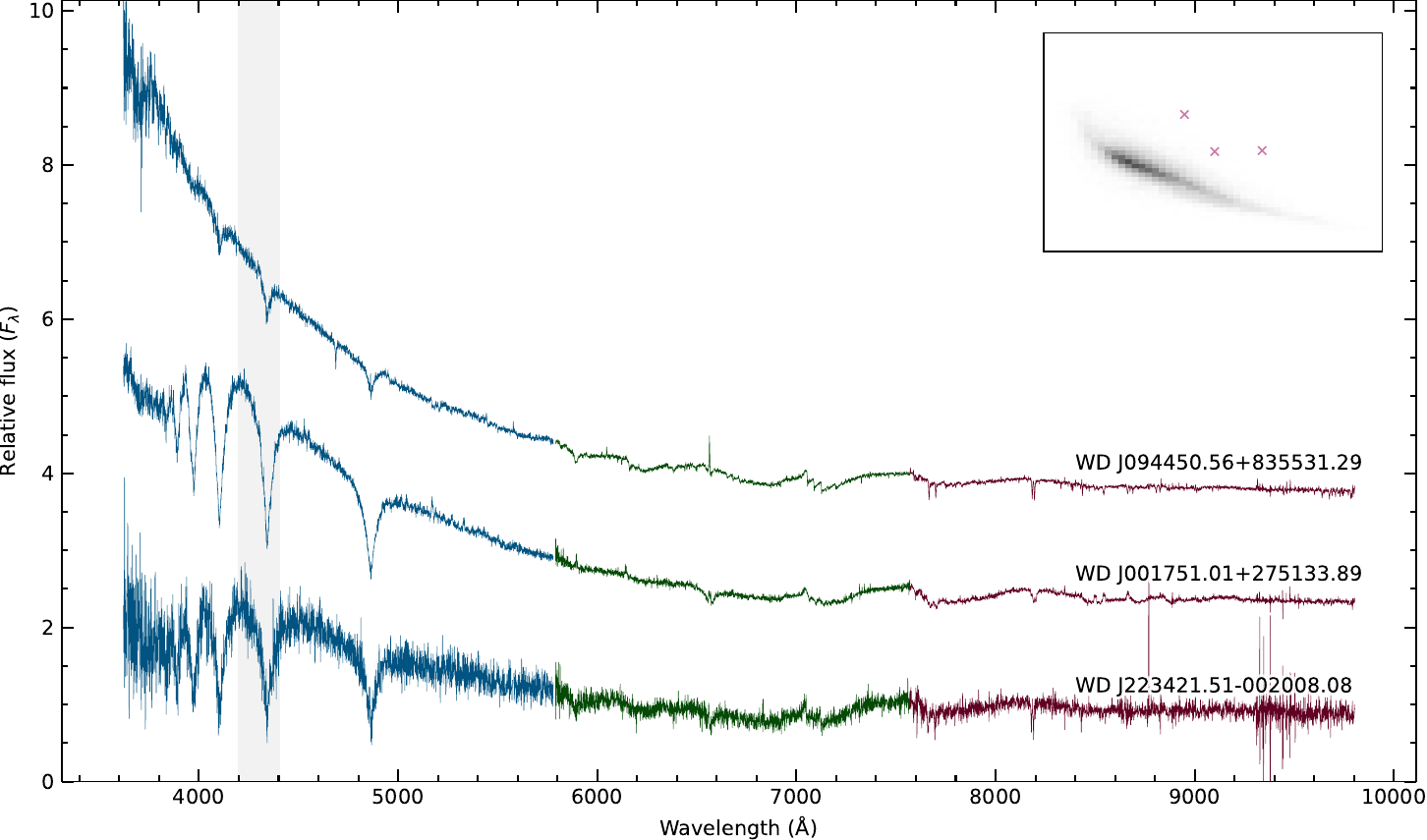}
\caption{\edit{Examples of white-dwarf--main-sequence binaries. The inset shows their positions relative to the white dwarf cooling track on the HR~diagram, using the same axes, bins and shading as Fig.~\ref{figureHRdiagrams}. Spectra are shown at full resolution, normalised and offset vertically. Grey shading indicates the region potentially affected by an instrumental artefact. The short periods measured from their ZTF photometry (0.134 to 0.997\,d, see Section\,\ref{subsubsectionWDMS}) identify these systems as post-common envelope binaries.}}
\label{figurePCEBs}
\end{figure*}

\edit{Fig.~\ref{figurePCEBs} shows three examples from the 73 WD+MS systems for which we measured periods from the ZTF data. WD\,J001751.01+275133.89 is a DA+dM binary with a period of $0.138055083\pm0.000000083$\,d, whose DESI spectrum shows a host of emission lines, indicating significant heating of the companion. The system was previously reported as eclipsing with a period consistent with our value \citep{MeirShani2025}, however we find that the ZTF light curve does not show any sign of eclipses (Fig.~\ref{figurePCEB0017}). WD\,J223421.51$-$002008.08 was identified by \citet{Silvestri2007} on the basis of its spectrum in SDSS; here we report the eclipsing nature and the orbital period of this system (Fig.~\ref{figurePCEB2234}). Finally, WD\,J094450.56+835531.29 is a particularly interesting case with a DAO+dM spectral type. The ZTF light curve appears to show irregular long-term variability. The periodogram computed from the data shows a strong signal near $1\,\mathrm{d}^{-1}$ and its harmonics and sub-harmonics, which is typical of single-site ground-based data, and such a signal is usually disregarded as an artefact. However, inspecting the TESS \citep{Ricker2015} 120-s cadence data obtained in sectors 19, 20, 26, 40, 47 and 53 confirms the presence of a large-amplitude modulation with a period of $0.9970029\pm0.0000083$\,d  (Fig.~\ref{figurePCEB0944}). We identify this as the orbital period of this system, illustrating the need for space-based photometry to unambiguously measure orbital periods which are very close to 1\,d, or harmonics thereof.}

\edit{An important point to bear in mind is that the sample of WD+MS binaries presented here is strongly biased towards systems with a dominant contribution of the white dwarf, as they were targeted as single white dwarf candidates (see Section~\ref{subsectionDESIwhiteDwarfs}), and it is likely that more WD+MS binaries, with stronger contributions of the companion, can be identified among the full set of DESI spectroscopy.}

\subsubsection{Unusual spectra}
\begin{figure*}
\centering
\includegraphics[width=\textwidth]{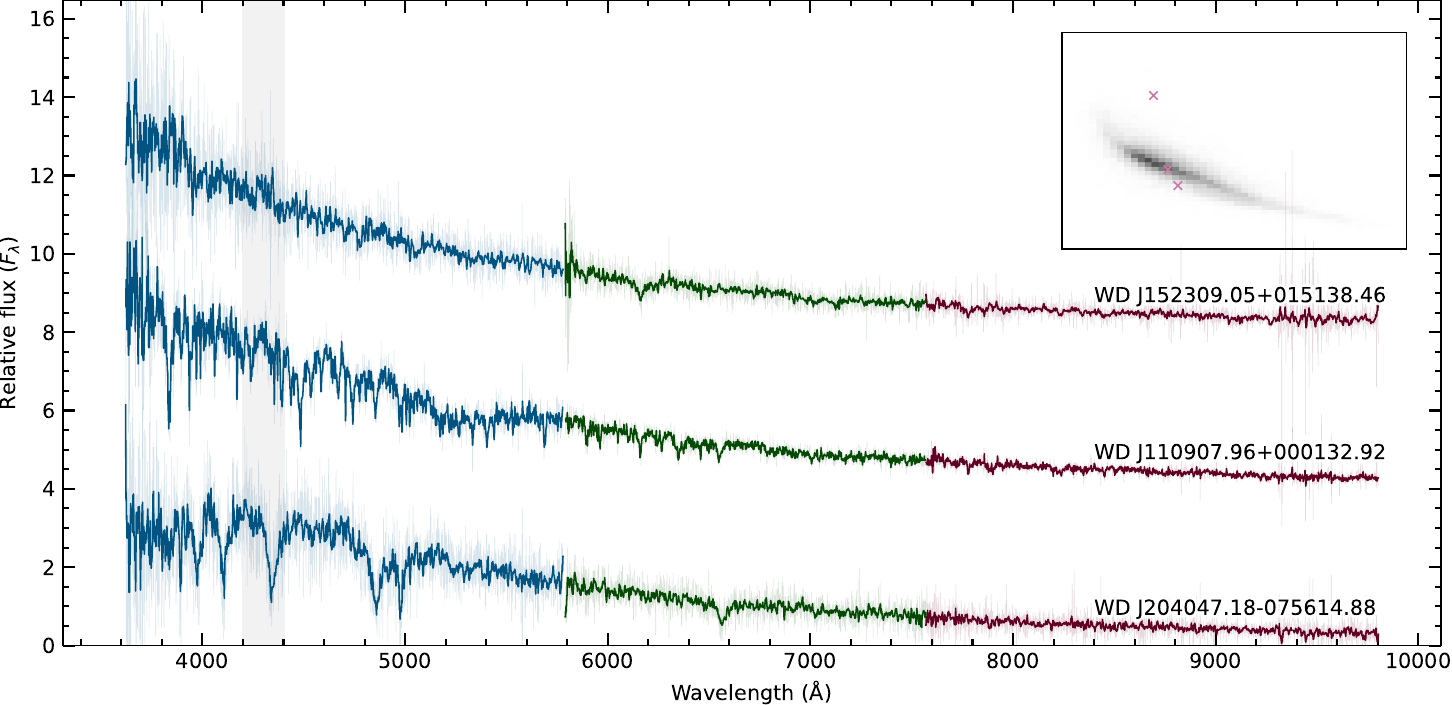}
\caption{White dwarfs with unusual absorption lines. The inset shows their positions relative to the white dwarf cooling track on the HR~diagram, using the same axes, bins and shading as Fig.~\ref{figureHRdiagrams}. Spectra are shown smoothed with a 5- or 10-pixel boxcar median, normalised and offset vertically. Grey shading indicates the region potentially affected by an instrumental artefact.}
\label{figureUnusual}
\end{figure*}

A few white dwarfs in DESI DR1 have unidentified absorption features in their spectra but no evidence for flux contamination, shown in Fig.~\ref{figureUnusual}. WD\,J204047.18$-$075614.88 has a prominent line near 4980\,\AA, which we could not identify. WD\,J152309.05+015138.46 is a white dwarf with an O-dominated atmosphere \citep{Kepler2015}, a rare class whose progenitors are thought to have had masses close to the Chandrasekhar limit \citep{Gansicke2010,Kepler2016oxygen}. WD\,J110907.96+000132.92 is a member of the LP~40-365 class \citep{ElBadry2023}, which are thought to be the partially-burned remnants of the surviving member of a binary that was unbound by a Type~Ia supernova caused by mass transfer \citep{Vennes2017,Raddi2018}. Such stars are ejected at high velocity ($\sim1000$\,\kms), similar to the participants in a dynamically-driven double-degenerate double-detonation that form a related class \citep{Shen2018, Hollands2025D6}.

\begin{figure}
\centering
\includegraphics[width=\columnwidth]{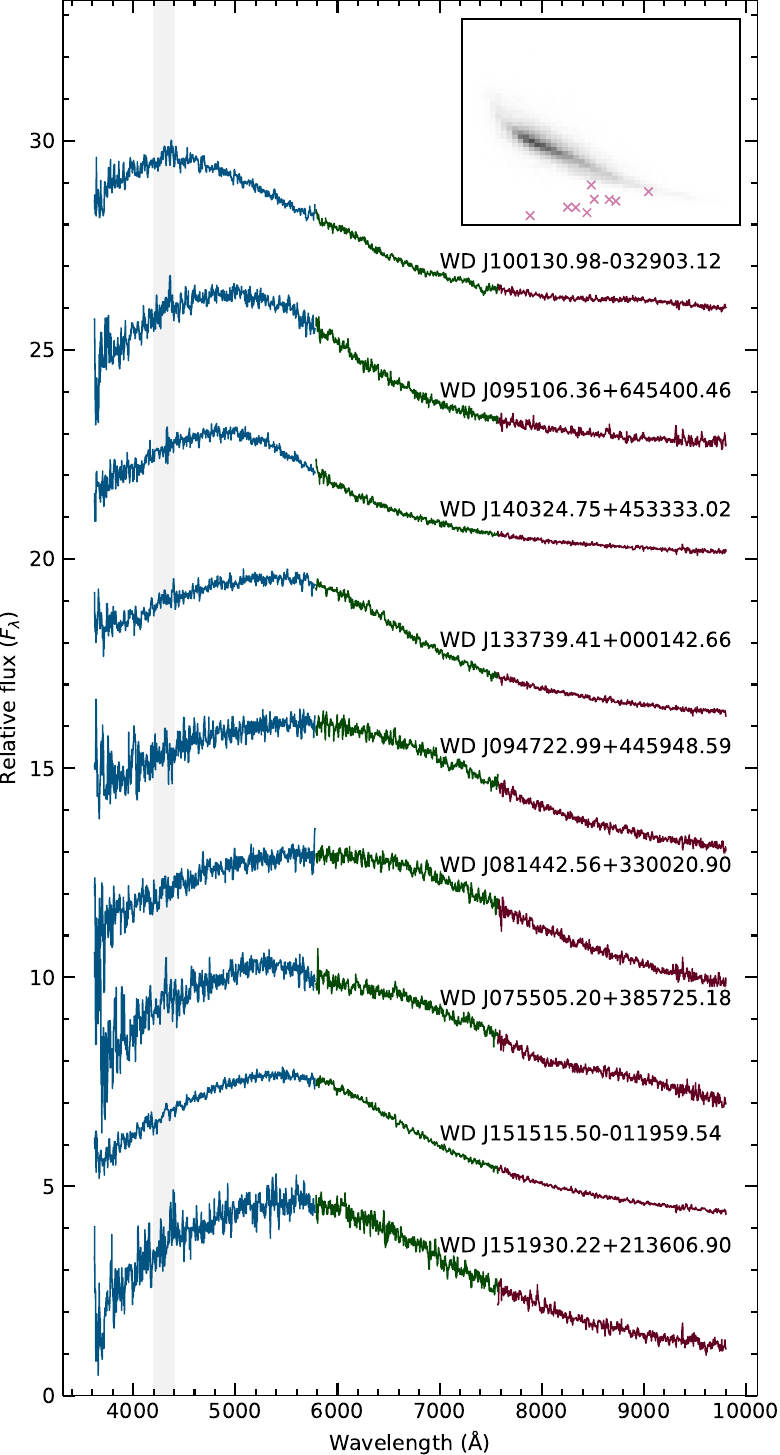}
\caption{Cool white dwarfs whose spectra show the signature of collisionally induced absorption. The inset shows their positions relative to the white dwarf cooling track on the HR~diagram, using the same axes, bins and shading as Fig.~\ref{figureHRdiagrams}. Spectra are shown smoothed with a 15-pixel boxcar median, normalised and offset vertically. Grey shading indicates the region potentially affected by an instrumental artefact.}
\label{figureCIA}
\end{figure}

The mixed He+H atmospheres of ultra-cool white dwarfs can be sufficiently transparent that the photosphere is at a depth where densities approach those typical of liquids \citep{Saumon2022}. Collision-induced absorption involving H\textsubscript{2} molecules manifests strongly in the infrared, causing these stars to be unusually blue for their temperature. One was present in the DESI EDR sample \citep{Manser2024DESIEDR}, and DR1 contains several more (Fig.~\ref{figureCIA}).

\subsubsection{Circumstellar gas}
\begin{figure}
\centering
\includegraphics[width=\columnwidth]{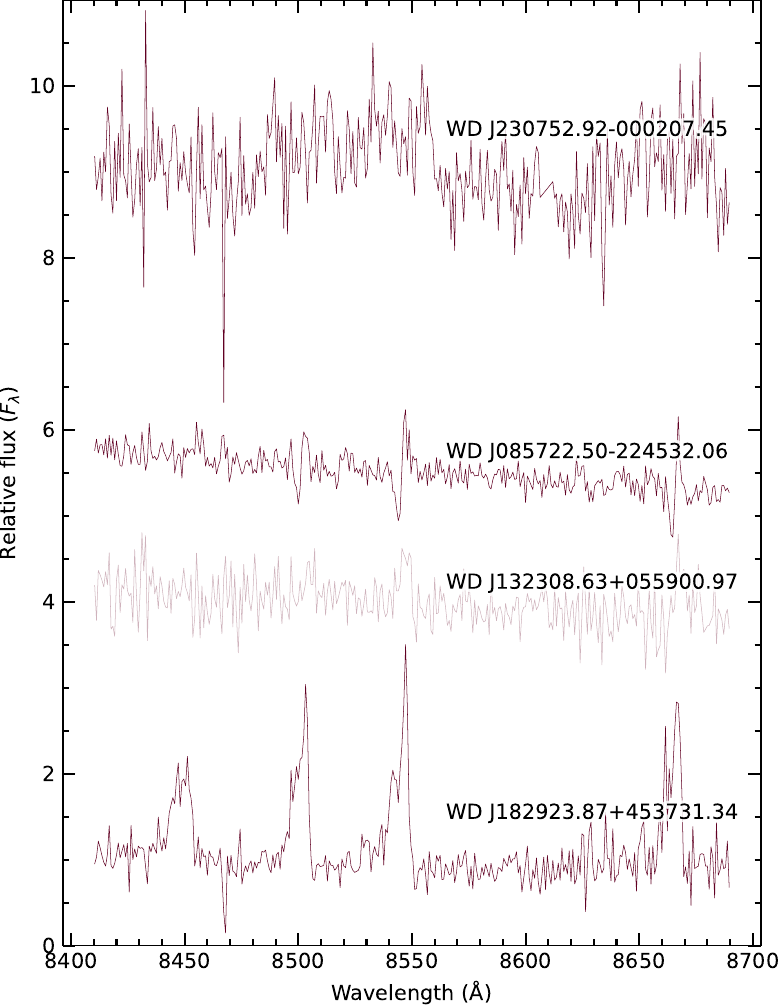}
\caption{\ion{Ca}{ii} lines. Three stars (darker lines) appear to host circumstellar gas, while one (lighter line) is an example of emission from a companion to demonstrate the potential for misidentification. Spectra are shown at full resolution, normalised and offset vertically. Vertical scaling is exaggerated in some cases to highlight the features.}
\label{figureCaIIemission}
\end{figure}

Remnant planetary systems are the source of the metals that are routinely observed in white dwarf photospheres \citep{Jura2014review}. Accretion of metals is mediated by debris discs, formed when an asteroid is perturbed onto a highly-eccentric orbit and tidally disrupted as it crosses the stellar Roche radius \citep{Jura2003}. These discs are most often seen in the infrared due to thermal emission from warm dust \citep{Farihi2016}, but can also exhibit optical transits or spectral lines from circumstellar gas \citep[e.g.][]{vanderburg2015, Gansicke2006}. The latter most commonly manifests via \ion{Ca}{ii} triplet emission, and DESI EDR spectra have already been examined for this rare phenomenon \citep{Ma2025}. While we did not conduct a targeted search of DESI DR1, we identified three systems of interest during the classification exercise. The are shown in Fig.~\ref{figureCaIIemission}, alongside an `imposter' where similar features arise from a companion.

The 24 systems already known to exhibit circumstellar \ion{Ca}{ii} emission have stellar temperatures in the range $13\,000\lesssim\Teff\lesssim29\,000$\,K \citep{Dennihy2020gas,Melis2020gas,GentileFusillo2021GasDiscs,Bhattacharjee2025ZTF}. One of the new discoveries presented here (WD\,J182923.87+453731.34) is the hottest yet known at 34\,500\,K. At such temperatures, dust within the stellar Roche limit is sublimated \citep{Steckloff2021}, so gas may remain the only signature of closely-orbiting planetary debris. Meanwhile, WD\,J085722.50$-$224532.06 is cooler than any confirmed system at 11\,300\,K. Finally, at 15\,300\,K, WD\,J230752.92$-$000207.45 would lie within the temperature range of known systems if its tentative features can be confirmed.

The strong emission from WD\,J182923.87+453731.34 is unambiguously circumstellar, showing double-peaked lines from gas on Keplerian orbits with velocities of several hundred \kms. Features from Ca, O, Mg, and Fe are immediately obvious on visual inspection, as are some unidentified lines that have been seen at another white dwarf hosting circumstellar gas \citep[WD\,J052914.32$-$340108.11;][]{GentileFusillo2021GasDiscs}.

The subtle features at WD\,J085722.50$-$224532.06 are unusual for showing both absorption and emission. This system is optically variable, potentially due to transiting debris \citep{Guidry2021}. If circumstellar gas is associated with transiting solids, it would be viewed close to edge-on, which could account for the absorption component. It is the second such system to be identified \citep{Bhattacharjee2025ZTF}, and its relative brightness makes it a valuable observational target.

The \ion{Ca}{ii} features at WD\,J132308.63+055900.97 appear similar to those at the candidate circumstellar disc hosts just mentioned. There is also H$\upalpha$ emission, which is unusual but not unprecedented for circumstellar debris \citep{GentileFusillo2021GasDiscs}. However, we attribute these features to an irradiated companion instead, as the system is photometrically variable. Its ZTF lightcurve shows modulation on a period of $0.11178495\pm0.00000015$\,d, with higher amplitude (around 0.1\,mag) in the \textit{r} band than in \textit{g} (Fig.~\ref{figureCaIIZTF}). Such behaviour is consistent with the presence of a companion irradiated by the white dwarf.

\subsubsection{Na absorption}
\begin{figure}
\centering
\includegraphics[width=\columnwidth]{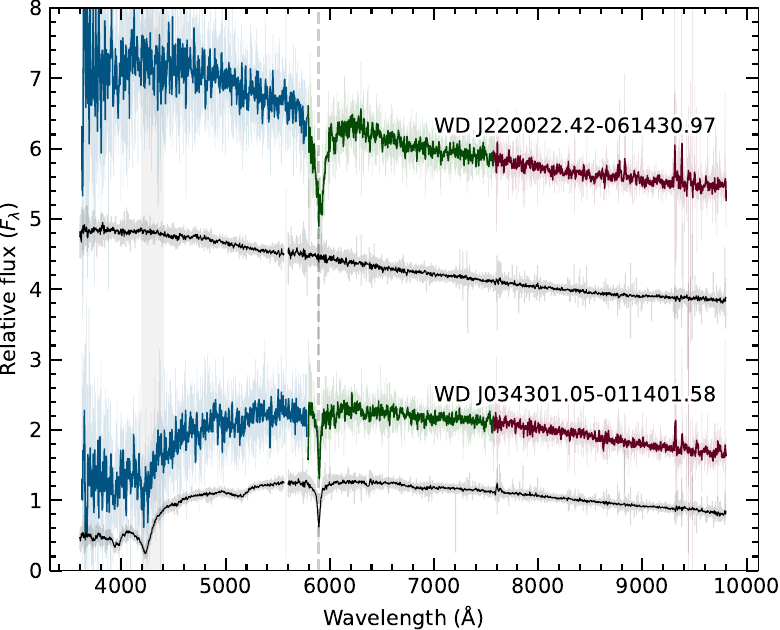}
\caption{Two stars with strong Na\,D absorption in their DESI spectra (colours), only one of which is confirmed by X-Shooter (black). DESI spectra are shown smoothed with a 10-pixel boxcar median, normalised and offset vertically. X-shooter spectra are similarly presented, smoothed to about the same resolution. Grey shading indicates the region potentially affected by an instrumental artefact.}
\label{figureSodium}
\end{figure}

Some white dwarfs have broad and deep absorption around the Na~D doublet (Fig.~\ref{figureSodium}). Such features can be seen in cool white dwarfs with He-dominated atmospheres that have accreted planetary material \citep[e.g.~][]{Blouin2019cool}. We obtained follow-up spectra of two apparently Na-rich stars using X-Shooter on the Very Large Telescope at Paranal, confirming the feature at WD\,J034301.05$-$011401.58, but not at WD\,J220022.42$-$061430.97. Our fit to the DESI spectrum of the latter yields a temperature of 6400\,K, where metal sinking timescales are several Myr. Thus, a photospheric feature would not have evolved between observations unless the absorbing material were unevenly distributed across the surface and had rotated out of sight when observed with X-Shooter. The feature appears in both DESI exposures, but they were taken consecutively, so a transient problem could have impacted both, though other white dwarfs observed in the same exposures appear unaffected.

\subsubsection{Rare spectral types with metals}
\begin{figure}
\centering
\includegraphics[width=\columnwidth]{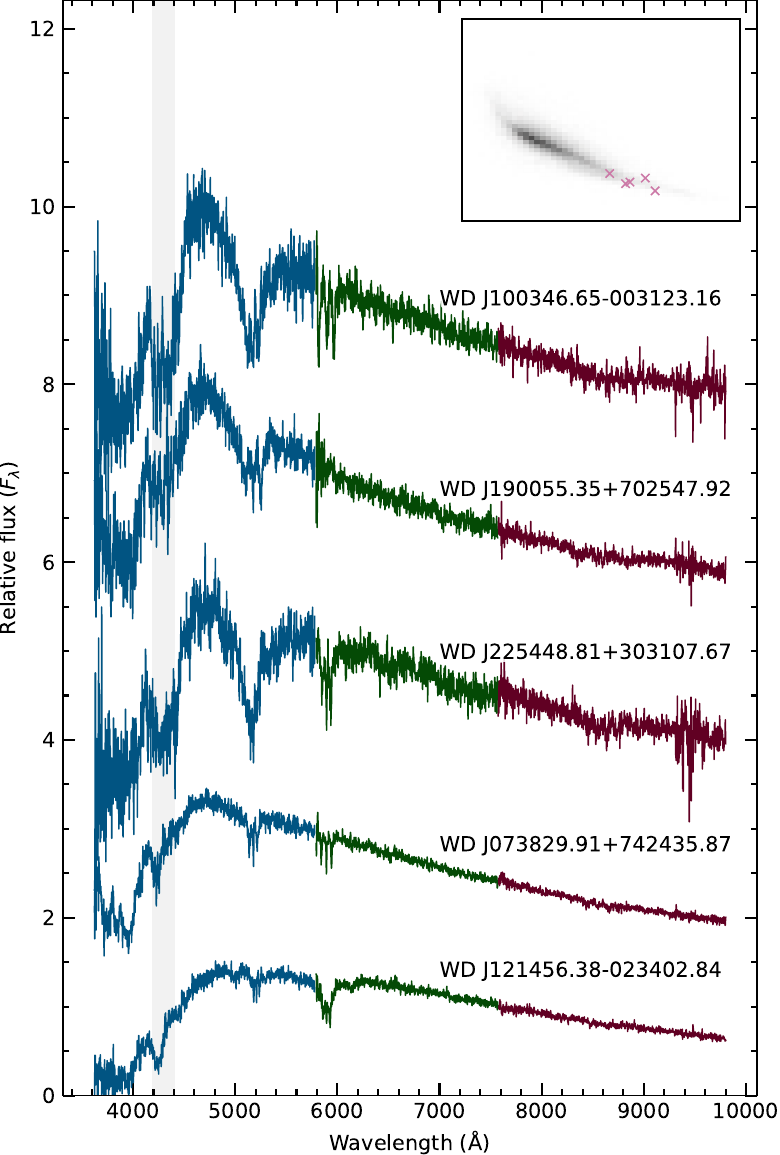}
\caption{DZH stars, with spectra dominated by Zeeman-split metal lines. The inset shows their positions relative to the white dwarf cooling track on the HR~diagram, using the same axes, bins and shading as Fig.~\ref{figureHRdiagrams}. Spectra are shown smoothed with a 5-pixel boxcar median, normalised and offset vertically. Grey shading indicates the region potentially affected by an instrumental artefact.}
\label{figureDZH}
\end{figure}

\begin{figure}
\centering
\includegraphics[width=\columnwidth]{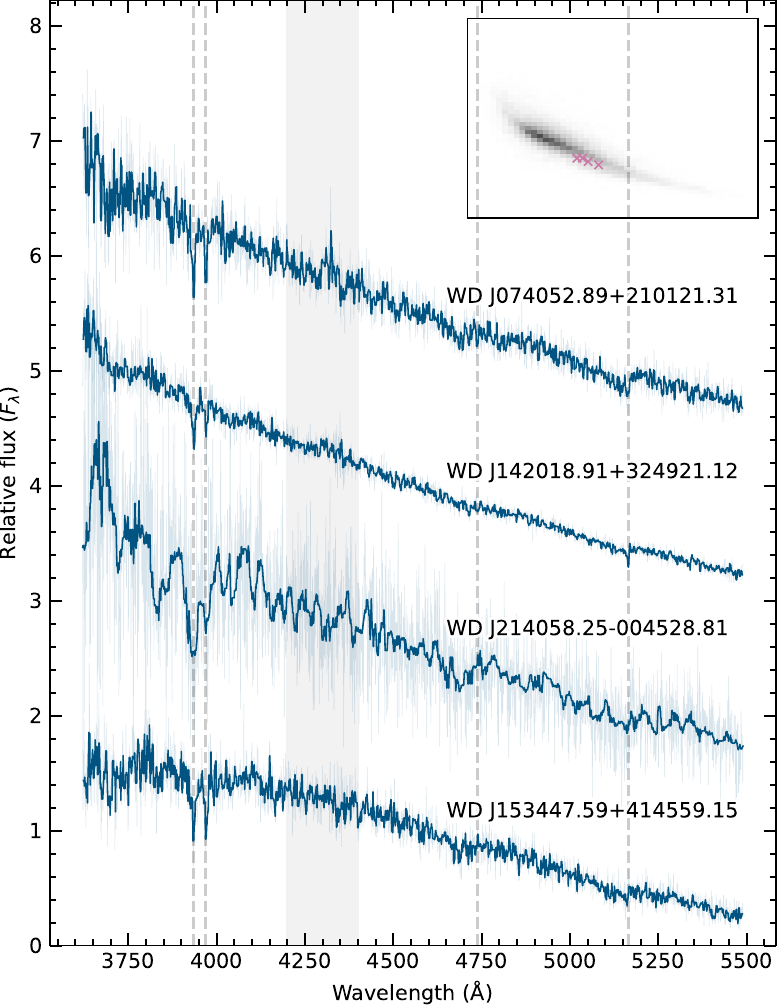}
\caption{DQZ and DZQ stars. Vertical dashed lines indicate the Ca\,H and K resonance lines and two of the C\textsubscript{2} Swan bands. The inset shows their positions relative to the white dwarf cooling track on the HR~diagram, using the same axes, bins and shading as Fig.~\ref{figureHRdiagrams}. Spectra are shown smoothed with a 5-, 20-pixel boxcar median, normalised and offset vertically. Grey shading indicates the region potentially affected by an instrumental artefact.}
\label{figureDQZ}
\end{figure}

The catalogue contains a handful of DQZ, DZQ and DZH stars, most of which are new identifications, albeit at only moderate confidence in some cases. The incidence of magnetism in cool white dwarfs is challenging to quantify \citep{Bagnulo2020}, owing to their typical lack of spectral features, making the DZH stars (Fig.~\ref{figureDZH}) useful targets for measuring field strengths \citep{Hollands2015}. The DQZ and DZQ stars in Fig.~\ref{figureDQZ} are notable because while carbon and metal features are commonplace in white dwarf spectra, they are rarely observed together. The presence of metals in the atmosphere can suppress carbon features \citep{Blouin2022,Hollands2022}, but that may not be sufficient to fully account for the low incidence of DQZ/DZQ stars \citep{Farihi2024}.

\subsubsection{A new subclass?}
\begin{figure}
\centering
\includegraphics[width=\columnwidth]{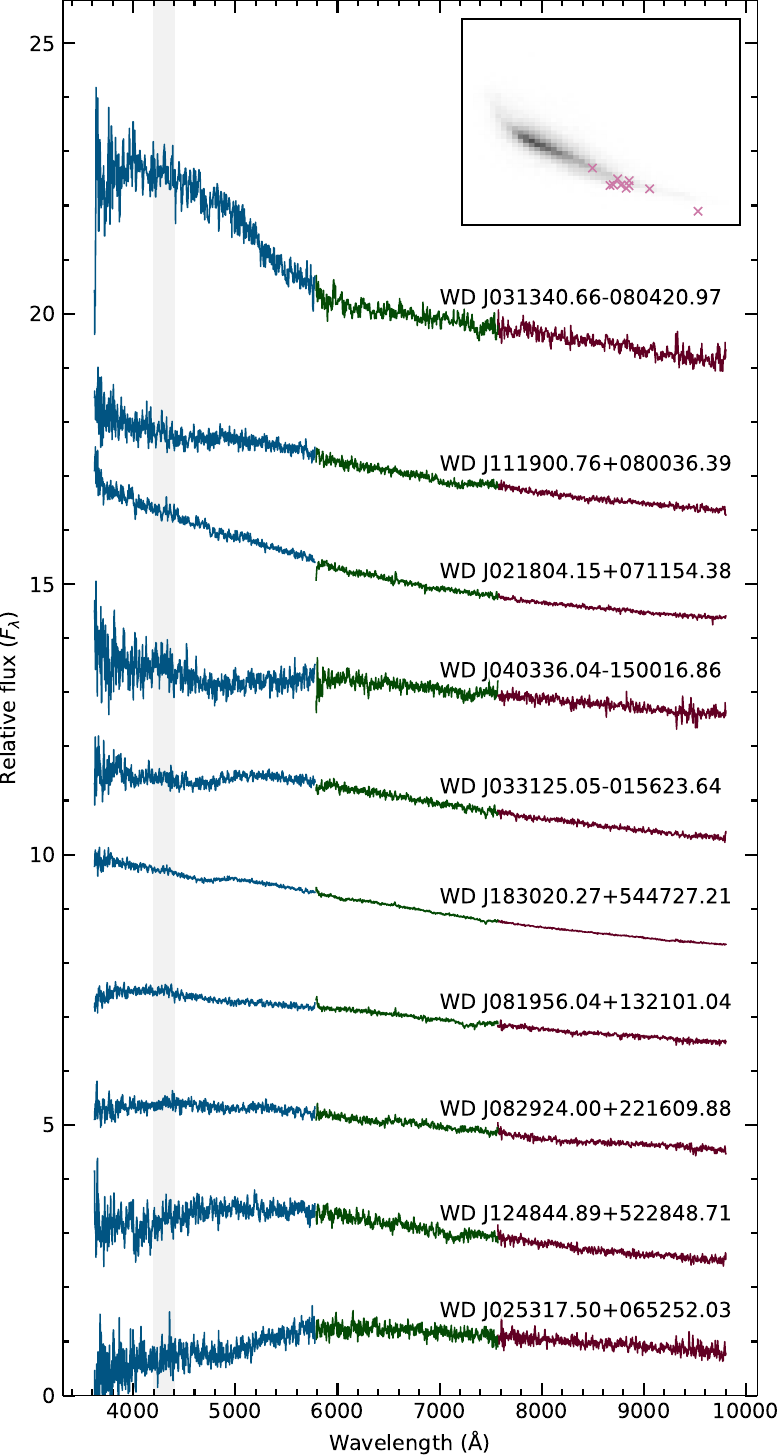}
\caption{Objects with shallow dips in their spectra. The inset shows their positions relative to the white dwarf cooling track on the HR~diagram, using the same axes, bins and shading as Fig.~\ref{figureHRdiagrams}. Spectra are shown smoothed with a 10-pixel boxcar median, normalised and offset vertically. Grey shading indicates the region potentially affected by an instrumental artefact.}
\label{figureWigglers}
\end{figure}

Several objects stood out as being unusual but superficially similar to each other. The spectra shown in Fig.~\ref{figureWigglers} have in common shallow dips, some quite subtle. Several occupy a small region on the HR~diagram (centred around $\bprp\approx0.75$, $G_{\text{abs}}\approx14.4$). These objects might have strong magnetic fields, as has been suggested for similarly mysterious objects (GF19, fig.~19). Spectropolarimetry would be required to test that hypothesis.

\section{Cautions and recommendations}
\label{sectionCautionsRecommendations}

The main catalogue represents our best attempt at classifying and characterising each object, but we stress that we have not performed a bespoke fit to each object, and that all entries carry uncertainty. Set out below are some caveats on interpretation, and then some recommendations for use.

In summary, users should consider the following points when interpreting results or selecting samples from the catalogue:

\begin{itemize}
    \item Our primary metric for identifying white dwarfs is $\pwdDESI$, which combines both visual classifications and model fit results (\texttt{pwd} in Table~\ref{tableMainCatalogue}).
    \item The \texttt{fluxContamination\_DESI} column in Table~\ref{tableMainCatalogue} flags potentially problematic data. Contamination at the 0.01 level is strong enough to alter the spectral type (Section~\ref{subsectionCatalogueFluxContamination}), but the threshold for concern will depend on the application.
    \item Visual classifications become increasingly unreliable at $\SN<4$ (Fig.~\ref{figureConfidenceVsSN}).
    \item Objects targeted as white dwarfs in the \texttt{dark} and \texttt{bright} surveys of the \texttt{main} program (the \textit{primary sample}) have a straightforward selection function, and are thus suitable for statistical studies. We flag these using the \texttt{primary} column of Table~\ref{tableMainCatalogue} for convenience.
    \item When selecting objects by spectral type, we recommend $\texttt{specType\_confidence}>0.8$ to achieve a good balance between completeness and misidentifications.
    \item If our inferred $\Teff$ or $\log{g}$ are close to the boundaries of the model grids, or the distance is strongly discrepant with the \textit{Gaia} parallax, that may indicate problematic data, or a stellar type incompatible with our H and He models.
\end{itemize}

\edit{Our automated target crossmatching will have missed some objects that fell outside our 1.5-arcsec search radius, but also allows false positive matches to slip through. One example is discussed in Section~\ref{subsubsectionUHE}, where an exposure targeting an M~dwarf has been assigned to its white dwarf companion 1\,arcsec away, but such cases may be identifiable by inspection of the target bitmasks, which in that particular case show that the rogue exposure was not associated with a white dwarf target.}

Classification and model fitting were performed independently, and thus do not always agree. Indeed, they \textit{should not} always agree, as physical characteristics do not map uniquely onto spectral types (e.g. He-atmosphere stars with large H/He abundances can appear as DA types), and feature visibility depends on both resolution and S/N. Research continues into white dwarf modelling, so the synthetic spectra are subject to systematics that may influence fitting. Similarly, our fits are constrained by the bounds of the parameter space of the mass--radius relations, and will not be appropriate for low- or high-mass objects whose internal compositions deviate from those assumed in the evolutionary models. 

Only three types of model atmosphere were considered: pure H, pure He, and a $\text{H}+\text{He}$ mixture. While appropriate for most objects in the catalogue, results will deteriorate where other elements are likely to be present, e.g. for the carbon-rich DQ types, especially those with strong molecular features, and for the metal-enriched DZ stars.

\edit{Our model fits assume a single star, but that will not always be valid. We provide masses calculated from the mass--radius relation used during fitting, which may help to identify unresolved binary candidates at $M\lesssim0.55\,\Msun$, where single star evolution has not yet had time to produce white dwarfs \citep[e.g.][]{Cummings2018}. However, masses for stars whose spectra deviate from the models we used will be unreliable. Separately, some spectral types may indicate binarity, for example DAQs \citep[e.g.][]{Vennes2012,AdamanePallathadka2026}.}

Our probabilistic catalogue allows samples to be selected in a flexible manner. For example, a high-purity sample of DAs can be obtained by selecting objects with that classification where $\pwdDESI=1.0$. Conversely, all stars where there was even the smallest suspicion of metal lines could be extracted from supplementary Table~\ref{tableClassifications}, which lists all classifications considered for each star. Alternative confidence metrics could be explored by re-analysing the raw classification logs (Table~\ref{tableClassificationLogs}).

It is particularly challenging to distinguish hot white dwarfs from subdwarfs, as each may have Balmer, \ion{He}{ii}, or metal lines, with only subtle differences in line widths between the two classes. Hotter white dwarfs are brighter, and thus observable at greater distances, where parallax errors become large enough that white dwarfs overlap with subdwarfs on the \textit{Gaia} HR~diagram. Our distance prior is tuned to white dwarfs, and we did not use subdwarf models, so our photometric fits are not intended to resolve the two populations, although objects near the lower $\log{g}$ grid boundary may well prove to be subdwarfs. With these issues in mind, we have assigned each object a probability of being a white dwarf based on visual classifications and~--~for objects potentially having DA types at any level of confidence~--~the results of spectroscopic fitting to the extended $\log{g}$ model grid. Users requiring a pure sample should consider filtering by $\pwdDESI$, where a high threshold such as 0.99 will exclude most problematic sources.

Among the most common metal lines in optical spectra are Ca\,H and K. They are subtle features in warm DAZ stars at the resolution of DESI, and as noted above there are residuals from the same features in the standard stars used in flux calibration. Contamination from interstellar Ca~H and K lines is a risk for stars outside the local bubble at $d\gtrsim100$\,pc, although the presence of the Na\,D doublet may help to identify such cases. Spectral classifications are based on appearance, not physical properties, so we caution that not every star with a Z in its spectral type has photospheric metals.

A residual H$\upalpha$ feature is often visible in DESI spectra, again due to the flux-calibration standard stars used in the data reduction pipeline. H$\upalpha$ often appears in emission, risking confusion with the rare DAe stars that exhibit Balmer emission \citep{Elms2023}.

Fig.~\ref{figureConfidenceVsSN} shows that human classification becomes increasingly unreliable at $\SN \lesssim 4$. Users should consider their $\SN$ requirements when selecting samples based on spectral type, or use the confidence scores as weights in their analyses. Further, some features may have been missed by the human classifiers: most spectra were examined only for a few seconds in total, so users interested in rare and unusual objects should not assume that they have all been identified.

Main-sequence stars can be included by the selection function for the white dwarf survey, because \textit{Gaia} colours for faint red objects can be unreliable \citep{Riello2021}. Their spectra are typically low-$\SN$, and thus it can be challenging to distinguish the sharp lines of a main-sequence star from the noisy continuum of a white dwarf. The presence of strong metal lines was often a clue that an object was a main-sequence star, but a definitive separation of the two types of star is challenging from visual classification alone.

White dwarfs typically have only a few broad lines, which makes measuring radial velocities a challenge. The sharp cores of the Balmer lines are not resolved in DESI spectra, and H$\upgamma$ is in the wavelength range affected by the collimator reflectivity issue. While we report velocities and any potential variability, the uncertainties are large, and we caution against over-interpretation of results for individual stars without closer scrutiny. There may be scope to improve on our results by incorporating data from other surveys, such as SDSS.

The DESI white dwarf survey has a well-defined selection function and is thus suited to statistical analyses, as discussed in Section~\ref{subsectionDESIwhiteDwarfs}. A largely unbiased and magnitude-limited sample can be assembled by selecting objects targeted as white dwarfs in the main survey and observed under the \texttt{dark} or \texttt{bright} programs. We identify such objects in our catalogue using the \texttt{primary} column (Table~\ref{tableMainCatalogue}). White dwarfs captured only by the \texttt{backup} program or observed incidentally will have a more complex selection function.

\section{Outlook}
\label{sectionOutlook}

This paper presents the largest catalogue of spectroscopically confirmed white dwarfs to date. Crucially, they provide a minimally-biased sample, whose magnitude-limited selection function is convenient for statistical studies. However, DESI operations continue, and the second data release will include almost two years' additional data, increasing the tally of observed white dwarf candidates to around 100\,000. Meanwhile, the SDSS-V survey is well underway, the 4MOST project has begun science operations, and commissioning of the WEAVE instrument is almost complete. Between them, they will increase the spectroscopic sample of white dwarfs by an order of magnitude within a few years.

Classifying and characterising so many objects with traditional methods would be a daunting task. Fortunately, automated techniques are in a phase of rapid development. Several avenues are being explored for machine classification and characterisation of white dwarfs \citep{Byrne2024,Kao2024,Vincent2024,PerezCouto2025,Zhang2025,GarciaZamora2025,Munday2026}. There are good prospects for efficiently processing the majority of data with automated methods, allowing researchers to focus on the outliers that require more careful analysis or yield new discoveries. We hope that the homogeneous observations and human-assigned labels presented in this catalogue will prove a useful resource for those developing new analysis techniques.

\section*{Acknowledgements}

We thank the anonymous referee for constructive feedback that improved the paper and the catalogue, and Ingrid Pelisoli for discussions on subdwarf spectra.

This project has received funding from the European Research Council (ERC) under the European Union’s Horizon 2020 research and innovation programme (Grant agreement No. 101020057). CJM acknowledges financial support from Imperial College London through an Imperial College Research Fellowship grant. LKR is supported by NOIRLab, which is managed by the Association of Universities for Research in Astronomy (AURA) under a cooperative agreement with the U.S. National Science Foundation. SX is supported by the international Gemini Observatory, a program of NSF NOIRLab, which is managed by the Association of Universities for Research in Astronomy (AURA) under a cooperative agreement with the U.S. National Science Foundation, on behalf of the Gemini partnership of Argentina, Brazil, Canada, Chile, the Republic of Korea, and the United States of America.

This material is based upon work supported by the U.S. Department of Energy (DOE), Office of Science, Office of High-Energy Physics, under Contract No. DE–AC02–05CH11231, and by the National Energy Research Scientific Computing Center, a DOE Office of Science User Facility under the same contract. Additional support for DESI was provided by the U.S. National Science Foundation (NSF), Division of Astronomical Sciences under Contract No. AST-0950945 to the NSF’s National Optical-Infrared Astronomy Research Laboratory; the Science and Technology Facilities Council of the United Kingdom; the Gordon and Betty Moore Foundation; the Heising-Simons Foundation; the French Alternative Energies and Atomic Energy Commission (CEA); the National Council of Humanities, Science and Technology of Mexico (CONAHCYT); the Ministry of Science, Innovation and Universities of Spain (MICIU/AEI/10.13039/501100011033), and by the DESI Member Institutions: \url{https://www.desi.lbl.gov/collaborating-institutions}. Any opinions, findings, and conclusions or recommendations expressed in this material are those of the author(s) and do not necessarily reflect the views of the U.S. National Science Foundation, the U.S. Department of Energy, or any of the listed funding agencies. The authors are honoured to be permitted to conduct scientific research on I'oligam Du'ag (Kitt Peak), a mountain with particular significance to the Tohono O’odham Nation.

This work has made use of data from the European Space Agency (ESA) mission \textit{Gaia} (\url{https://www.cosmos.esa.int/gaia}), processed by the \textit{Gaia} Data Processing and Analysis Consortium (DPAC, \url{https://www.cosmos.esa.int/web/gaia/dpac/consortium}). Funding for the DPAC has been provided by national institutions, in particular the institutions participating in the \textit{Gaia} Multilateral Agreement.

This research has used data, tools or materials developed as part of the EXPLORE project that has received funding from the European Union’s Horizon 2020 research and innovation programme under grant agreement No~101004214.

This work is based on observations obtained with the Samuel Oschin Telescope 48-inch and the 60-inch Telescope at the Palomar Observatory as part of the Zwicky Transient Facility project. ZTF is supported by the National Science Foundation under Grants No. AST-1440341 and AST-2034437 and a collaboration including current partners Caltech, IPAC, the Oskar Klein Center at Stockholm University, the University of Maryland, University of California, Berkeley, the University of Wisconsin at Milwaukee, University of Warwick, Ruhr University, Cornell University, Northwestern University and Drexel University. Operations are conducted by COO, IPAC, and UW

This research is based on observations collected at the European Organisation for Astronomical Research in the Southern Hemisphere under ESO programme 115.28GM.002

\textit{Author contributions}: AS wrote the paper and produced the figures, with BG, PI, CM, LR and SX contributing. BG managed the project. BG and CM designed the selection function for the white dwarf survey, and handled data logistics. BG, SX and AS cross-matched DESI objects against external catalogues. BG used a code developed by DK to produce the white dwarf models that AS fitted to the data. AS, BG, PI and CM visually classified spectra. LR calculated the flux contamination metrics. SX compared human-assigned and machine-learned classifications. The other authors have made significant contributions to the DESI Collaboration and the white dwarf survey underpinning this work.

\textit{Software used}: \href{https://astroquery.readthedocs.io/en/latest/}{\textsc{Astroquery}} \citep{Ginsburg2019}, \href{https://www.astropy.org}{\textsc{Astropy}} \citep{Astropy2013, Astropy2018, Astropy2022}, \href{https://samreay.github.io/ChainConsumer/}{\textsc{ChainConsumer}} \citep{Hinton2016}, \href{https://cmasher.readthedocs.io/}{\textsc{cmasher}} \citep{vanderVelden2020}, \href{https://dustmaps.readthedocs.io/en/latest/}{\textsc{dustmaps}} \citep{Green2018dustmaps}, 
\href{https://matplotlib.org/}{\textsc{Matplotlib}} \citep{Hunter2007}, 
\href{https://numpy.org/}{\textsc{NumPy}} \citep{Harris2020}, 
\href{https://opencv.org/}{\textsc{OpenCV}} \citep{Bradski2000}, 
\href{https://pocomc.readthedocs.io/en/latest/}{\textsc{pocoMC}} \citep{Karamanis2022pocoMC, Karamanis2022PMC}, 
\href{https://scikit-image.org/}{\textsc{scikit-image}} \citep{VanDerWalt2014}, 
\href{https://scikit-learn.org/}{\textsc{scikit-learn}} \citep{Pedregosa2011}, 
\href{https://scipy.org/}{\textsc{SciPy}} \citep{Virtanen2020}, 
\href{https://zenodo.org/records/13942238}{\textsc{specutils}} \citep{specutils1.18}
. Plots use the \href{https://jfly.uni-koeln.de/color/}{Okabe--Ito} colour palette.

\section*{Data Availability}

All data analysed here are available in public archives, including those of \href{https://data.desi.lbl.gov/doc/releases/dr1/}{DESI}, the \href{https://www.legacysurvey.org/}{DESI Legacy Imaging Surveys}, \href{https://gea.esac.esa.int/archive/}{\textit{Gaia}}, \href{https://outerspace.stsci.edu/display/PANSTARRS/}{Pan-STARRS}, \href{https://www.sdss.org/}{SDSS} and \href{https://irsa.ipac.caltech.edu/Missions/ztf.html}{ZTF}.

We provide the data described here at \href{https://cygnus.astro.warwick.ac.uk/phsdaj/WhiteDwarfs/}{\url{https://cygnus.astro.warwick.ac.uk/phsdaj/WhiteDwarfs/}} and will deposit the core dataset at Zenodo upon journal publication. The \texttt{Tables} folder contains the main catalogue (Table~\ref{tableMainCatalogue}) and auxiliary files (Tables~\ref{tableExposures}--\ref{tableKinematics}). The \texttt{J$xxxx$} folders contain per-object files, including spectra extracted from the DESI cframes and their per-arm coadds, samples and plots from the model-fitting exercise, and the data required to reconstruct our extinction priors.

In compliance with the DESI data management plan, coordinates for points shown in the figures in this paper are available at \href{https://doi.org/10.5281/zenodo.20642254}{doi:10.5281/zenodo.20642254}.



\bibliographystyle{mnras}
\bibliography{DESI_DR1} 



\appendix

\section{Comparison with DESI DR1 stellar catalogue}
\label{appendixMWS}

Stellar parameters, abundances and radial velocities are available for over 4\,million stars from the DESI DR1 Milky Way Survey \citep{Koposov2026}. They include results from the Stellar Parameters (SP) pipeline \citep{Koposov2024}, whose templates included earlier generations of the white dwarf models used here \citep{Koester2010}. We compare the contents of that catalogue to our own by cross-matching against the SP output table using \textit{Gaia} designations if available, or DESI \texttt{TARGETID}s otherwise. Multiple \texttt{TARGETID}s can be assigned to the same object \citep{Myers2023}, so there are fewer objects than catalogue table rows. The \texttt{BESTGRID} field identifies which kind of model gave the best fit, where codes \texttt{m\_rdesi6--9} represent white dwarfs, and the rest are stellar atmosphere models with $\log{g}<6.5$ \citep{Kurucz2005}.

\begin{figure}
\centering
\includegraphics[width=\columnwidth]{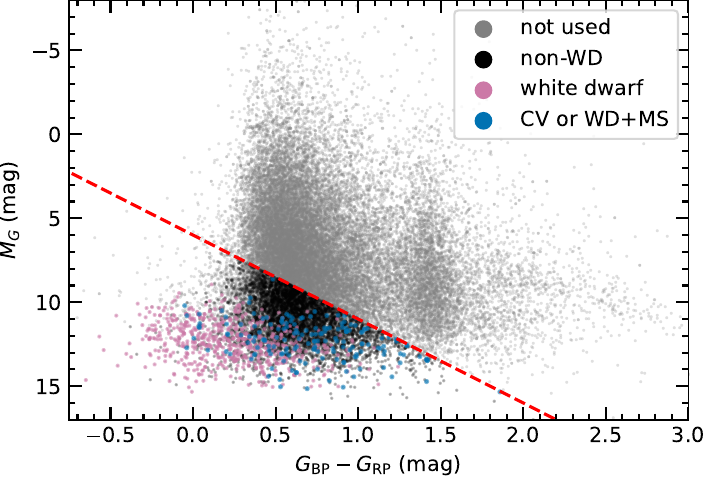}
\caption{\textit{Gaia} HR diagram with objects whose best fits were to white dwarf models in the DESI DR1 stellar catalogue \citep{Koposov2026}. The red dashed line corresponds to eqn.~1 from GF21, which we used to select objects for inclusion in our catalogue. Points are coloured based on our visual classifications (Section~\ref{sectionSpectralClassification}).}
\label{FigureMWScatalogueHRD}
\end{figure}

There are 389\,347 rows in the SP table whose \texttt{BESTGRID} field contained a white dwarf code. Of these, 68\,879 corresponded to H-atmosphere models and 320\,468 to He-atmosphere models, contrary to expectations (H-atmosphere white dwarfs dominate the observed population). Of those with \textit{Gaia} data, 42 per cent have no parallax measurements, and the remainder have a median parallax of 0.3\,mas, and a median brightness of $G=19.7$\,mag. Therefore, we suspect that the majority of these SP catalogue entries are spurious, but investigating the nature of so many objects is beyond the scope of this paper. We thus consider a more tractable subset: 5521 objects not already in our catalogue lie within the region of the \textit{Gaia} HR~diagram defined by equations~1 and 2 of GF21, as shown in Fig.~\ref{FigureMWScatalogueHRD}. We added them to our catalogue as white dwarf candidates, as mentioned in Section~\ref{subsectionDESIwhiteDwarfs}, although our visual classifications subsequently revealed that 85~per~cent of them are extragalactic sources.

Next, we inspected the 67\,425 rows that matched objects already in our catalogue, and found moderate agreement: 92 per cent of the rows where we found $\pwdDESI<0.5$ were listed in the SP catalogue with non-white-dwarf model types, but so were 38 per cent of those where we found $\pwdDESI\ge0.5$. However, where both our own and the SP catalogues agreed that an object was a white dwarf, they also agreed on the atmosphere type in 98 per cent of cases.

\begin{figure}
\centering
\includegraphics[width=\columnwidth]{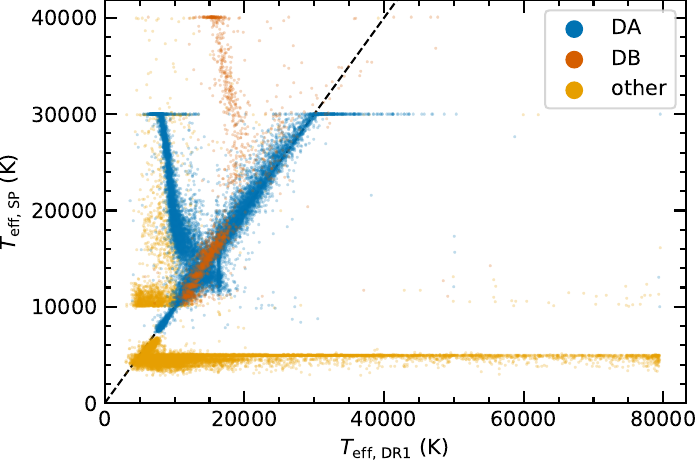}
\caption{Comparison between effective temperatures for white dwarfs in our catalogue and those determined by the SP pipeline from the DESI DR1 stellar catalogue. Points are coloured according to the spectral type we assigned by visual classification. The dashed line indicates a 1:1 correspondence between the axes.}
\label{FigureMWScatalogueTeff}
\end{figure}

Fig.~\ref{FigureMWScatalogueTeff} compares effective temperatures for objects with $\pwdDESI>0.5$ in our catalogue against those from the SP pipeline. While there is good agreement in many cases, there are several obvious artefacts. First, those objects best fit by Kurucz stellar models in the SP pipeline are distributed across the whole temperature range, with spectral types in proportions broadly similar to those in our catalogue, i.e. there is no obvious pattern. Second, many of the SP pipeline fits are at the edge of their respective model grids, which have narrower temperature ranges than ours. Third, there are bifurcations in both the DA and DB populations, almost certainly caused by degeneracy between `hot' and `cold' solutions in fits to normalised spectra only, as performed by the SP pipeline \citep{RebassaMansergas2007}. Our fits include spectra, photometry and parallaxes, lifting that degeneracy.

\onecolumn
\FloatBarrier
\section{Duplicated objects}
\label{appendixDuplicates}

As explained in Section~\ref{subsectionDataIssues}, some objects have the potential to cause confusion as they share the same DESI fibres, or because they have been assigned more than one WD\,J name. They are listed in Tables~\ref{tableDuplicateExposures} and \ref{tableDuplicateWDJnames}, respectively.

\begin{table}
\centering
\caption{Objects with separate WD\,J names and \textit{Gaia} designations that fall within the same DESI fibres}
\label{tableDuplicateExposures}
\begin{tabular}{lll}
\hline
WD\,J name & Gaia designation & System type \\
\hline
WD\,J020242.48+260512.22 & Gaia DR3 105617304552984576 & WD+WD+MS\\
WD\,J020242.57+260513.04 & Gaia DR3 105617338913173760 & \citep{Heintz2022} \vspace{5 pt}\\

WD\,J074852.96+302543.44 & Gaia DR3 879036662822920448 & DA+DAH\\
WD\,J074853.08+302543.59 & Gaia DR3 879036662822100224 & \citep{Dobbie2012} \vspace{5 pt}\\

WD\,J171233.31+425049.20 & Gaia DR3 1354425672030028416 & WD+WD\\
WD\,J171233.34+425047.87 & Gaia DR3 1354425667735822848 & \citep{Heintz2022} \vspace{5 pt}\\

WD\,J232301.02+013757.25 & Gaia DR3 2645720882496091648 & WD+WD\\
WD\,J232301.00+013756.14 & Gaia DR3 2645720886790968832 & \citep{Heintz2022}\\
\hline
\end{tabular}
\end{table}

\begin{table}
\centering
\caption{Objects with the same \textit{Gaia} designation but multiple WD\,J~names in the GF19 and GF21 catalogues}
\label{tableDuplicateWDJnames}
\begin{tabular}{ll}
\hline
GF19 (retained) & GF21 (discarded) \\
\hline
WD\,J011418.65$-$011957.08 & WD\,J011418.64$-$011957.08 \\
WD\,J014058.65+033525.05 & WD\,J014058.65+033525.00 \\
WD\,J025905.43+044538.21 & WD\,J025905.42+044538.20 \\
WD\,J084228.26+173248.58 & WD\,J084228.26+173248.57 \\
WD\,J084500.00+174138.22 & WD\,J084500.01+174138.22 \\
WD\,J091913.69+332455.76 & WD\,J091913.69+332455.75 \\
WD\,J092414.86+330727.12 & WD\,J092414.87+330727.11 \\
WD\,J100317.06+025510.33 & WD\,J100317.06+025510.32 \\
WD\,J102506.06+120022.36 & WD\,J102506.06+120022.35 \\
WD\,J105714.85+802810.20 & WD\,J105714.85+802810.19 \\
WD\,J121409.46+004341.29 & WD\,J121409.46+004341.28 \\
WD\,J125409.21+252108.06 & WD\,J125409.21+252108.05 \\
WD\,J132630.70$-$033402.84 & WD\,J132630.70$-$033402.83 \\
WD\,J135540.91+005654.09 & WD\,J135540.90+005654.09 \\
WD\,J135841.00$-$032017.65 & WD\,J135841.00$-$032017.64 \\
WD\,J141556.68$-$005814.84 & WD\,J141556.68$-$005814.86 \\
WD\,J143348.77+022422.33 & WD\,J143348.77+022422.31 \\
WD\,J144023.58+135454.66 & WD\,J144023.58+135454.67 \\
WD\,J144603.03+515455.44 & WD\,J144603.03+515455.45 \\
WD\,J154645.64+422955.44 & WD\,J154645.64+422955.36 \\
WD\,J163007.90+311318.06 & WD\,J163007.90+311318.05 \\
WD\,J213508.47+021732.63 & WD\,J213508.47+021732.64 \\
WD\,J224624.78+005040.98 & WD\,J224624.78+005040.97 \\
\hline
\end{tabular}
\end{table}

\FloatBarrier
\section{Spectral energy distribution fits}
\label{appendixFits}

\begin{figure*}
\centering
\begin{minipage}{0.64\textwidth}
    \includegraphics[width=\linewidth]{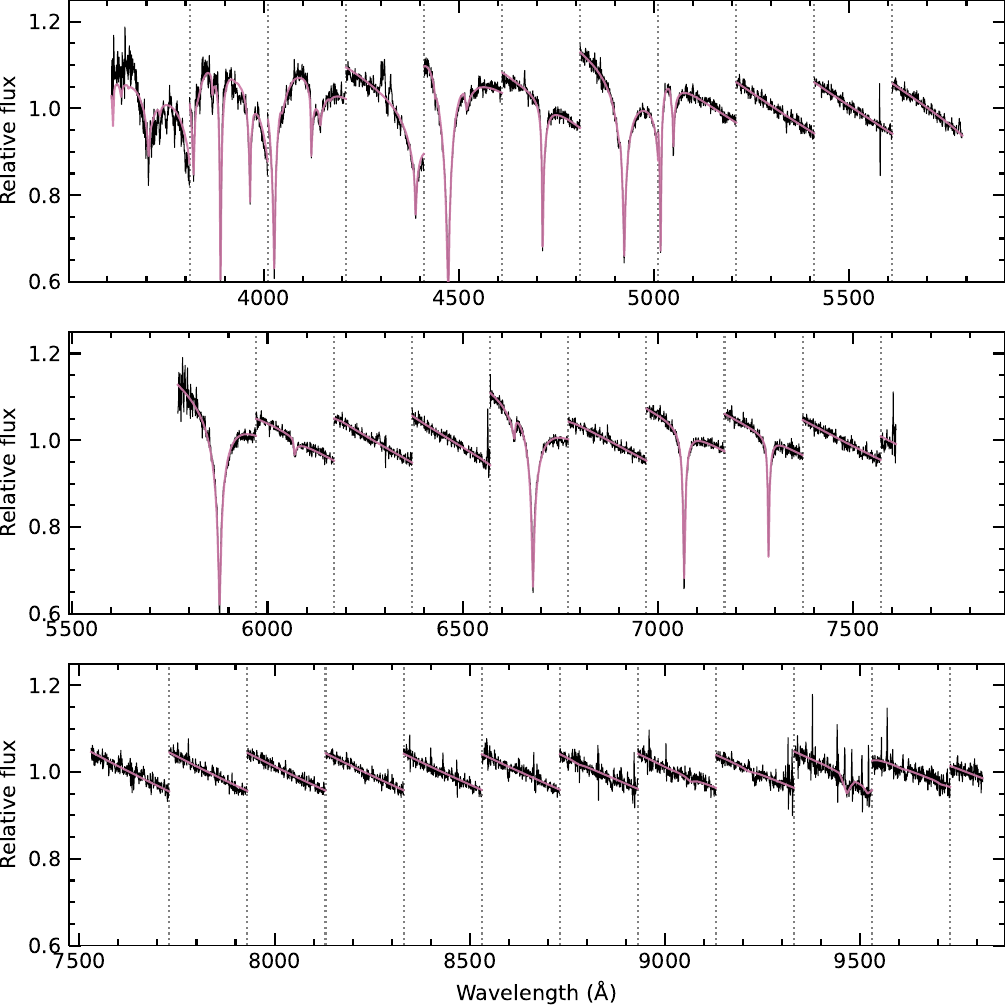}
\end{minipage}
\hfill
\begin{minipage}{0.33\textwidth}
    \includegraphics[width=\linewidth]{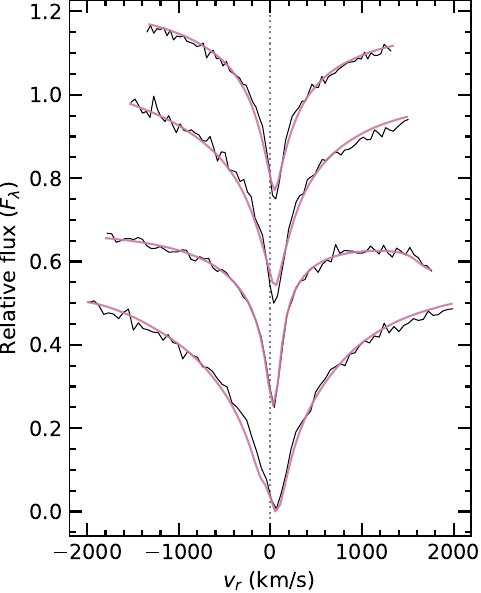}
    
    \vspace{0.5em}
    
    \includegraphics[width=\linewidth]{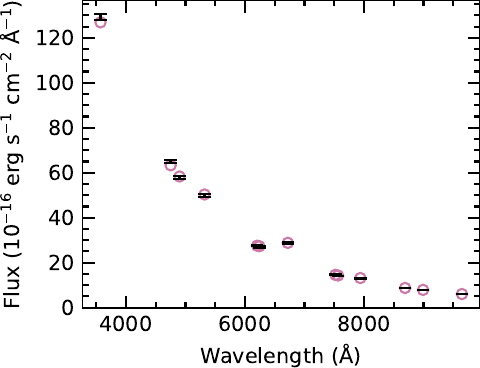}
\end{minipage}
\caption{Joint fit of the pure-He atmosphere model to WD\,J071959.42+402122.13, a DB white dwarf with $\Teff\approx17\,000$\,K, following Fig.~\ref{figureSpectralFitsH} but showing \ion{He}{i} lines at 4471.9, 5016.1, 5876.1 and 6678.7\,\AA\ in the upper-right panel.}
\label{figureSpectralFitsHe}
\end{figure*}

\begin{figure*}
\centering
\begin{minipage}{0.64\textwidth}
    \includegraphics[width=\linewidth]{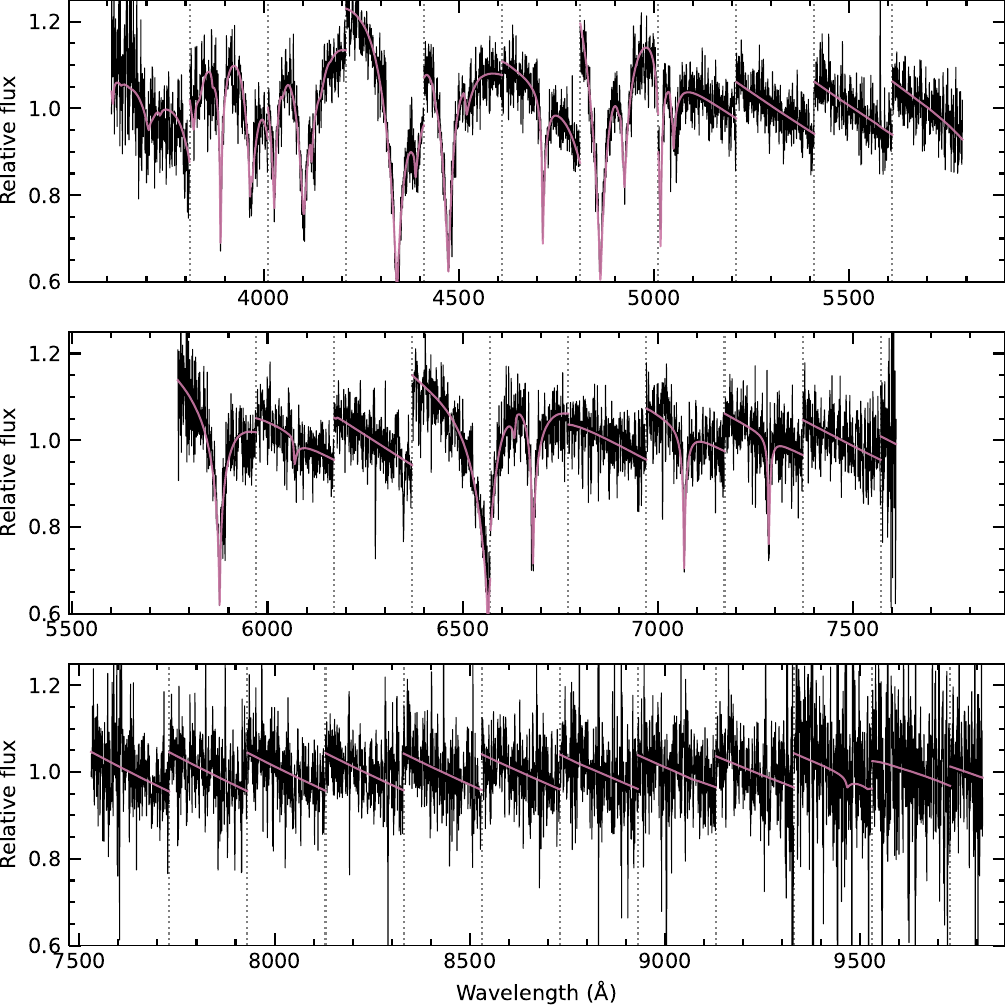}
\end{minipage}
\hfill
\begin{minipage}{0.33\textwidth}
    \includegraphics[width=\linewidth]{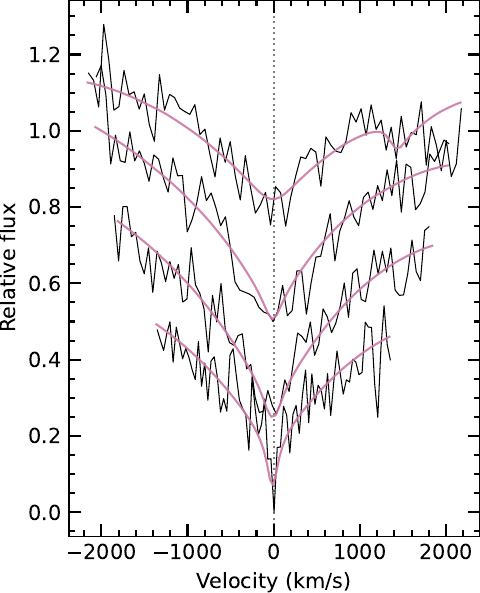}
    
    \vspace{0.5em}
    
    \includegraphics[width=\linewidth]{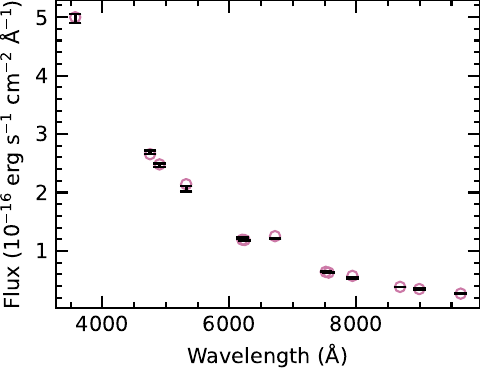}
\end{minipage}
\caption{Joint fit of the mixed He+H atmosphere model to WD\,J155627.91+012624.36, a DAB white dwarf with $\Teff\approx20\,000$\,K, following Fig.~\ref{figureSpectralFitsH}.}
\label{figureSpectralFitsHe+H}
\end{figure*}

The methods for fitting white dwarf models to the spectroscopic, photometric and astrometric data are described in Section~\ref{sectionModelFitting}, where Fig.~\ref{figureSpectralFitsH} shows a fit to a warm ($\Teff=30\,300$\,K) DA white dwarf using a pure H atmosphere model. For completeness, we show fits to stars using the He and He+H models in Figs.~\ref{figureSpectralFitsHe} and \ref{figureSpectralFitsHe+H}, respectively.

\begin{figure*}
\centering
\includegraphics[width=\textwidth]{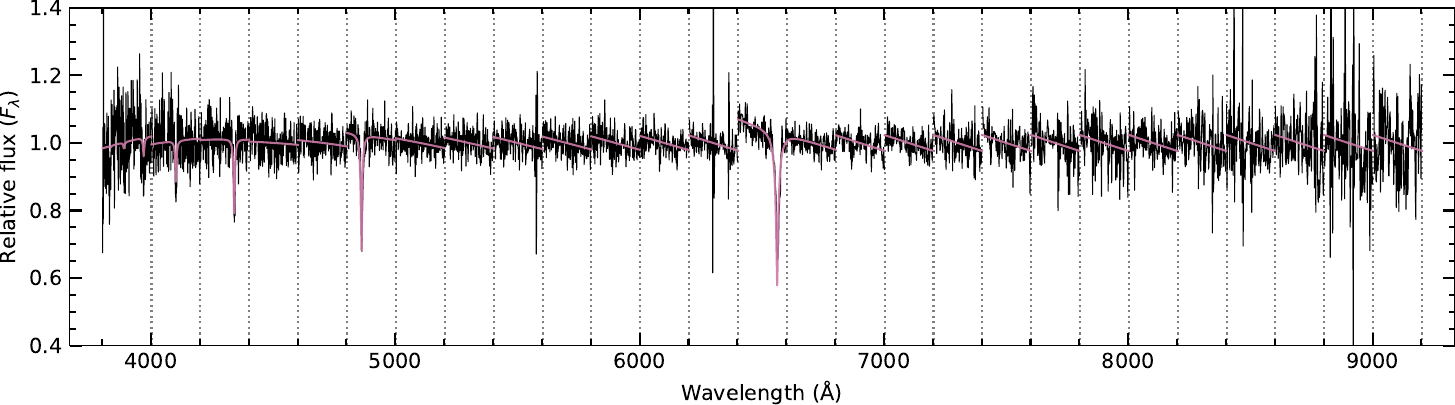}
\includegraphics[width=\textwidth]{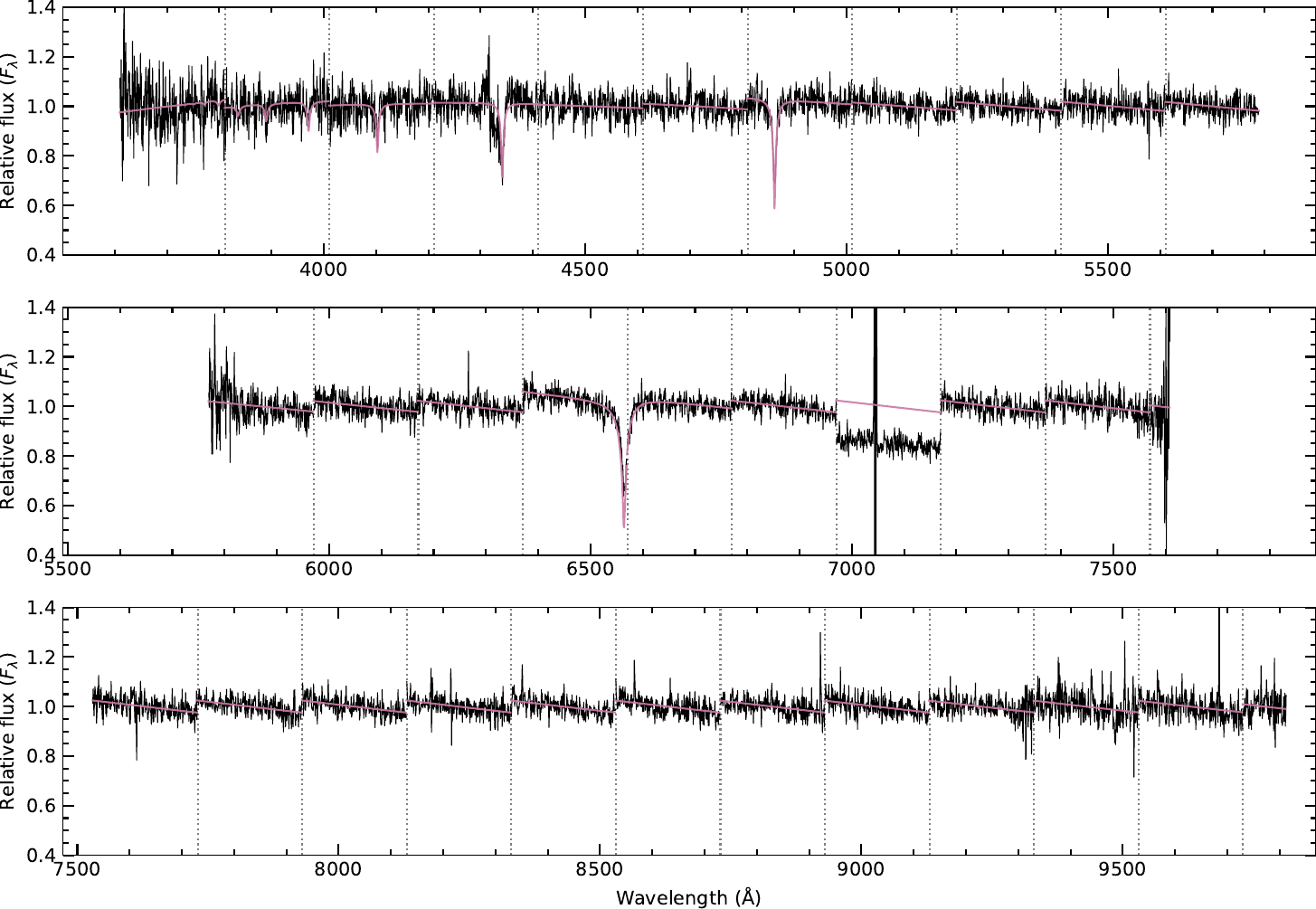}
\caption{Spectroscopic fits of a pure-H atmosphere model to WD\,J170401.50+200700.65, a DA white dwarf with $\Teff\approx6200$\,K. From top to bottom, the panels show the SDSS spectrum and the DESI~\textit{B}, \textit{R} and \textit{Z} spectra. Data are shown in black, with the best-fit model in red.}
\label{figureSpectralFit_WDJ170401.50+200700.65}
\end{figure*}

\begin{figure*}
\centering
\includegraphics[width=\textwidth]{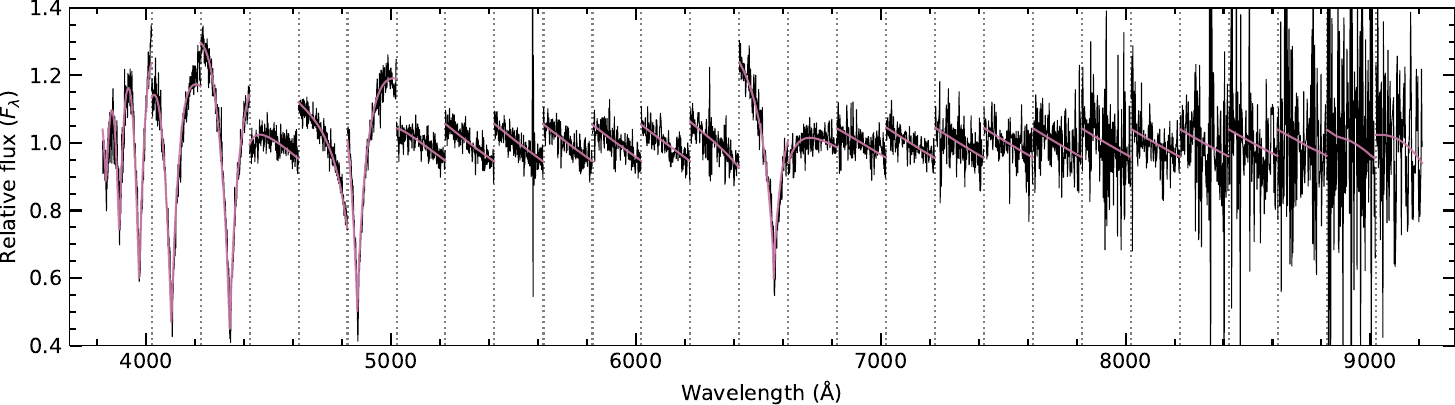}
\includegraphics[width=\textwidth]{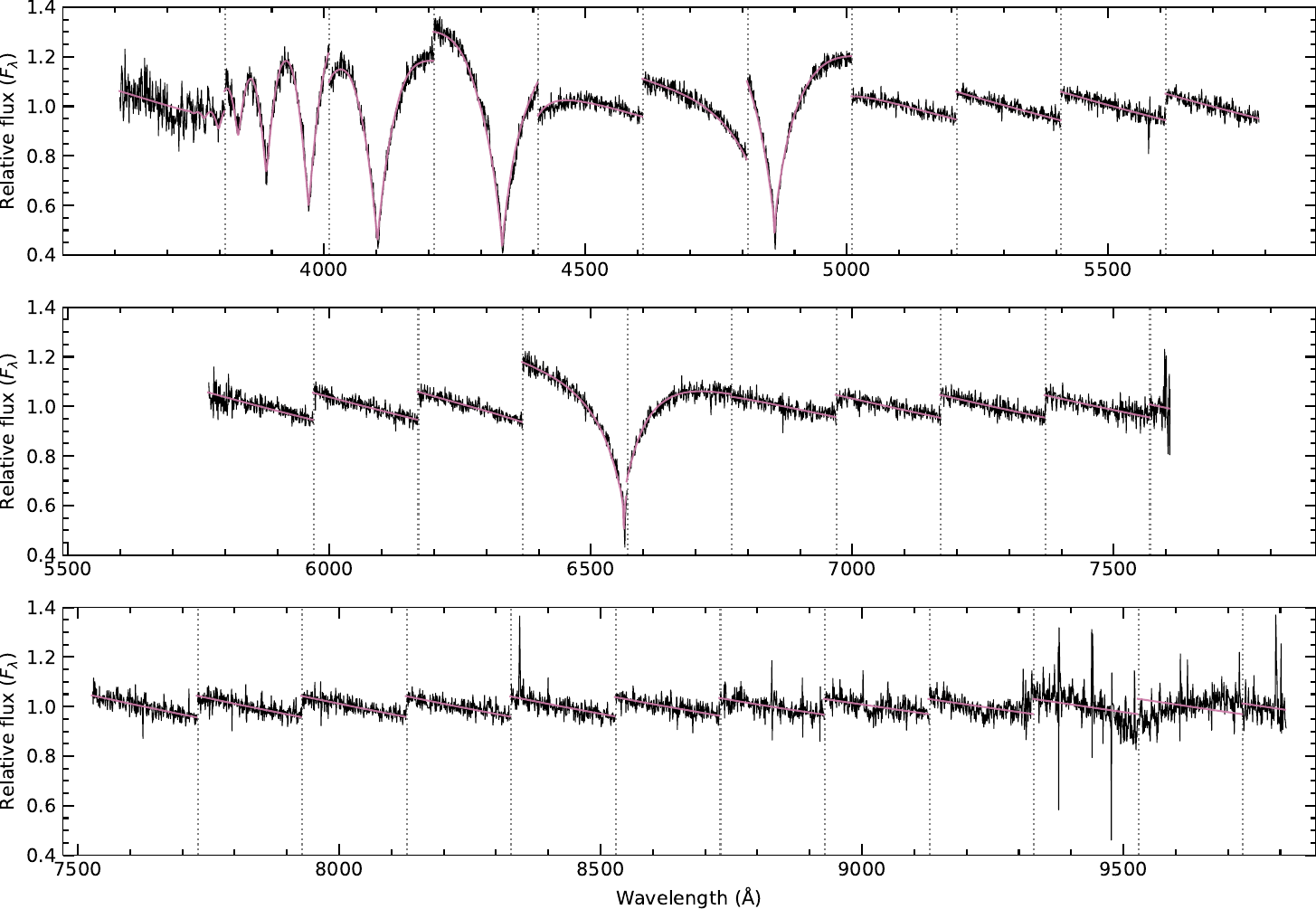}
\caption{Spectroscopic fits of a pure-H atmosphere model to WD\,J094844.84+003516.71, a DA white dwarf with $\Teff\approx13\,000$\,K, following Fig.~\ref{figureSpectralFit_WDJ170401.50+200700.65}.}
\label{figureSpectralFit_WDJ094844.84+003516.71}
\end{figure*}

\begin{figure*}
\centering
\includegraphics[width=\textwidth]{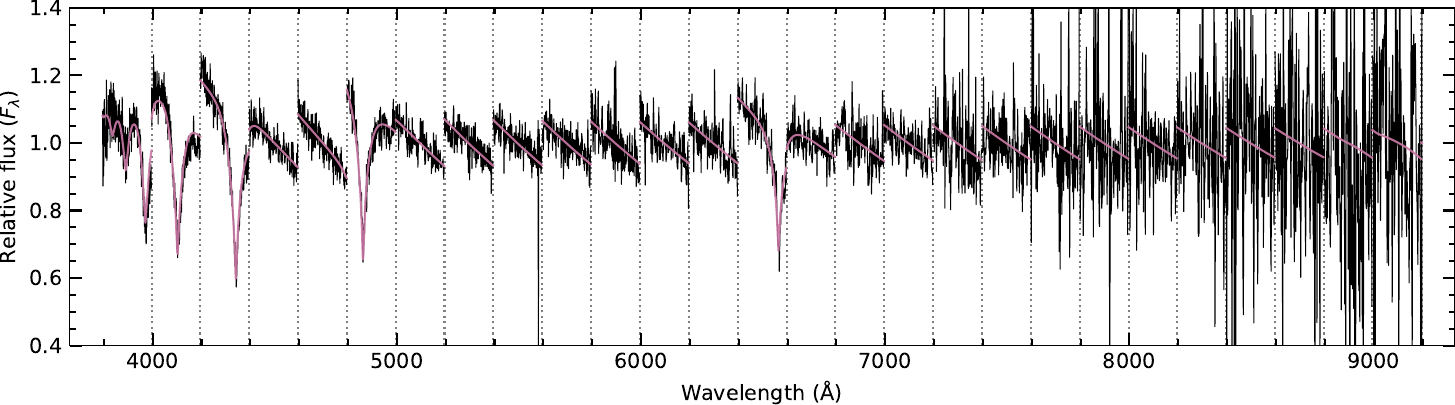}
\includegraphics[width=\textwidth]{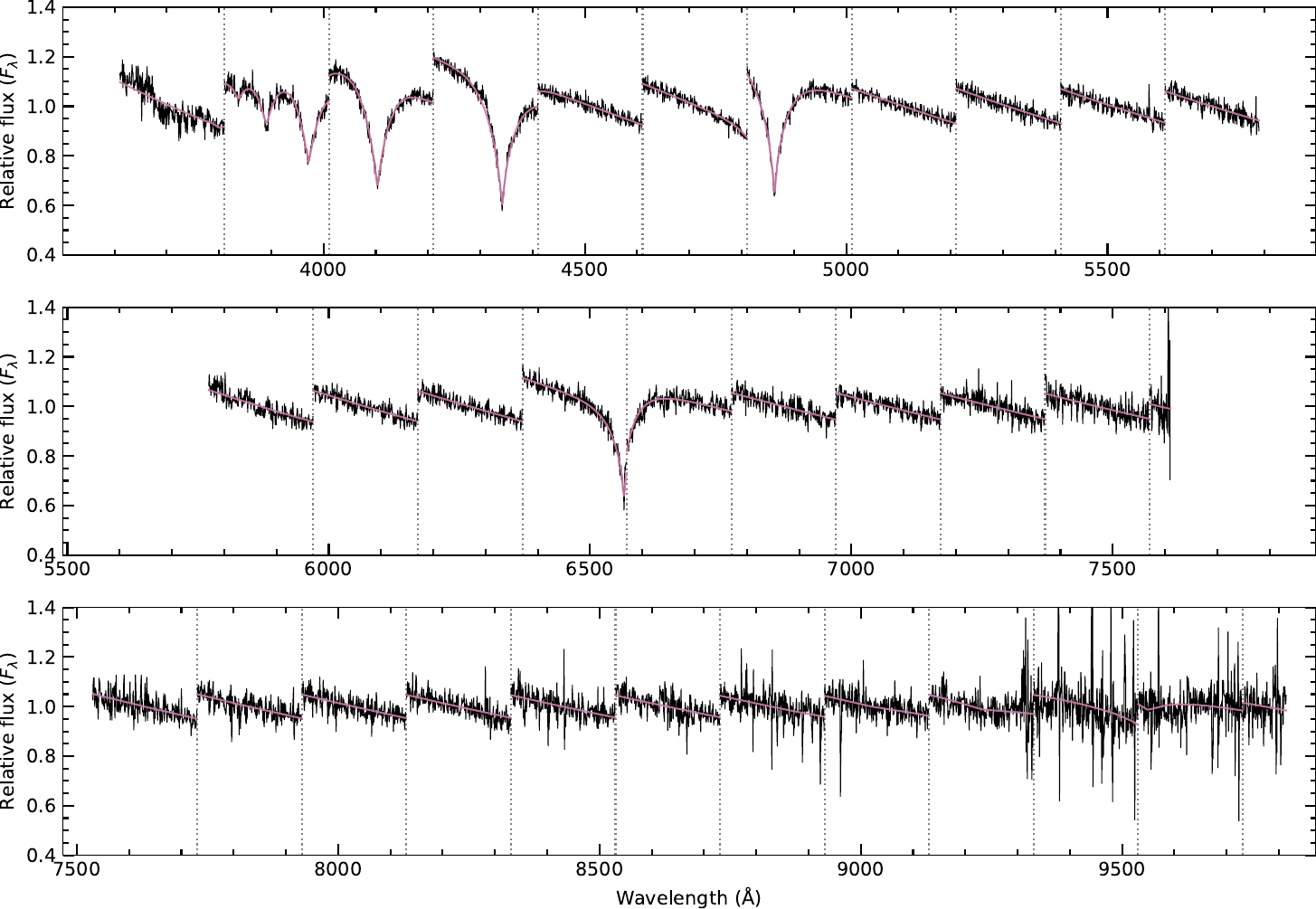}
\caption{Spectroscopic fits of a pure-H atmosphere model to WD\,J114608.73$-$002439.15, a DA white dwarf with $\Teff\approx30\,000$\,K, following Fig.~\ref{figureSpectralFit_WDJ170401.50+200700.65}.}
\label{figureSpectralFit_WDJ114608.73-002439.15}
\end{figure*}

\begin{figure*}
\centering
\includegraphics[width=\textwidth]{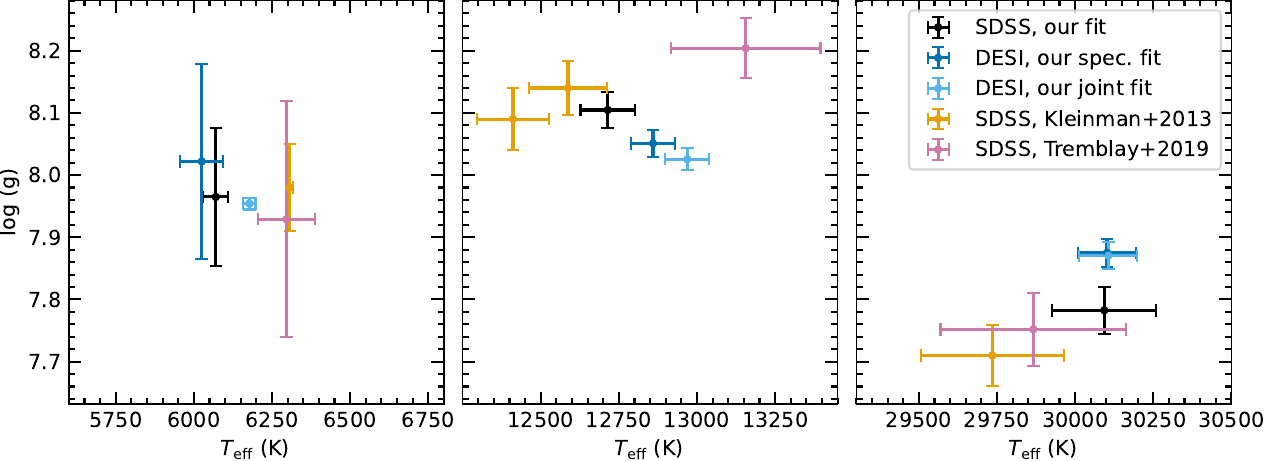}
\caption{Results from our H-atmosphere model fits to SDSS and DESI data for three DA stars, compared against values from the literature \citep{Kleinman2013,Tremblay2019DADB}. Our spectroscopic fits used coadded SDSS spectra, whereas \citet{Kleinman2013} fitted individual exposures. Uncertainties on our fits are purely statistical.}
\label{figureSDSSfits}
\end{figure*}

To demonstrate the efficacy of our spectral normalisation method we present fits to some DA stars that also have SDSS spectra. A deeper comparison was performed for the DESI EDR white dwarf catalogue, and our methods are similar here, especially for the spectroscopic fits, so we do not repeat that exercise in full. We select three stars whose Balmer lines have different profiles, namely a cool object with narrow lines (WD\,J170401.50+200700.65), a warm one with broad lines (WD\,J094844.84+003516.71), and a warmer one with intermediate line widths (WD,J114608.73$-$002439.15). To facilitate comparison with previous work we examine purely spectroscopic fits to the DESI and SDSS data, shown in Figs.~\ref{figureSpectralFit_WDJ170401.50+200700.65}--\ref{figureSpectralFit_WDJ114608.73-002439.15}. While not all formally consistent within their statistical errors, the results are broadly in line with both those from our joint fits presented in the main catalogue and with previous studies using the SDSS data (Fig.~\ref{figureSDSSfits}).

There is a glitch at around 7040\AA\ in the R~arm of the DESI spectrum for WD\,J170401.50+200700.65 that has not been masked by the reduction pipeline. It results in a substantial offset between the data and model continua within the 200-\AA\ normalisation window, but its impact is confined to that window and thus does not exert a strong influence on the overall result.

\FloatBarrier
\twocolumn
\section{Radial velocity variation with robust loss functions}
\label{appendixRVs}

For a measurement $y$ with expected value $\hat{y}$ and uncertainty $\sigma$, the normalised residual is:

\begin{equation}
\label{equationResidual}
r=\frac{|y-\hat{y}|}{\sigma}
\end{equation}

The variability of a set of $N$ measurements with Gaussian uncertainties can be  quantified using the standard statistic:

\begin{equation}
\label{equationChiSquared}
\chi^2=\sum_{i=1}^N{r_i^2}
\end{equation}

Under the null hypothesis that the quantity being measured is constant, this statistic follows a $\chi^2$ distribution with $N-1$ degrees of freedom, and a corresponding $p$-value can be calculated. We can then define a threshold for $p$ below which we reject the null hypothesis in favour of variability.

This breaks down if the data contain spurious outliers, whose residuals dominate the $\chi^2$ statistic. Here, that might arise from a failed radial velocity fit, or an artefact in {low-\SN} data. To reduce sensitivity to outliers, we explore robust modifications to the $\chi^2$ test, using the Huber loss \citep{Huber1964} and Tukey's biweight \citep{Beaton1974}. This involves estimating a robust mean, and then calculating a $\chi^2$ value using weighted residuals.

To estimate the robust mean $\hat{y}_\rho$, we minimise the sum of a loss function:

\begin{equation}
\label{equationRobustMean}
\hat{y}_\rho=\mathop{\mathrm{arg\,min}}_\mu\sum_{i=1}^N\rho(\frac{y_i-\mu}{\sigma_i})
\end{equation}

\noindent where the loss function $\rho(r)$ is either of:

\begin{equation}
\label{equationHuberLoss}
\rho_{\mathrm{Huber}}(r)=\begin{cases}
    \frac{r^2}{2} & \text{for } |r|\leq\delta\\
    \delta(|r|-\frac{\delta}{2}) & \text{for } |r|>\delta
    \end{cases}
\end{equation}

\begin{equation}
\label{equationTukeyLoss}
\rho_{\mathrm{Tukey}}(r)=\begin{cases}
    \frac{c^2}{6}(1-[1-(\frac{r}{c})^2]^3) & \text{for } |r|\leq c\\
    \frac{c^2}{6} & \text{for } |r|>c
    \end{cases}
\end{equation}

For the Huber loss the tuning parameter $\delta$ controls the transition between linear and quadratic behaviour, and for the more conservative Tukey biweight $c$ sets the scale beyond which outliers are fully suppressed. We adopt $\delta=1.345$ and $c=4.685$, which are standard choices that achieve roughly 95~per cent efficiency for Gaussian uncertainties.

We then modify equations~\ref{equationResidual} and \ref{equationChiSquared} as follows:

\begin{equation}
\label{equationResidualRobust}
r_i=\frac{|y_i-\hat{y}_\rho|}{\sigma_i}
\end{equation}

\begin{equation}
\label{equationChiSquaredRobust}
\chi^2_{\text{robust}}=\sum_{i=1}^N{w(r_i)r_i^2}
\end{equation}

\noindent using weights:

\begin{equation}
\label{equationHuberWeight}
w_{\mathrm{Huber}}(r)=\begin{cases}
    1 & \text{for } |r|\leq\delta\\
    \frac{\delta}{|r|} & \text{for } |r|>\delta
    \end{cases}
\end{equation}

\begin{equation}
\label{equationTukeyWeight}
w_{\mathrm{Tukey}}(r)=\begin{cases}
    (1-(\frac{r}{c})^2)^2 & \text{for } |r|\leq c\\
    0 & \text{for } |r|>c
    \end{cases}
\end{equation}

The robust $\chi^2$ statistics do not follow the standard $\chi^2$ distribution, complicating interpretation of the $p$-value returned by the standard $\chi^2$ survival function. We therefore calibrate the significance thresholds empirically using simulated non-variable data, so that our variability test remains consistent across all three formulations of the $\chi^2$ statistic. We generate a constant-velocity dataset by making 10 copies of the real data, setting the velocities to a constant, adding Gaussian noise based on the measured uncertainties, and repeating our $\chi^2$ analysis described in Section~\ref{subsectionRVvariation}. The distributions of the resulting $p$-values represent those expected when a non-varying population is observed by DESI.

\begin{figure}
\centering
\includegraphics[width=\columnwidth]{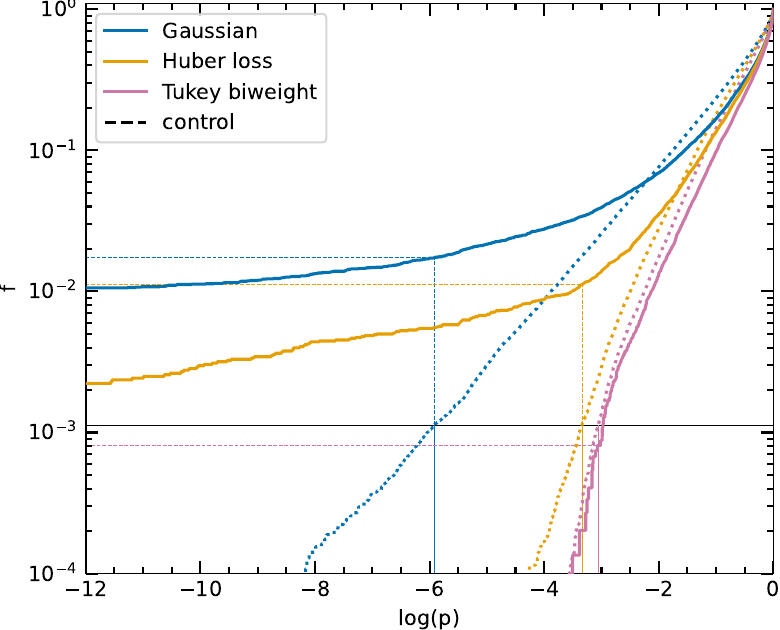}
\caption{Cumulative histogram of chi-squared test results for radial velocity variation in real data (solid lines) and simulated non-variable data (dotted lines). Horizontal and vertical lines provide a visual guide to how the 3-$\upsigma$ variability thresholds are calibrated against the simulated data, and how fraction of variable candidates in the real data can be read off, as described in detail in the text.}
\label{figureRVvariation}
\end{figure}

We then define detection thresholds by choosing an appropriate quantile in the $p$-values calculated for our simulated dataset. As expected, the analytical 3-$\upsigma$ threshold of $\ln{p}<-5.91$ for the standard $\chi^2$ analysis is recovered by taking the 0.0027 quantile of the $p$-values for the simulated data. The corresponding values for the other two functions are $\ln{p_{\text{Huber}}}<-3.42$  and $\ln{p_{\text{Tukey}}}<-3.09$. Fig.~\ref{figureRVvariation} illustrates this: the black horizontal line shows the 3-$\upsigma$ level on a cumulative histogram of $p$-values, and the thresholds are defined by where that line intercepts the histogram corresponding to each loss function. The vertical lines indicate those intercepts, and can be used to read off the 3-$\upsigma$ thresholds from the $x$-axis. Fig.~\ref{figureRVvariation} also shows the distributions measured from the real data, and extending the vertical lines to intercept the histograms for real data allow the fraction of variable sources to be estimated from the $y$-axis, as indicated by the dashed horizontal lines.

As an example of how robust statistics reduce false positives, WD\,J142446.25+322930.23 has 14 exposures, with a single velocity outlier in the R~arm that has a small uncertainty. The star scores $\ln{p}=-81$ based on Gaussian uncertainties, well beyond the 3-$\upsigma$ variability threshold, but that reduces to $\ln{p_{\text{Huber}}}=-1.9$ and $\ln{p_{\text{Tukey}}}=-0.01$, neither of which are close to their respective thresholds.

\onecolumn
\FloatBarrier
\section{Supplementary data files}
\label{appendixSupplementaryDataFiles}

Our main catalogue (Table~\ref{tableMainCatalogue}) presents our best estimates of the properties of each white dwarf candidate in DESI DR1. We also provide supplementary tables containing the intermediate products used to assemble the catalogue, and this Appendix lists them and their contents. We repeat the warning given in Section~\ref{subsectionCatalogueSupplementaryInfo}: these are \textit{not} science-ready outputs, and should be used with caution.

\begin{table}
\centering
\caption{Details of individual exposures. Provided as \texttt{exposures.csv}.}
\label{tableExposures}
\begin{tabular}{ll}
\hline
Column & Description \\
\hline
WDJname & White dwarf name as defined by GF19 \\
NIGHT & Exposure date (copied from DESI cframe file header) \\
EXPOSURE & Exposure number (copied from DESI cframe file header) \\
FIBER & Fibre number (copied from DESI cframe file header) \\
MJD & Time at start of exposure (copied from DESI cframe file header) \\
BMJD & Barycentric mid-exposure time \\
SURVEY & DESI survey name (copied from DESI cframe file header) \\
PROGRAM & DESI program name (copied from DESI cframe file header) \\
TARGETID & DESI target ID (copied from DESI cframe file header). NB some objects have multiple target IDs. \\
EXPTIME & Exposure time (copied from DESI cframe file header) \\
EFFTIME\_SPEC & Effective exposure time (copied from DESI cframe file header) \\
SEEING\_GFA & Seeing measured during exposure (copied from DESI cframe file header) \\
MWS\_WD & Boolean flag indicating a Milky Way Survey white dwarf target in the \texttt{main} survey. Specifically: $\texttt{MWS\_TARGET} \mathbin{\&} \texttt{MWS\_WD} \neq 0$ \\
SV1\_MWS\_WD & As above, for the \texttt{SV1} validation survey: $\texttt{SV1\_MWS\_TARGET} \mathbin{\&} \texttt{MWS\_WD} \neq 0$ \\
SV2\_MWS\_WD & As above, for the \texttt{SV2} validation survey: $\texttt{SV2\_MWS\_TARGET} \mathbin{\&} \texttt{MWS\_WD} \neq 0$ \\
SV3\_MWS\_WD & As above, for the \texttt{SV3} validation survey: $\texttt{SV3\_MWS\_TARGET} \mathbin{\&} \texttt{MWS\_WD} \neq 0$ \\
STD\_WD & Boolean flag: targeted as a white dwarf flux standard in the \texttt{main} survey: $\texttt{DESI\_TARGET} \mathbin{\&} \texttt{STD\_WD} \neq 0$ \\
SV1\_STD\_WD & As above, for the \texttt{SV1} validation survey: $\texttt{SV1\_DESI\_TARGET} \mathbin{\&} \texttt{GAIA\_STD\_WD} \neq 0$ \\
SV2\_STD\_WD & As above, for the \texttt{SV2} validation survey: $\texttt{SV2\_DESI\_TARGET} \mathbin{\&} \texttt{GAIA\_STD\_WD} \neq 0$ \\
SV3\_STD\_WD & As above, for the \texttt{SV3} validation survey: $\texttt{SV3\_DESI\_TARGET} \mathbin{\&} \texttt{GAIA\_STD\_WD} \neq 0$ \\
CMX\_WD & As above, for the \texttt{CMX} commissioning survey: $\texttt{CMX\_TARGET} \mathbin{\&} \texttt{SV0\_WD} \neq 0$ \\
SN\_B & Median $\SN$ calculated from flux and uncertainty in blue arm \\
SN\_R & Median $\SN$ calculated from flux and uncertainty in red arm \\
SN\_Z & Median $\SN$ calculated from flux and uncertainty in near-infrared arm \\
SNmeasured\_B & $\SN$ measured statistically in blue arm \\
SNmeasured\_R & $\SN$ measured statistically in red arm \\
SNmeasured\_Z & $\SN$ measured statistically in infrared arm \\
fractionMasked\_B & Fraction of pixels masked in blue arm \\
fractionMasked\_R & Fraction of pixels masked in red arm \\
fractionMasked\_Z & Fraction of pixels masked in near-infrared arm \\
\hline
\end{tabular}
\end{table}

\begin{table}
\centering
\caption{External data used for fitting white dwarf models to each object. Provided as \texttt{externalData.csv}.}
\label{tableExternalData}
\begin{tabular}{ll}
\hline
Column & Description \\
\hline
WDJname & White dwarf name as defined by GF19 \\
Gaia.DR2.G & electron s$^{-1}$ \\
Gaia.DR2.G\_error & electron s$^{-1}$ \\
Gaia.DR2.BP & electron s$^{-1}$ \\
Gaia.DR2.BP\_error & electron s$^{-1}$ \\
Gaia.DR2.RP & electron s$^{-1}$ \\
Gaia.DR2.RP\_error & electron s$^{-1}$ \\
Gaia.DR3.G & electron s$^{-1}$ \\
Gaia.DR3.G\_error & electron s$^{-1}$ \\
Gaia.DR3.BP & electron s$^{-1}$ \\
Gaia.DR3.BP\_error & electron s$^{-1}$ \\
Gaia.DR3.RP & electron s$^{-1}$ \\
Gaia.DR3.RP\_error & electron s$^{-1}$ \\
Pan-STARRS.g & mag \\
Pan-STARRS.g\_error & mag \\
Pan-STARRS.r & mag \\
Pan-STARRS.r\_error & mag \\
Pan-STARRS.i & mag \\
Pan-STARRS.i\_error & mag \\
Pan-STARRS.z & mag \\
Pan-STARRS.z\_error & mag \\
Pan-STARRS.y & mag \\
Pan-STARRS.y\_error & mag \\
SDSS.u & mag \\
SDSS.u\_error & mag \\
SDSS.g & mag \\
SDSS.g\_error & mag \\
SDSS.r & mag \\
SDSS.r\_error & mag \\
SDSS.i & mag \\
SDSS.i\_error & mag \\
SDSS.z & mag \\
SDSS.z\_error & mag \\
fluxContamination\_PanSTARRS & Expected fractional contamination of Pan-STARRS photometry by field stars based on \textit{Gaia} data \\
fluxContamination\_SDSS & As above, for SDSS \\
fluxContamination\_Gaia\_G & As above, for \textit{Gaia G} \\
fluxContamination\_Gaia\_BPRP & As above, for \textit{Gaia BP} and \textit{RP} \\
parallax & mas \\
parallax\_error & mas \\
distance\_prior\_scale & Scale factor for distance prior (pc) \\
\hline
\end{tabular}
\end{table}

\begin{table}
\centering
\caption{Classification(s) for each star, with associated confidence values. Provided as \texttt{classifications.csv}.}
\label{tableClassifications}
\begin{tabular}{ll}
\hline
Column & Description \\
\hline
WDJname & White dwarf name as defined by GF19 \\
votesFor & Number of classifications for this type \\
unsure & Number of uncertain classifications or validations \\
votesAgainst & Number of no votes during validation \\
specType\_confidence & Confidence in spectral type \\
specType & Spectral type \\
\hline
\end{tabular}
\end{table}

\begin{table}
\centering
\caption{Log of all classification and validation responses. Provided as \texttt{classificationLog.csv}.}
\label{tableClassificationLogs}
\begin{tabular}{ll}
\hline
Column & Description \\
\hline
WDJname & White dwarf name as defined by GF19 \\
worker & Human worker ID \\
specType & Spectral type \\
weight & Weight assigned to this response \\
uncertain & Flag for whether a classification is uncertain \\
response & Validation of spectral type: yes, no, or unsure \\
\hline
\end{tabular}
\end{table}

\begin{table}
\centering
\caption{Full fitting results: all models, all fitting strategies. Provided as \texttt{fullFitResults.csv}.}
\label{tableFullFitResults}
\begin{tabular}{ll}
\hline
Column & Description \\
\hline
WDJname & White dwarf name as defined by GF19 \\
model & Atmosphere type used in fit \\
fitType & Type of fit: photometric, spectroscopic, or joint \\
TEFF & Effective temperature (K) \\
TEFF\_lo & Effective temperature, 16th percentile \\
TEFF\_hi & Effective temperature, 84th percentile \\
LOGG & Surface gravity (log [\cms])\\
LOGG\_lo & Surface gravity, 16th percentile \\
LOGG\_hi & Surface gravity, 84th percentile \\
H & Numerical H abundance as log(H/He) \\
H\_lo & Numerical H abundance, 16th percentile \\
H\_hi & Numerical H abundance, 84th percentile \\
distance & Distance (pc) \\
distance\_lo & Distance, 16th percentile \\
distance\_hi & Distance, 84th percentile \\
EBminusV & Reddening as $E_{B-V}$ \\
EBminusV\_lo & Reddening, 16th percentile \\
EBminusV\_hi & Reddening, 84th percentile \\
logEvidence & Bayesian evidence $\ln{\mathcal{Z}}$ \\
logEvidence\_err & Uncertainty on Bayesian evidence \\
\hline
\end{tabular}
\end{table}

\begin{table}
\centering
\caption{Full radial velocity results. Provided as \texttt{RVs.csv}. Where night--exposure--fiber values are not given, velocities are a joint measurement on all exposures for that object.}
\label{tableRVs}
\begin{tabular}{ll}
\hline
Column & Description \\
\hline
WDJname & White dwarf name as defined by GF19 \\
NIGHT & Exposure date (copied from DESI cframe file header) \\
EXPOSURE & Exposure number (copied from DESI cframe file header) \\
FIBER & Fibre number (copied from DESI cframe file header) \\
fitType & Type of fit: spectroscopic, or joint \\
model & Atmosphere type used in fit \\
RV\_B & Radial velocity in blue arm (\kms) \\
RV\_B\_error & Uncertainty on radial velocity in blue arm \\
RV\_R & Radial velocity in red arm (\kms) \\
RV\_R\_error & Uncertainty on radial velocity in red arm \\
\hline
\end{tabular}
\end{table}

\begin{table}
\centering
\caption{Full results from the kinematics model presented in Section~\ref{subsectionContentsKinematics}. Provided as \texttt{kinematics.csv}. Uncertainties for $UVW$ velocities are the standard deviations of the sample distributions.}
\label{tableKinematics}
\begin{tabular}{ll}
\hline
Column & Description \\
\hline
WDJname & White dwarf name as defined by GF19 \\
U & Galactocentric $U$ space velocity (\kms) \\
U\_error & Uncertainty on $U$ (\kms) \\
V & Galactocentric $V$ space velocity (\kms) \\
V\_error & Uncertainty on $V$ (\kms) \\
W & Galactocentric $W$ space velocity (\kms) \\
W\_error & Uncertainty on $W$ (\kms) \\
pHalo & Probability of halo membership (indicative only; see Section~\ref{subsectionContentsKinematics} for method used) \\
\hline
\end{tabular}
\end{table}

\FloatBarrier
\section{WD+MS photometric periods}

\edit{We list here details for WD+MS objects in whose ZTF light curves we detected a photometric period (Section~\ref{subsubsectionWDMS}).}

\begin{table}
\centering
\caption{\edit{Photometric periods identified from ZTF for WD+MS objects}}
\label{tableZTFperiods}
{\edit{
\begin{tabular}{llllp{6.25cm}}
\hline
WDJname & period (d) & light curve type & spectral types & previously known \\
\hline
WD\,J001751.01+275133.89 & 0.138055083(83) & reflection effect & DA+dM & WD+MS \& eclipses \& period \citep{MeirShani2025} \\
WD\,J003221.86+073934.50 & 0.15399437(23) & reflection effect & DA+dM & WD+MS \& period \citep{Li2024}\\
WD\,J010623.00$-$001456.30 & 0.085015297(18) & eclipses & DA+dM & WD+MS \& eclipses \& period \citep{Parsons2015}\\
WD\,J015841.01$-$062841.90 & 0.16696407(17) & eclipses & DA+dM & \\
WD\,J020528.60+324058.02 & 0.19965547(37) & reflection effect & DA+dM &  \\
WD\,J020724.06+071546.88 & 0.211581017(1) & reflection effect & DA+dM & infrared excess candidate \citep{Girven2011,Wang2026}     \\
WD\,J021616.13+013321.70 & 0.13855279(5) & reflection effect & DA+dM &  \\
WD\,J022317.92$-$024252.37 & 0.3094825(32) & reflection effect & DA+dM &  \\
WD\,J025301.62$-$013006.55 & 0.4264325(17) & reflection effect & DA+dM & WD+MS \citep{Ren2014}\\
WD\,J041016.80$-$083419.31 & 0.081109316(11) & eclipses & DA+dM & WD+MS \& eclipses \& period \citep{Brown2023}\\
WD\,J043741.99$-$085739.16 & 0.5215396(11) & reflection effect & DA+dM &  \\
WD\,J045414.83$-$085902.85 & 0.21978821(58) & reflection effect & DA+dM &  \\
WD\,J065940.37+624434.79 & 0.26721950(42) & reflection effect & DA+dM &  \\
WD\,J071452.27+662928.49 & 0.20784688(79) & reflection effect & DA+dM &  \\
WD\,J071709.79+740040.53 & 0.103064400(83) & reflection effect & DA+dM & WD+MS \& period \citep{Marsh1996} \\ 
WD\,J081126.68+053912.02 & 0.3788769(15) & reflection effect & DA+dM & WD+MS \citep{Silvestri2006}\\
WD\,J081449.78+405920.74 & 0.082273188(22) & eclipses & DA+dM & WD+MS \citep{Silvestri2006} \\
WD\,J083531.69+315503.31 & 0.184333083(18) & reflection effect & DA+dM & WD+MS \citep{Ren2018}\\
WD\,J083618.61+432651.50 & 0.19689455(2) & reflection effect & DA+dM & WD+MS \citep{Silvestri2006}, period \citep{Parsons2013}  \\
WD\,J083845.86+191416.78 & 0.130112292(25) & eclipses & DA+dM & WD+MS \citep{Heller2009}, eclipses \& period \citep{Drake2010} \\
WD\,J084957.41+464545.94 & 0.066798658(1) & eclipses & DA+dM & WD+MS \citep{Kepler2015}, eclipses \& period \citep{Murawski2022} \\
WD\,J085713.28+331843.10 & 0.106027354(4) & eclipses & DA+dM & WD+MS \& eclipses \& period \citep{vanRoestel2017}\\
WD\,J085746.18+034255.35 & 0.0650965550(83) & eclipses & DA+dM & WD+MS \& eclipses \& period \citep{Parsons2012} \\
WD\,J090451.94+453105.43 & 0.12037502(12) & ellipsoidal modulation? & DC+dM & WD+MS \citep{RebassaMansergas2016} \\
WD\,J094450.56+835531.29 & 0.9970029(83) & reflection & DAO+dM &  \\
WD\,J102743.35+503057.58 & 0.5646587(23) & partially eclipses  & DQ+dM &  \\
WD\,J102834.88$-$000029.39 & 0.5972500(87) & reflection effect & DA+dM & WD+MS \& period \citep{Saffer1993} \\ 
WD\,J110805.37+652211.45 & 0.32083429(24) & eclipses & DA+dM & WD+MS \citep{Silvestri2006}, eclipses \& period \citep{Kosakowski2022} \\
WD\,J114132.99+042028.86 & 0.06228890(14) & reflection effect & DA+dM & WD+MS \citep{Silvestri2006}, eclipses \& period \citep{Armstrong2016}\\
WD\,J122339.61$-$005631.21 & 0.090077983(3) & eclipses & DA+dM & WD+MS \citep{Raymond2003}, eclipses \& period \citep{Parsons2013}\\
WD\,J123159.51+670918.91 & 0.11299920(19) & eclipses & DA+dM & WD+MS \citep{RebassaMansergas2010}, eclipses \& period \citep{Kao2016}\\
WD\,J123309.09+592555.57 & 0.11505375(46) & reflection effect & DA+dM &  \\
WD\,J125715.01+014324.63 & 0.16249002(11) & eclipses & DA+dM & WD+MS \citep{Silvestri2006}, eclipses \citep{Murawski2022} \\
WD\,J125736.92+475521.73 & 0.17581071(58) & reflection effect? & DA+dM & WD+MS \citep{Silvestri2006}\\
WD\,J132341.90+541636.57 & 0.24822059(2) & eclipses & DA+dM & WD+MS \citep{Silvestri2006} \\
WD\,J133100.78+084404.62 & 0.08408152(17) & reflection effect & DA+dM & WD+MS \citep{Heller2009} \\
WD\,J134848.36$-$064720.90 & 0.21065283(2) & reflection effect & DA+dM & WD+MS \citep{Koester2009UVES} \\
WD\,J141057.74$-$020236.73 & 0.36349654(54) & eclipses & DA+dM & WD+MS \citep{Silvestri2006}, \& eclipses \& period \citep{Drake2010}\\
WD\,J141134.70+102839.82 & 0.16751007(12) & eclipses & DA+dM & WD+MS \citep{RebassaMansergas2010}, eclipses \& period \citep{Parsons2013}\\
WD\,J142632.83+012739.94 & 0.11415075(26) & reflection effect & DA+dM &  \\
WD\,J143547.87+373338.70 & 0.125631037(58) & partially eclipses & DA+dM & WD+MS \citep{Silvestri2006}, eclipses \& period \citep{Steinfadt2008}\\
\hline
\end{tabular}
}}
\end{table}

\begin{table}
\centering
\caption*{\edit{cont.}}
{\edit{
\begin{tabular}{llllp{6.25cm}}
\hline
WDJname & period (d) & light curve type & spectral types & previously known \\
\hline
WD\,J143947.62$-$010606.78 & 1.522610(26) & reflection effect & DA+dM & WD+MS \citep{Raymond2003}, period \citep{NebotGomezMoran2011} \\
WD\,J145958.42+384737.05 & 0.2646456(16) & reflection effect & DA+dM &  \\
WD\,J150210.76+475454.63 & 0.19122858(14) & eclipses & DA+dM & WD+MS \citep{RebassaMansergas2016} \\
WD\,J155904.63+035623.49 & 0.094347358(46) & reflection effect & DA+dM & WD+MS \citep{RebassaMansergas2010}, period \citep{NebotGomezMoran2011} \\
WD\,J160144.45+334433.93 & 0.11352954(24) & reflection effect & DA+dM & WD+MS \citep{Kepler2021}  \\
WD\,J161941.93+253314.91 & 0.21590008(46) & reflection effect & DA+dM & WD+MS \citep{RebassaMansergas2010}\\
WD\,J163320.67+014949.18 & 0.11255522(25) & reflection effect & DA+dM &  \\
WD\,J163858.80+010223.93 & 0.21392895(26) & eclipses & DA+dM &  \\
WD\,J164400.83+165246.82 & 0.11182891(35) & reflection effect & DA+dM &  \\
WD\,J165640.81+211812.89 & 2.78763(24) & reflection effect & DA+dM & WD+MS \citep{Li2014} \\
WD\,J172406.13+562003.08 & 0.33301573(2) & reflection effect & DA+dM & WD+MS \citep{Silvestri2006}, period \citep{Li2024} \\
WD\,J173002.49+333401.81 & 0.156947150(39) & reflection effect & DA+dM & WD+MS \& period \citep{Parsons2013}\\
WD\,J173021.14+235540.33 & 0.20173978(18) & eclipses & DA+dM &  \\
WD\,J174339.51+331123.27 & 0.20229387(33) & reflection effect & DA+dM &  \\
WD\,J174424.71+390216.23 & 0.112476519(14) & eclipses & DA+dM & WD+MS \& eclipses \& period \citep{Kosakowski2022} \\
WD\,J175016.77+223010.94 & 0.4583615(19) & reflection effect & DA+dM &  \\
WD\,J181143.50+445501.17 & 0.3571049(15) & reflection effect & DA+dM &  \\
WD\,J181200.93+483509.47 & 0.7858125(5) & reflection effect & DA+dM &  \\
WD\,J181824.35+465721.46 & 0.155592750(46) & reflection effect & DA+dM &  \\
WD\,J181928.96+385757.75 & 0.20849173(35) & reflection effect & DA+dM &  \\
WD\,J184434.41+485736.29 & 0.0726693221(62) & eclipses & DA+dM & WD+MS \& eclipses \& period \citep{Keller2022} \\
WD\,J213212.43+141952.80 & 0.21223777(11) & eclipses & DA+dM &  \\
WD\,J214454.89$-$040918.07 & 0.30317725(67) & reflection effect & DA+dM &  \\
WD\,J214656.86+143125.17 & 0.2846190(12) & reflection effect & DA+dM &  \\
WD\,J215037.02+234000.07 & 0.26618432(15) & eclipses & DA+dM & WD+MS \& eclipses \& period \citep{Kosakowski2022} \\
WD\,J220303.58+282500.56 & 0.2004349(1) & reflection effect & DA+dM & WD+MS (GF21) \\
WD\,J220823.66$-$011534.09 & 0.156505679(34) & eclipses & DA+dM & WD+MS \citep{Silvestri2006}, eclipses \& period \citep{Drake2010} \\
WD\,J222957.50+264304.42 & 0.91075825(1) & reflection effect & DA+dM &  \\
WD\,J223421.51$-$002008.08 & 0.187709437(67) & eclipses & DA+dM & WD+MS \citep{Silvestri2006} \\
WD\,J224524.37+285830.82 & 0.12594031(15) & reflection effect & DA+dM &  \\
WD\,J225814.72+090119.95 & 0.21095075(5) & reflection effect & DA+dM &  \\
WD\,J233910.91+255205.81 & 0.12031935(28) & reflection effect & DA+dM & WD+MS \citep{RebassaMansergas2025} \\
\hline
\end{tabular}
}}
\end{table}

\FloatBarrier
\clearpage
\section{Supplementary figures}

This Appendix contains figures that are not critical to the main text, to avoid breaking the flow of the paper.

\FloatBarrier
\begin{figure}
\centering
\includegraphics[width=\textwidth]{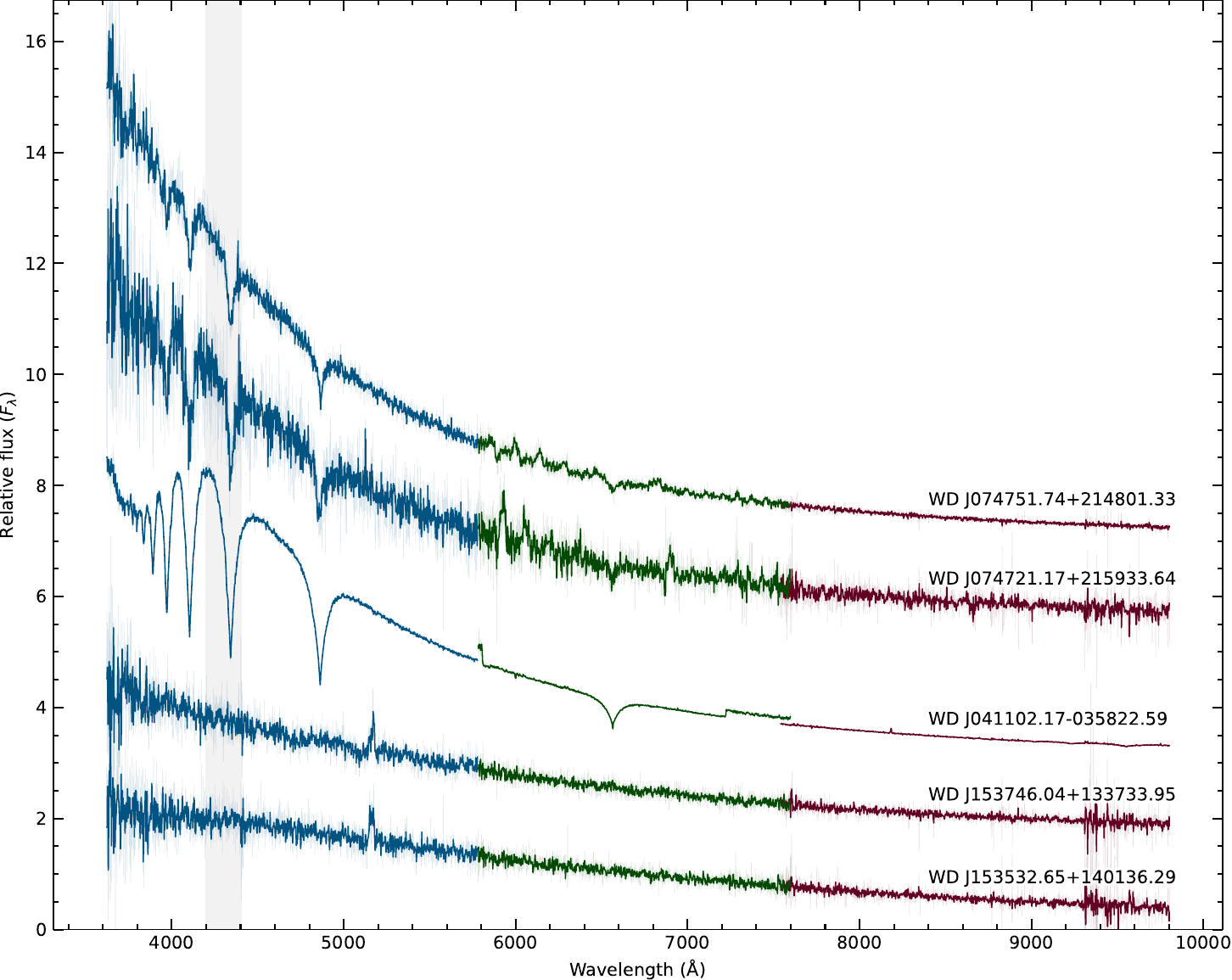}
\caption{Examples of spectral artefacts in DESI DR1. The top two spectra are from the same exposure, where several examples of the spurious emission features throughout the R~arm are seen. The middle spectrum has a discontinuity near 7200\,\AA, and the ends of the R~arm are offset from the B and Z~arms. The bottom two spectra were also taken together, and share a spurious feature near 5250\,\AA.}
\label{figureArtefacts}
\end{figure}

\clearpage
\twocolumn

\begin{figure*}
\centering
\includegraphics[height=0.925\textheight]{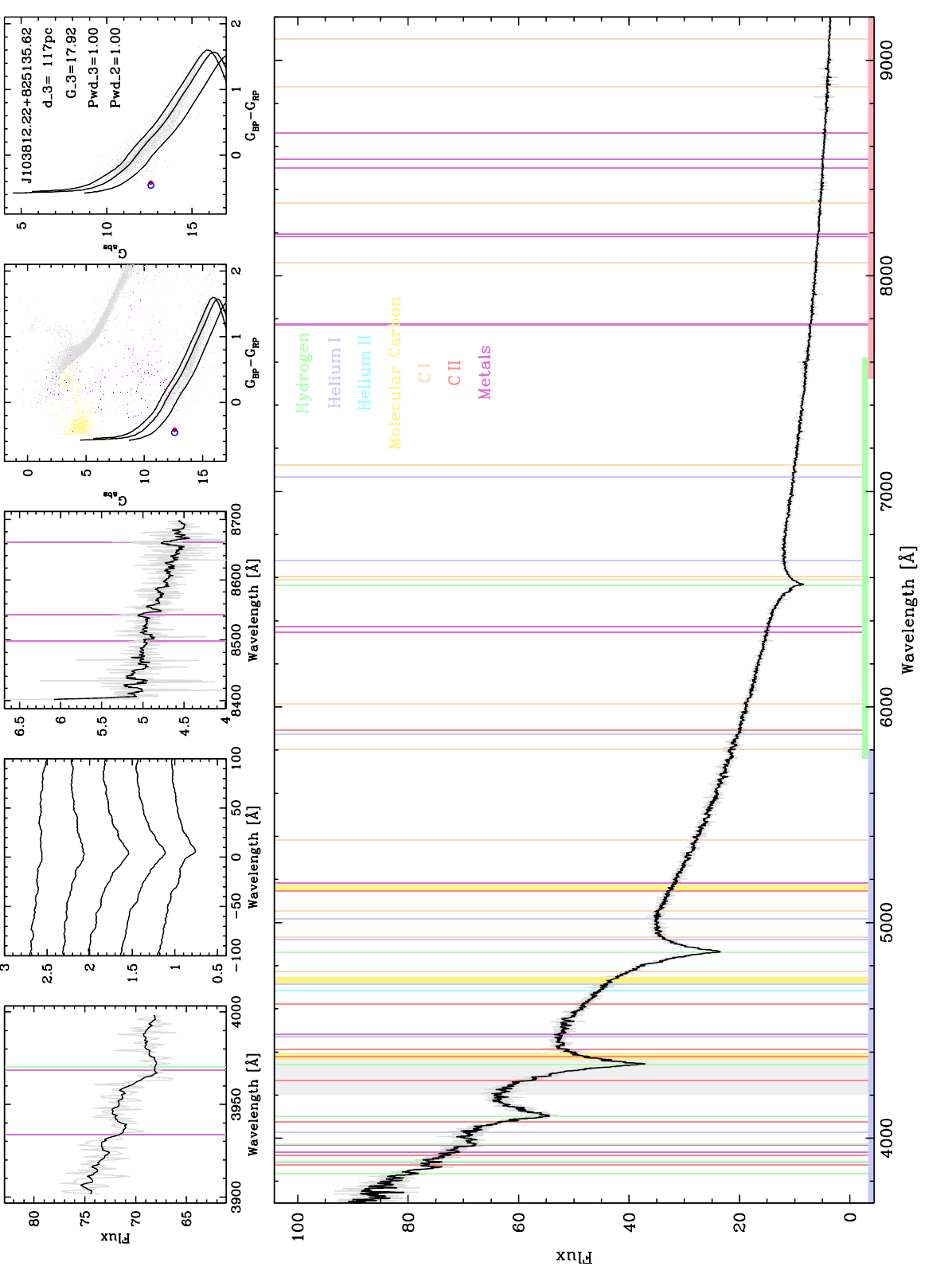}
\caption{Classification image for WD\,J103812.22+825135.62 (spectral fit shown in Fig.~\ref{figureSpectralFitsH}). The spectrum in the main panel is slightly smoothed to aid recognition, and the vertical lines indicate features often encountered in white dwarf spectra. From left to right, the top panels show: the \ion{Ca}{ii}~H and K lines; stacked Balmer lines H$\upalpha$--$\upepsilon$; the infrared \ion{Ca}{ii} triplet; an HR~diagram showing the locations of the target (red and black circles, based on \textit{Gaia}~DR2 and DR3 data, respectively), subdwarfs (yellow), cataclysmic variables (blue) and white-dwarf--main-sequence binaries (magenta); and a clean HR~diagram including \textit{Gaia} DR3 data and \pwdGaia scores.}
\label{figureClassificationImage}
\end{figure*}

\begin{figure*}
\centering
\includegraphics[width=\textwidth]{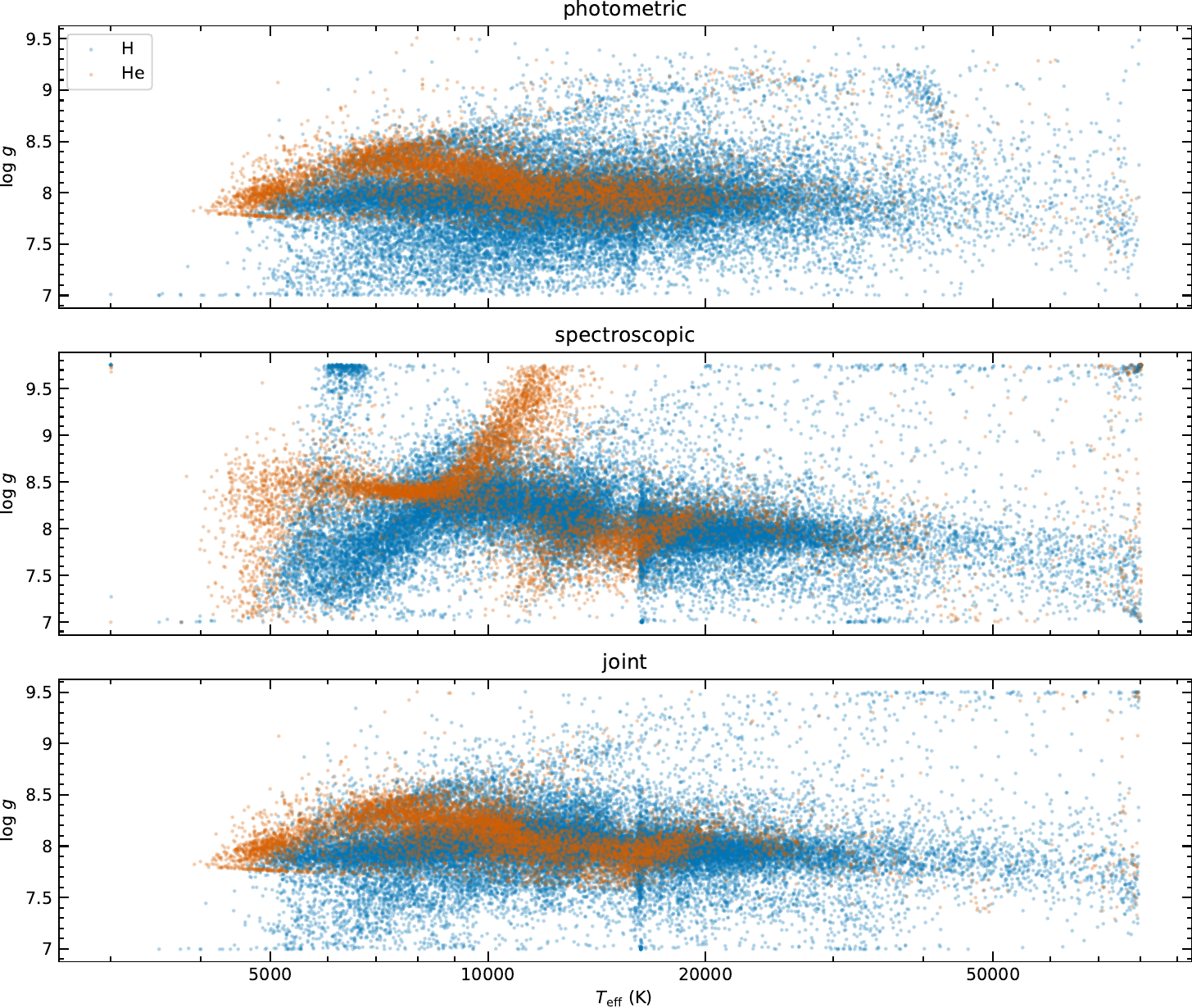}
\caption{\Teff and $\log{g}$ distributions of white dwarfs fitted using only photometry, only spectroscopy, and a joint fit of both. \edit{Only the best-fitting atmosphere type is shown for each star.}}
\label{figureTEFFlogg}
\end{figure*}

\begin{figure*}
\centering
\includegraphics[width=0.37\textwidth]{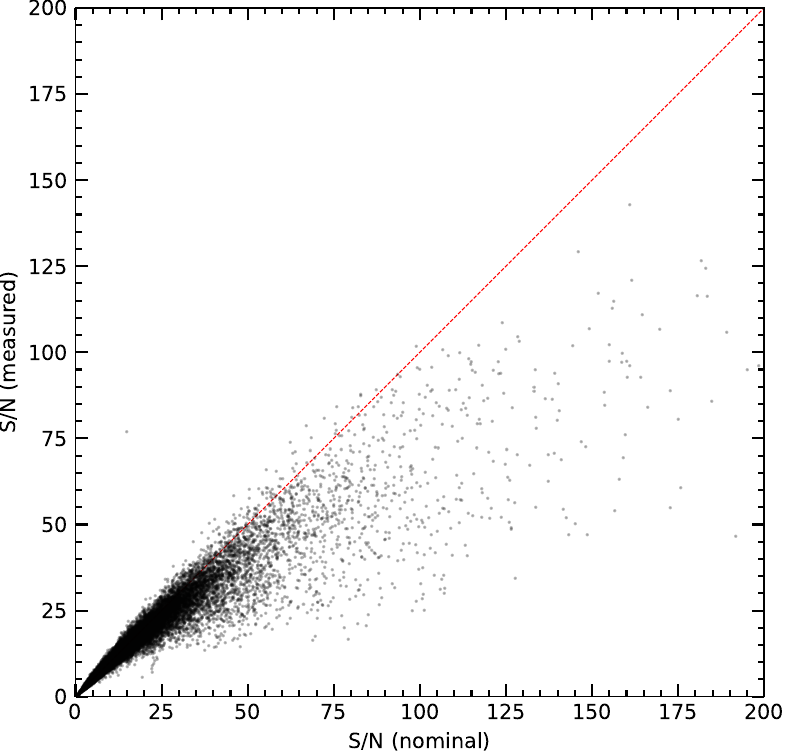}
\includegraphics[width=0.37\textwidth]{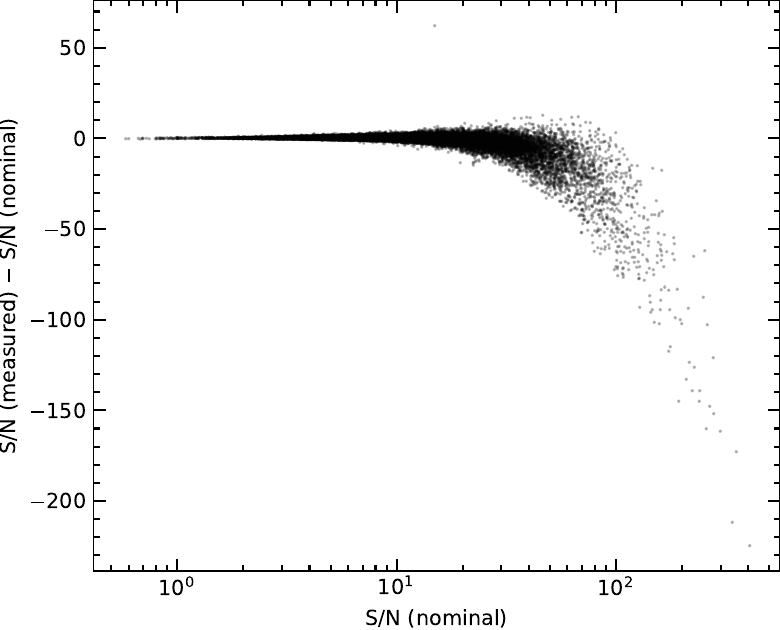}
\caption{Comparison of our \SN measurements against the nominal \SN from the reduction pipeline (\textit{left panel}), and their residuals (\textit{right panel}).}
\label{figureSN}
\end{figure*}

\begin{figure}
\centering
\includegraphics[width=\columnwidth]{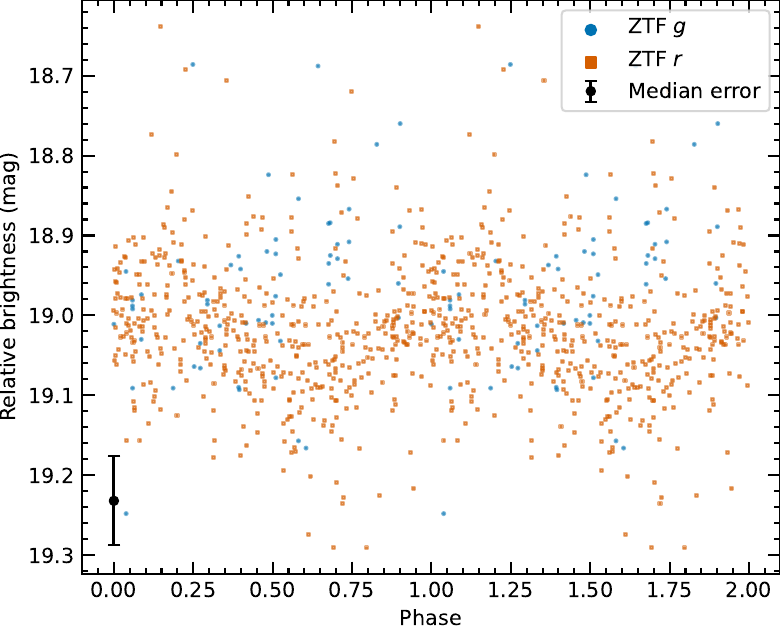} \vspace{6pt}\\
\includegraphics[width=\columnwidth]{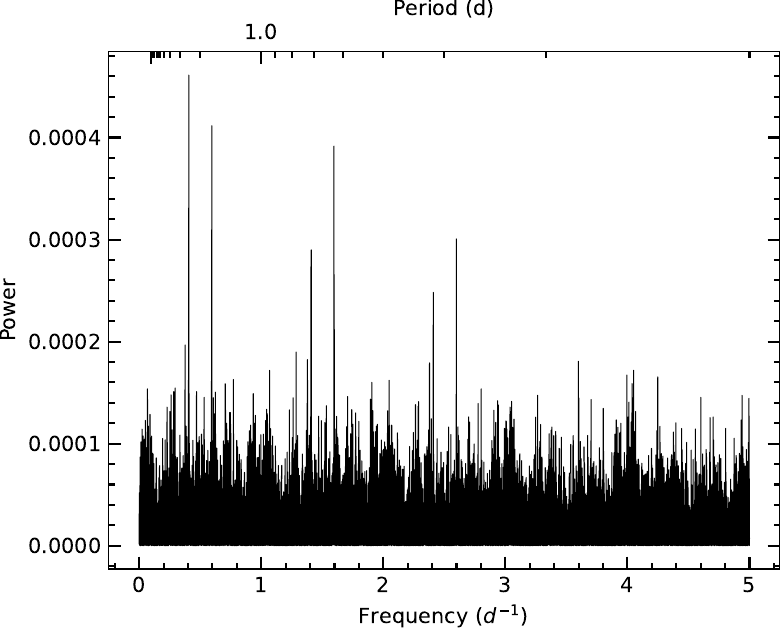}
\caption{\textit{Top:} Light curves for the for the candidate UHE star WD\,J081251.35+042744.09 in the ZTF \textit{g} and \textit{r} bands, folded to a period of $2.45759\pm0.00021$\,d. \textit{Bottom}: Discrete Fourier transform of the ZTF light curve.}
\label{figureUHEcandidate}
\end{figure}

\begin{figure}
\centering
\includegraphics[width=\columnwidth]{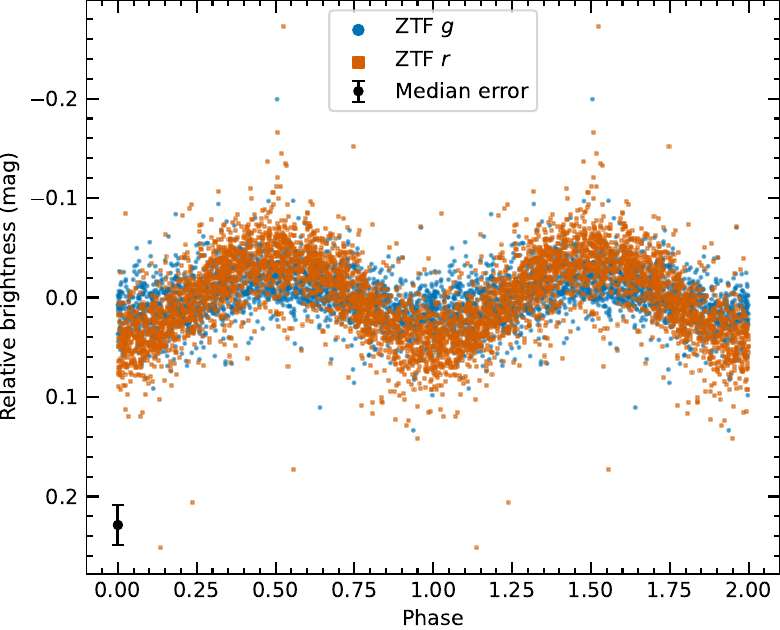} \vspace{6pt}\\
\includegraphics[width=\columnwidth]{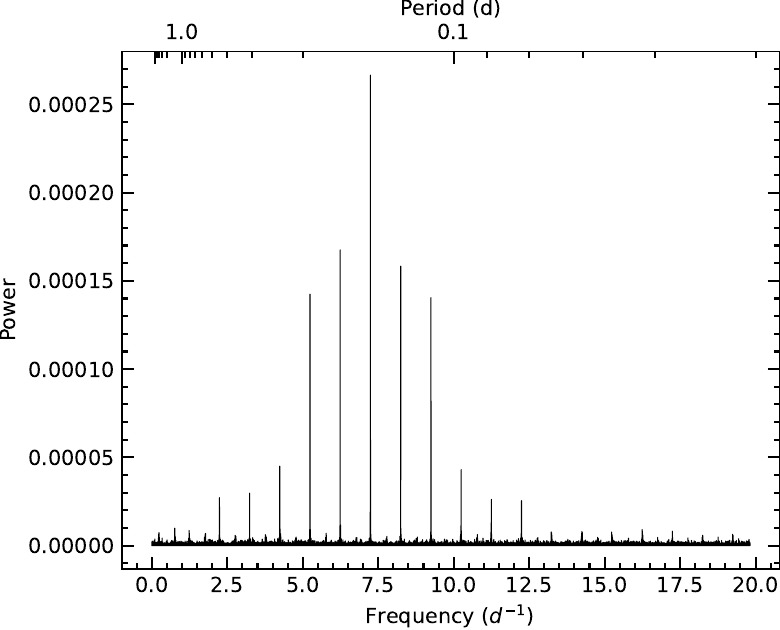}
\caption{\edit{\textit{Top:} Light curves for the for the likely post-common-envelope binary WD\,J001751.01+275133.89 in the ZTF \textit{g} and \textit{r} bands, folded to a period of $0.138055083\pm0.000000083$\,d. \textit{Bottom}: Discrete Fourier transform of the ZTF light curve.}}
\label{figurePCEB0017}
\end{figure}

\begin{figure}
\centering
\includegraphics[width=\columnwidth]{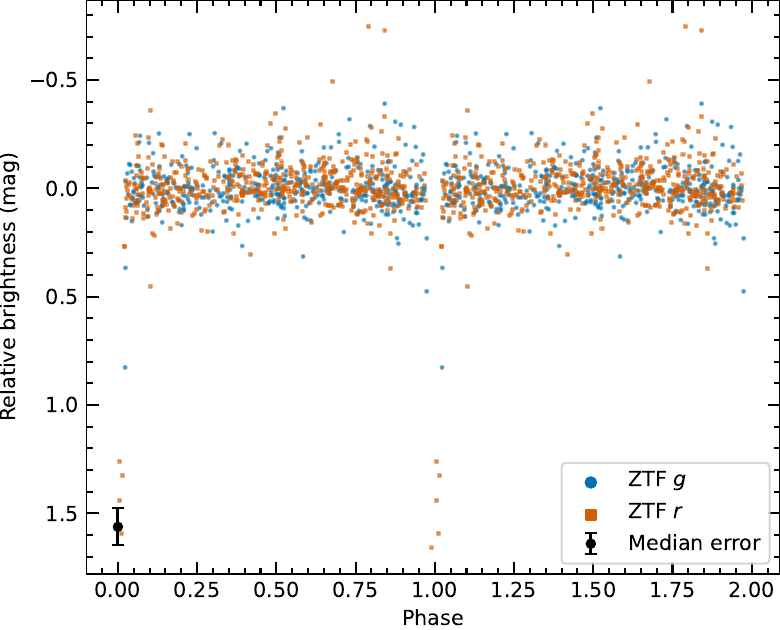} \vspace{6pt}\\
\includegraphics[width=\columnwidth]{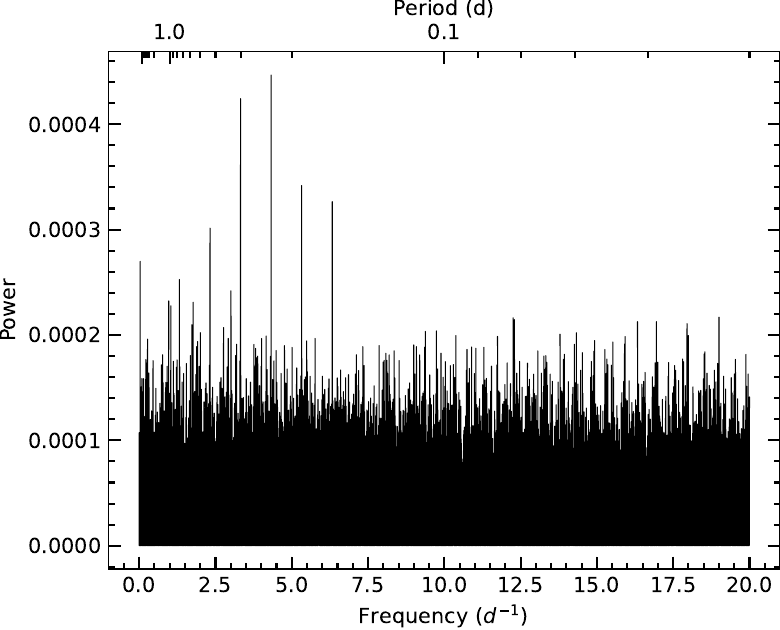}
\caption{\edit{\textit{Top:} Light curves for the for the likely post-common-envelope binary WD\,J223421.51$-$002008.08 in the ZTF \textit{g} and \textit{r} bands, folded to a period of $0.187709437\pm0.000000067$\,d. \textit{Bottom}: Discrete Fourier transform of the ZTF light curve.}}
\label{figurePCEB2234}
\end{figure}

\begin{figure}
\centering
\includegraphics[width=\columnwidth]{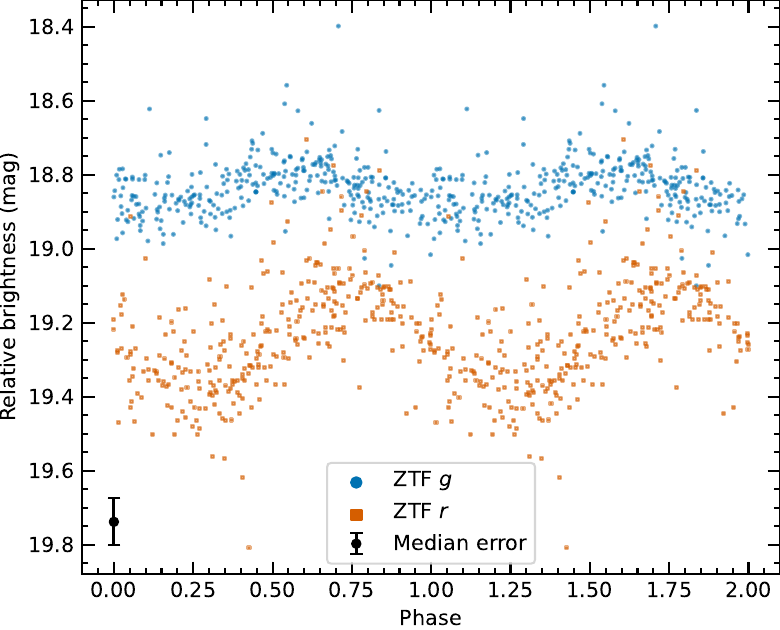} \vspace{6pt}\\
\includegraphics[width=\columnwidth]{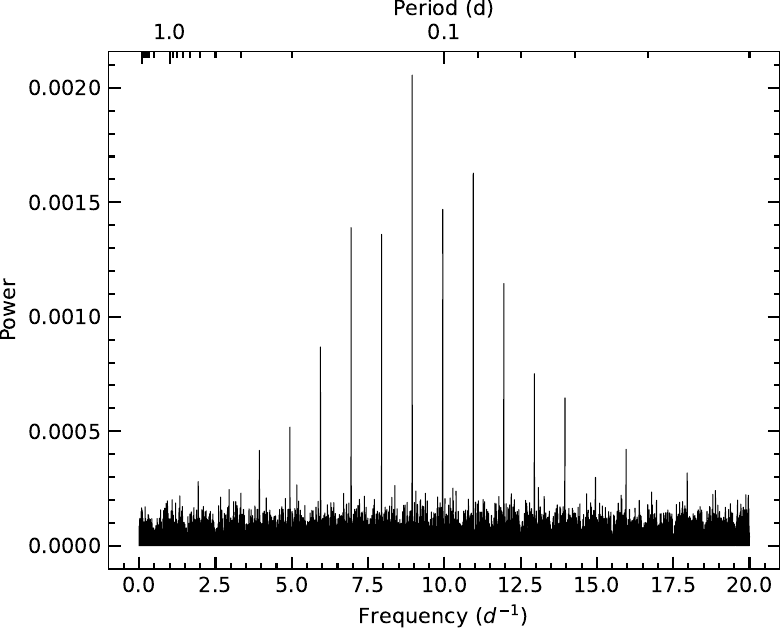}
\caption{\textit{Top:} Light curves for the binary system WD\,J132308.63+055900.97 in the ZTF \textit{g} and \textit{r} bands, folded to a period of $0.11178495\pm0.00000015$\,d. \textit{Bottom}: Discrete Fourier transform of the combined ZTF light curve.}
\label{figureCaIIZTF}
\end{figure}

\begin{figure*}
\centering
\includegraphics[width=0.49\textwidth]{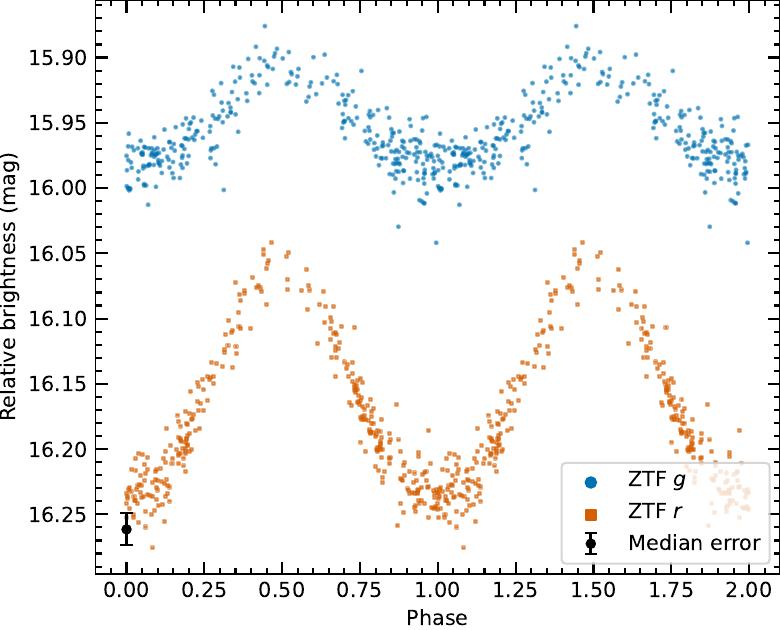}
\includegraphics[width=0.49\textwidth]{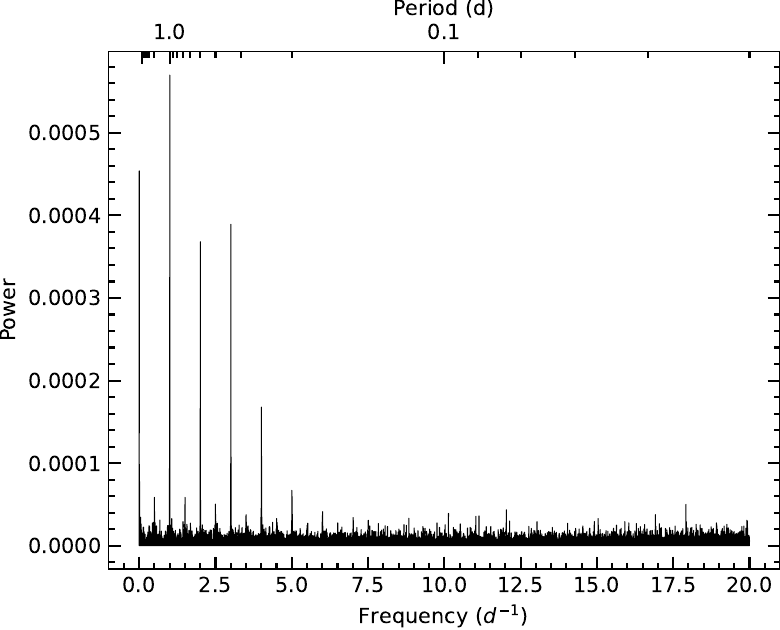} \vspace{6pt}\\
\includegraphics[width=\textwidth]{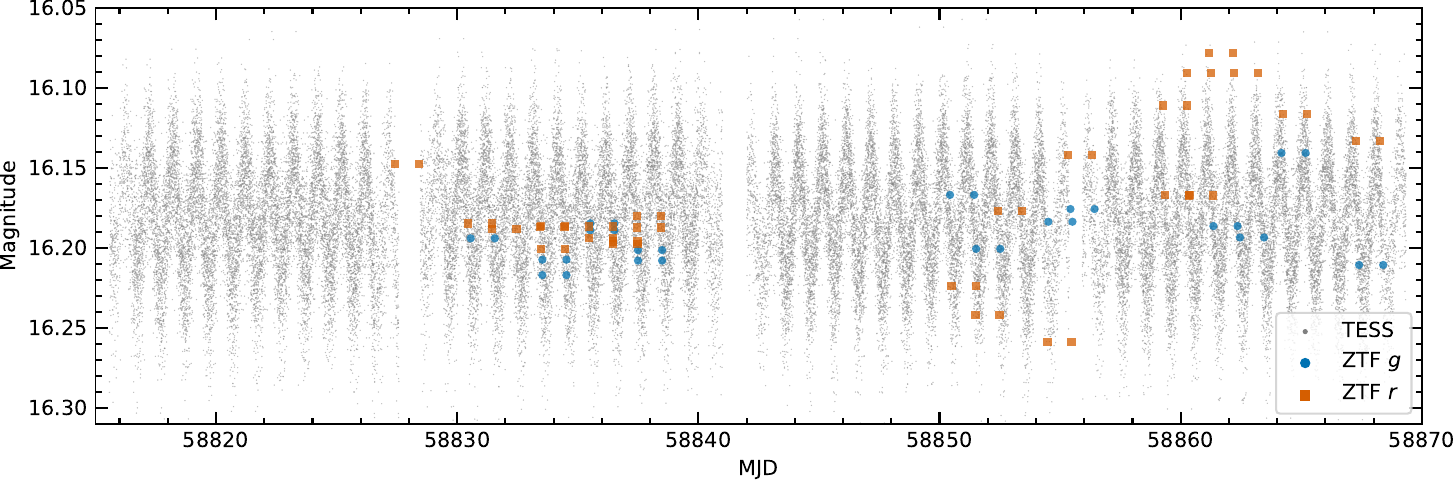} \vspace{6pt}\\
\caption{\edit{\textit{Top left:} Light curves for the for the likely post-common-envelope binary WD\,J094450.56+835531.29 in the ZTF \textit{g} and \textit{r} bands, folded to a period of  $0.9970029\pm0.0000083$\,d. \textit{Top right}: Discrete Fourier transform of the ZTF light curve. \textit{Bottom}: Light curve during TESS sectors 19 and 20, confirming that the periodic signal is genuine.}}
\label{figurePCEB0944}
\end{figure*}

\clearpage
\onecolumn
\section*{Affiliations}
$^{1}$Department of Physics, University of Warwick, Coventry CV4~7AL, UK\\
$^{2}$Institut f\"ur Theoretische Physik und Astrophysik, University of Kiel, 24098 Kiel, Germany\\
$^{3}$Astrophysics Group, Department of Physics, Imperial College London, Prince Consort Rd, London SW7~2AZ, UK\\
$^{4}$NOIRLab, 950 N Cherry Ave, Tucson, AZ, 85719, USA\\
$^{5}$Lawrence Berkeley National Laboratory, 1 Cyclotron Road, Berkeley, CA 94720, USA\\
$^{6}$Department of Physics, Boston University, 590 Commonwealth Avenue, Boston, MA 02215 USA\\
$^{7}$Departamento de Astrof\'{\i}sica, Universidad de La Laguna (ULL), E-38206, La Laguna, Tenerife, Spain\\
$^{8}$Instituto de Astrof\'{\i}sica de Canarias, C/ V\'{\i}a L\'{a}ctea, s/n, E-38205 La Laguna, Tenerife, Spain\\
$^{9}$Observat\'orio Nacional, Rio de Janeiro - RJ, 20921-400, Brasil\\
$^{10}$Dipartimento di Fisica ``Aldo Pontremoli'', Universit\`a degli Studi di Milano, Via Celoria 16, I-20133 Milano, Italy\\
$^{11}$INAF-Osservatorio Astronomico di Brera, Via Brera 28, 20122 Milano, Italy\\
$^{12}$Department of Physics \& Astronomy, University College London, Gower Street, London WC1E 6BT, UK\\
$^{13}$Institut d'Estudis Espacials de Catalunya (IEEC), c/ Esteve Terradas 1, Edifici RDIT, Campus PMT-UPC, 08860 Castelldefels, Spain\\
$^{14}$Institute of Space Sciences, ICE-CSIC, Campus UAB, Carrer de Can Magrans s/n, 08913 Bellaterra, Barcelona, Spain\\
$^{15}$Instituto de F\'{\i}sica, Universidad Nacional Aut\'{o}noma de M\'{e}xico,  Circuito de la Investigaci\'{o}n Cient\'{\i}fica, Ciudad Universitaria, Cd. de M\'{e}xico  C.~P.~04510,  M\'{e}xico\\
$^{16}$Instituci\'{o} Catalana de Recerca i Estudis Avan\c{c}ats, Passeig de Llu\'{\i}s Companys, 23, 08010 Barcelona, Spain\\
$^{17}$Institut de F\'{i}sica d'Altes Energies (IFAE), The Barcelona Institute of Science and Technology, Edifici Cn, Campus UAB, 08193, Bellaterra (Barcelona), Spain\\
$^{18}$Departamento de F\'isica, Universidad de los Andes, Cra. 1 No. 18A-10, Edificio Ip, CP 111711, Bogot\'a, Colombia\\
$^{19}$Observatorio Astron\'omico, Universidad de los Andes, Cra. 1 No. 18A-10, Edificio H, CP 111711 Bogot\'a, Colombia\\
$^{20}$Institute of Cosmology and Gravitation, University of Portsmouth, Dennis Sciama Building, Portsmouth PO1 3FX, UK\\
$^{21}$Department of Physics, Universit\`a degli Studi di Trieste, Via A. Valerio 2, I-34127 Trieste, Italy\\
$^{22}$University of Virginia, Department of Astronomy, Charlottesville, VA 22904, USA\\
$^{23}$Fermi National Accelerator Laboratory, PO Box 500, Batavia, IL 60510, USA\\
$^{24}$Center for Cosmology and AstroParticle Physics, The Ohio State University, 191 West Woodruff Avenue, Columbus, OH 43210, USA\\
$^{25}$Department of Physics, The Ohio State University, 191 West Woodruff Avenue, Columbus, OH 43210, USA\\
$^{26}$The Ohio State University, Columbus, 43210 OH, USA\\
$^{27}$Department of Physics, The University of Texas at Dallas, 800 W. Campbell Rd., Richardson, TX 75080, USA\\
$^{28}$Institute for Astronomy, University of Edinburgh, Royal Observatory, Blackford Hill, Edinburgh EH9 3HJ, UK\\
$^{29}$Institute of Astronomy, University of Cambridge, Madingley Road, Cambridge CB3 0HA, UK\\
$^{30}$Sorbonne Universit\'{e}, CNRS/IN2P3, Laboratoire de Physique Nucl\'{e}aire et de Hautes Energies (LPNHE), FR-75005 Paris, France\\
$^{31}$Department of Astronomy \& Astrophysics, University of Toronto, Toronto, ON M5S 3H4, Canada\\
$^{32}$Departament de F\'{i}sica, Serra H\'{u}nter, Universitat Aut\`{o}noma de Barcelona, 08193 Bellaterra (Barcelona), Spain\\
$^{33}$Department of Physics and Astronomy, Siena University, 515 Loudon Road, Loudonville, NY 12211, USA\\
$^{34}$Departament de F\'isica, EEBE, Universitat Polit\`ecnica de Catalunya, c/Eduard Maristany 10, 08930 Barcelona, Spain\\
$^{35}$Department of Physics and Astronomy, University of Waterloo, 200 University Ave W, Waterloo, ON N2L 3G1, Canada\\
$^{36}$Perimeter Institute for Theoretical Physics, 31 Caroline St. North, Waterloo, ON N2L 2Y5, Canada\\
$^{37}$Waterloo Centre for Astrophysics, University of Waterloo, 200 University Ave W, Waterloo, ON N2L 3G1, Canada\\
$^{38}$Instituto de Astrof\'{i}sica de Andaluc\'{i}a (CSIC), Glorieta de la Astronom\'{i}a, s/n, E-18008 Granada, Spain\\
$^{39}$Department of Physics and Astronomy, Sejong University, 209 Neungdong-ro, Gwangjin-gu, Seoul 05006, Republic of Korea\\
$^{40}$CIEMAT, Avenida Complutense 40, E-28040 Madrid, Spain\\
$^{41}$Department of Physics, University of Michigan, 450 Church Street, Ann Arbor, MI 48109, USA\\
$^{42}$University of Michigan, 500 S. State Street, Ann Arbor, MI 48109, USA\\
$^{43}$Department of Physics \& Astronomy, Ohio University, 139 University Terrace, Athens, OH 45701, USA\\
$^{44}$University of Michigan, 500 S. State Street, Ann Arbor, MI 48109, USA\\
$^{45}$National Astronomical Observatories, Chinese Academy of Sciences, A20 Datun Road, Chaoyang District, Beijing, 100101, P.~R.~China\\
\\


\bsp	
\label{lastpage}
\end{document}